\documentclass[twocolumn,tighten,twocolappendix]{aastex631} %
\usepackage{graphicx}
\usepackage{amssymb}
\usepackage{amsmath}
\usepackage{hyperref}
\usepackage{booktabs}
\usepackage{xcolor}
\usepackage{comment}
\usepackage{capt-of}

\def\gtrsim{\mathrel{\hbox{\rlap{\hbox{\lower4pt\hbox{$\sim$}}}\hbox{$>$}}}}
\def\lesssim{\mathrel{\hbox{\rlap{\hbox{\lower4pt\hbox{$\sim$}}}\hbox{$<$}}}}
\def\gtrsim{\mathrel{\hbox{\rlap{\hbox{\lower4pt\hbox{$\sim$}}}\hbox{$>$}}}}

\shorttitle{Classifying Chandra Sources using Machine Learning}
\shortauthors{Yang et al.}

\begin{document}

\title{Classifying Unidentified X-ray Sources in the Chandra Source Catalog Using a \\ Multiwavelength Machine-learning Approach}

\author{Hui Yang}
\affiliation{Department of Physics, The George Washington University, 725 21st St. NW, Washington, DC 20052, USA}
\email{huiyang@gwmail.gwu.edu}

\author{Jeremy Hare}
\affiliation{NASA Goddard Space Flight Center, Greenbelt, MD, 20771, USA}
\affiliation{NASA Postdoctoral Program Fellow}

\author{Oleg Kargaltsev}
\affiliation{Department of Physics, The George Washington University, 725 21st St. NW, Washington, DC 20052, USA}

\author{Igor Volkov}
\affiliation{Department of Physics, The George Washington University, 725 21st St. NW, Washington, DC 20052, USA}

\author{Steven Chen}
\affiliation{Department of Physics, The George Washington University, 725 21st St. NW, Washington, DC 20052, USA}

\author{Blagoy Rangelov}
\affiliation{Department of Physics, Texas State University, 601 University Drive, San Marcos, TX 78666, USA}

\begin{abstract}
The rapid increase in serendipitous X-ray source detections requires the development of novel approaches to efficiently explore the nature of X-ray sources. If even a fraction of these sources could be reliably classified, it would enable population studies for various astrophysical source types on a much larger scale than currently possible. Classification of large numbers of sources from multiple classes characterized by multiple properties (features) must be done automatically and supervised  machine learning (ML) seems to provide the only feasible approach. 
We perform classification of Chandra Source Catalog version 2.0 (CSCv2) sources to explore the potential of the ML approach and identify various biases, limitations, and bottlenecks that present themselves in these kinds of studies. We establish the framework and present a flexible and expandable Python pipeline, which can be used and improved by others. We also release the training data set of 2941 X-ray sources with confidently established classes. 
In addition to providing probabilistic classifications of 66,369 CSCv2 sources (21\% of the entire CSCv2 catalog), we perform several narrower-focused case studies (high-mass X-ray binary candidates and X-ray sources within the extent of the H.E.S.S. TeV sources) to demonstrate some possible applications of our ML approach. We also discuss future possible modifications of the presented pipeline, which are expected to lead to substantial improvements in classification confidences.
\end{abstract}

\keywords{Catalogs (205), X-ray sources (1822), Classification (1907), Random Forests (1935), X-ray binary stars (1811), Active galactic nuclei (16), X-ray stars (1823), Young stellar objects (1834), Cataclysmic variable stars (203), Astrostatistics tools (1887), X-ray surveys (1824), Compact objects (288)}

\section{Introduction} \label{sec:intro}

X-ray astrophysics is currently in an unprecedented era with observatories such as {\sl Chandra}, The X-ray Multi-Mirror
Mission ({\sl XMM-Newton}), and {\sl Swift} X-Ray Telescope (XRT), all viewing the sky.  Data are continuously being produced in large quantities, and this amount will continue to increase as more sensitive observatories begin functioning and/or reach their design sensitivity (e.g., {\sl eROSITA} has already detected roughly one million sources in its first all-sky survey). This has resulted in a substantial growth in the number of detected X-ray sources, the vast majority of which have been detected serendipitously. Consequently, most of these sources have not been studied or classified. This implies that for every observation there are large amounts of data not being utilized to their fullest potentials. Developing and testing methods that facilitate the automated classification of these sources is important because it will enable population studies (e.g., evolution, spatial distribution) with much larger samples. It will also help find remarkable outliers, which may represent new classes of high-energy objects or rare cases of already known objects that push current models to their limits. Additionally, higher energy observatories (e.g., {\sl Fermi}, H.E.S.S., HAWC) are also scanning the sky across many decades in energy and discovering tens of thousands of high-energy $\gamma$-ray sources. One strategy to understand the nature of these extreme particle accelerators is to explore the classifications of all X-ray sources in the extent of the $\gamma$-ray  source to  find a lower-wavelength counterpart. This is often difficult as the extent of the $\gamma$-ray sources are often several arcminutes or more in size and can contain many potential X-ray counterparts.

Traditional multiwavelength (MW) classification methods typically rely on examining various two-parameter plots in the MW parameter space (e.g., color-color diagrams; color-magnitude diagrams, hereafter CMDs; X-ray hardness ratios, and hardness ratios are hereafter HRs) to conceive simple criteria (e.g., a single dividing line) for differentiating between source types \citep[e.g.,][]{2006ApJS..163..344K,2010ApJ...725..931M}. 
Besides being very time consuming and tedious, the traditional classification approach does not provide any prescription on how to assign the confidence in these classifications. 
One solution to these problems is to develop an efficient, automated classification of a large number of astronomical sources using machine-learning (ML) methods. 
In relation to X-ray sources, the ML approach was pioneered by \cite{2004ApJ...616.1284M}, who applied a supervised ML algorithm known as oblique decision trees \citep{1994cs........8103M} to classify $\sim$80,000 sources from the {\sl ROSAT} all sky survey (RASS) into six distinct classes, i.e., stars, white dwarfs, X-ray binaries (XRBs), galaxies, active galactic nuclei (AGNs), and galaxy clusters. In addition to the RASS catalog \citep{1999A&A...349..389V}, the ML algorithm used optical \citep[USNO-B;][]{2003AJ....125..984M} and radio data   \citep[SUMMS and NVSS;][]{2007MNRAS.382..382M,1998AJ....115.1693C} to define nine source attributes (positions, X-ray fluxes, two HRs, source extent, $B$ and $R$ magnitudes, and radio counterpart flags). However, the limited positional accuracy of {\sl ROSAT} sources ($\sim$10$\arcsec$--30$\arcsec$) resulted in a large degree of confusion with only $\sim$50\% of optical associations being true counterparts in the Galactic plane.

Recently, \cite{2014ApJ...786...20L}, \cite{2015ApJ...813...28F} used the random forest \citep[RF;][]{BreimanML} ML algorithm to classify 411 and 2876 variable sources in the {\sl XMM-Newton} source serendipitous  catalogs (the second XMM-Newton serendipitous source catalog Data Release (DR) 2; \citeauthor{2009A&A...493..339W} \citeyear{2009A&A...493..339W} and the third XMM-Newton serendipitous source catalog DR4; \citeauthor{2015ASPC..495..319R} \citeyear{2015ASPC..495..319R}) achieving an overall classification accuracy of 92$\%$. 
The sources were classified into seven different classes including AGNs, cataclysmic variables (CVs), XRBs, stars, gamma-ray bursts (GRBs), super soft sources, and ultraluminous X-ray sources (ULXs), all known to be variable, using a training data set (TD) consisting of $\sim$870 sources. 
The other recent large-scale study by \cite{2022AA...657A.138T} used a more interpretable ML algorithm, known as the naive Bayes classifier \citep{2001ISR...69...385}, to classify 315,378 sources (55\% of the whole fourth XMM-Newton serendipitous source DR10 catalog, hereafter 4XMM-DR10). The authors defined four classes (AGNs, stars, XRBs, and CVs), built a large ($\sim$25,000 sources) TD, and achieved good performance with $>90\%$ of AGNs, stars, and XRBs expected to be accurately classified, while only 34\% of CVs were classified correctly. 
\cite{2022AA...657A.138T} also tried an RF algorithm and found that it performed equally well, but they preferred the naive Bayes classifier as it provides a more straightforward interpretation of each classification. A similar (naive Bayes) approach was implemented earlier by \cite{2011ApJS..194....4B} in a narrower study to classify X-ray sources from the Chandra Carina Complex Project and to identify young stellar objects (YSOs) belonging to a starburst region.    
\cite{2021MNRAS.503.5263Z} used various ML methods to classify X-ray sources from the fourth XMM-Newton serendipitous source catalog DR9 into three broad classes (quasars, galaxies and stars). 
ML methods have also been used to classify X-ray sources in a few interesting but small fields, using the {\sl Chandra} \citep{2020MNRAS.492.5075A} and {\sl XMM-Newton} data \citep{2016ApJ...821...54S,2016ApJ...816...52H,2017ApJ...841...81H,2020ApJ...901..157K}. 

X-ray sources detected by {\sl Chandra} during its first $\sim$15 yr of the mission (i.e., before the end of 2014) are compiled in the Chandra Source Catalog \citep[currently version 2.0, hereafter CSCv2;][]{2010ApJS..189...37E,2020AAS...23515405E}. The CSCv2 contains $\sim$317,000 unique sources and covers $\sim$550\,deg$^2$ of the sky down to fluxes as low as 10$^{-18}$\,erg\,s$^{-1}$\,cm$^{-2}$ in the 0.5--7.0\,keV band for the deepest fields. 
It also provides a large number of properties for each source (e.g., positions, fluxes in multiple bands, HRs, variability properties). 
The sub-arcsecond {\sl Chandra} source localizations greatly reduce the degree of confusion with the MW counterparts in the crowded regions of the Galactic plane, compared to the several arcsecond localizations of {\sl Swift-XRT} and {\sl XMM-Newton} EPIC. To date there was no published attempt to classify all or a large fraction of CSCv2 sources.

In this paper, we present the automated MUlti-Wavelength CLASSification pipeline (MUWCLASS) of the X-ray sources and a couple of its potential applications. 
Section \ref{sec:Data}  discusses our TD, consisting of eight source classes that are commonly detected in X-rays, as well as the X-ray and MW properties extracted for the X-ray sources. 
Section \ref{sec:methods} describes the pipeline workflow,  the data processing, and the  choice of the algorithm. We also detail the approach we use to account for the  uncertainties of various measured source properties. 
In Section \ref{sec:algorithm_performance}, we discuss the optimization of the chosen ML algorithm and MUWCLASS's performance. We then apply the pipeline to classify 66,369 well-characterized CSCv2 sources  
in Section \ref{sec:results} and discuss the classification outcomes as well as several related consistency tests. 
In Section \ref{sec:examples}, we explore several  high-mass X-ray binary (HMXB) candidates and a number of  interesting X-ray sources within the extent of unidentified TeV sources  from the H.E.S.S. Galactic plane survey (HGPS). Finally, Section \ref{sec:discuss} discusses  current  limitations and future developments. We conclude with a summary in Section \ref{sec:summary}.

\section{Data}
\label{sec:Data}

\subsection{Training Data Set (TD)}
\label{sec:TDCats}

Supervised ML relies on a TD to train (fit) a classification model. All sources in the TD must have known, confidently assigned labels (classes). To ensure the reliability of the classifications used in the TD, we only used sources from peer-reviewed publications that were confidently classified by detecting a feature (or a set of features), which is unique to a particular kind of source (e.g., the redshift for an AGN, the pulsation period for a pulsar, the orbital period, and the donor star type for an XRB, the associated star-forming region (SFR), and the infrared (IR) excess for a YSO).  
We have compiled a TD of 2941 literature-verified sources\footnote{These are sources with well-established robust classifications based on traditional, non-ML-based methods, and they have been studied extensively in the literature.} belonging to eight classes:
AGNs\footnote{These include quasars, AGNs and BL Lac objects.}, CVs, high-mass stars (HM-STARs)\footnote{These include Wolf-Rayet, O, B stars.}, HMXBs, low-mass stars (LM-STARs), low-mass X-ray binaries (LMXBs)\footnote{This class also includes nonaccreting X-ray binaries, such as wide-orbit binaries with millisecond pulsars, as well as red-back and black widow systems.}, pulsars and isolated neutron stars (NSs)\footnote{This class includes 11 magnetars.}, and YSOs of various kinds. Admittedly, the definitions of the classes are broad, and some of the classes are rather heterogeneous (i.e., include sources with quite different properties). However, we have tried other class definitions and found these to be a reasonable compromise between having an astrophysically meaningful class and having a large enough number of confidently identified members that we could assign to each of the classes. Using more detailed classes would result in a very small number of sources in some classes, which would lead to poor performance for that class.

\begin{deluxetable}{l c}
\tablecaption{Training Data Set  Breakdown by Source Class\label{tab:tdsourcebreak}}
\tablewidth{0pt}
\tablehead{
\colhead{Source Type} & \colhead{Number of CSCv2 sources}
}
\startdata
Active galactic nucleus (AGN) & 1390 \\
Cataclysmic variable (CV) & 44 \\
High-mass star (HM-STAR) &  118 \\
High-mass X-ray binary (HMXB) &  26 \\
Low-mass star (LM-STAR) &  207 \\
Low-mass X-ray binary (LMXB) &  65 \\
Pulsar and isolated neutron star (NS) & 87\\
Young stellar object (YSO) &  1004 \\
Total & 2941 \\
\enddata
\end{deluxetable}

To construct the TD we first select catalogs for each source type and then crossmatch sources from those catalogs with CSCv2 using a circular region with $3\arcsec$ radius (which we later reduce; see below). 
The selected catalogs are as follows: 

\begin{enumerate}
\item AGNs from Veron Catalog of Quasars $\&$ AGN \citep[thirteenth edition;][]{2010A&A...518A..10V};
\item CVs from Cataclysmic Variables Catalog \citep[2006 edition;][]{2001PASP..113..764D};
\item stars from Catalog of Stellar Spectral Classifications \citep{2014yCat....1.2023S}\footnote{We remove faint sources with magnitudes $>$23, and Orion type stars to avoid mixing of LM-STARs and YSOs, and sources with ambiguous information in the SpType or remark columns of this catalog.} with O, B or W (e.g., WN, WR stars) types are labeled as HM-STARs and A, F, G, K, or M types are labeled as LM-STARs; 
\item LM-STARs from the APOGEE-2 data in Sloan Digital
Sky Survey (SDSS) DR16 \citep{2020AJ....160..120J}\footnote{We also  remove stars that lack reliable effective temperature or surface gravity measurements or show evidence of binary with VSCATTER $>$ 1 km\,s$^{-1}$ and/or VSCATTER $>$ 5 VERR\_MED or are not flagged as a star based on Washington/DDO 51 photometry.};
\item HM-STARs from the VIIth Catalog of Galactic Wolf-Rayet Stars \citep{2001NewAR..45..135V} and its annex catalog \citep{2006A&A...458..453V};
\item HMXBs from the Catalog of HMXBs in the Galaxy \citep[fourth edition;][]{2006A&A...455.1165L};
\item LMXBs from the Low-Mass X-ray Binary Catalog \citep{2007A&A...469..807L};
\item CVs and LMXBs from Catalog of CVs, LMXBs and related objects \citep[seventh edition;][]{2003A&A...404..301R};
\item NSs and nonaccreting XRBs from ATNF Pulsar Catalog \citep{2005AJ....129.1993M}; 
\item YSOs from multiple molecular clouds and open clusters \citep{2005A&A...429..963O,2007A&A...463..275G,2011A&A...531A.141D,2011ApJS..194...14P,2011ApJS..196....4R,2012AJ....144..192M};
\item LMXBs, HMXBs and CVs from the INTEGRAL General Reference Catalog (version 43) where LMXBs and HMXBs have been confirmed by \cite{2020NewAR..8801536S}, \cite{2015A&ARv..23....2W};
\item HMXBs from the BeSS catalog with their SIMBAD \citep{2000A&AS..143....9W} types classified as HMXBs. 
\end{enumerate}

Sources from populous classes (AGNs, HM-STARs, LM-STARs and YSOs) are omitted if their class-specific catalog and X-ray combined 2$\sigma$ positional uncertainties\footnote{The X-ray positional uncertainties are approximated as circles with the radius equal to the semimajor axis of the 2$\sigma$ error ellipse in the CSCv2.} (PUs) are $>1\arcsec$ or if the separations of the class-specific catalog and the CSCv2 coordinates exceed the 2$\sigma$ PUs. 
There are several cases within underpopulated classes (CVs, HMXBs, LMXBs, and NSs) where the CSCv2 positions of the sources are offset by $>1\arcsec$ from their class-specific catalog coordinates, likely due to poor absolute astrometry, limited angular resolution of the instrument used in the class-specific catalog, or large proper motion. 
For some of these sources, we manually confirm the classifications and matches by reviewing the literature (besides the catalog itself) and/or by inspecting the X-ray and MW images. If the associations are deemed to be credible, we add them to our TD. 
Next, the CSCv2 coordinates of the sources remaining after the above-described vetting procedure are matched to SIMBAD \citep{2000A&AS..143....9W}, and sources with classifications conflicting with the main SIMBAD class are omitted from the TD (unless a mistake in the SIMBAD class is obvious from looking at the original publications), while keeping those stars that are classified as Orion variable, T Tauri star, or YSOs from SIMBAD as YSOs.  Sources that are classified as candidates in the peer-reviewed publications and/or SIMBAD are also omitted. 
We also omit sources from our TD residing in some crowded environments such as globular clusters, the Large Magellanic Cloud (LMC), the Small Magellanic Cloud (SMC), and the Galactic center as well as sources strongly affected by complex diffuse emission around them, e.g., sources within bright pulsar wind nebulae (PWNe) or supernova remnants (SNRs). 
Finally, we omit stars (i.e., HM-STARs, LM-STARs, and YSOs) if they have no crossmatched MW counterpart  (see Section \ref{sec:MWdata}). The final content of the TD is summarized in Table \ref{tab:tdsourcebreak}. 
We note that our TD is not all-encompassing and its scope is limited by the time and efforts we could allocate for its creation. It is certainly possible to find reliable classifications for more CSCv2 sources in the published literature. We will continue updating our TD, and we hope, in future, to turn it into a community-driven effort with the web-based open database of classified X-ray sources.  

For each source, our pipeline extracts and calculates up to a total of 29 MW features (i.e., attributes or parameters to be used by an ML algorithm) with X-ray features described in Section \ref{sec:Xdata}, and MW features described in Section \ref{sec:MWdata}. All of the features used in our TD can be found in Table \ref{table:MWfeat}. 
We provide the Python-Jupyter notebook in the GitHub reporitory\footnote{\url{https://github.com/huiyang-astro/MUWCLASS_CSCv2}} which has more details and  can be used to reconstruct the TD from scratch. The TD is also available in the electronic (machine-readable) format, and a subset of the whole TD is shown in Table \ref{tab:TDMRT} (see Appendix \ref{sec:MRTs}). 

\begin{figure*}
\begin{center}
\hspace*{-0.9cm} 
\includegraphics[width=350pt,trim=0 0 0 0]{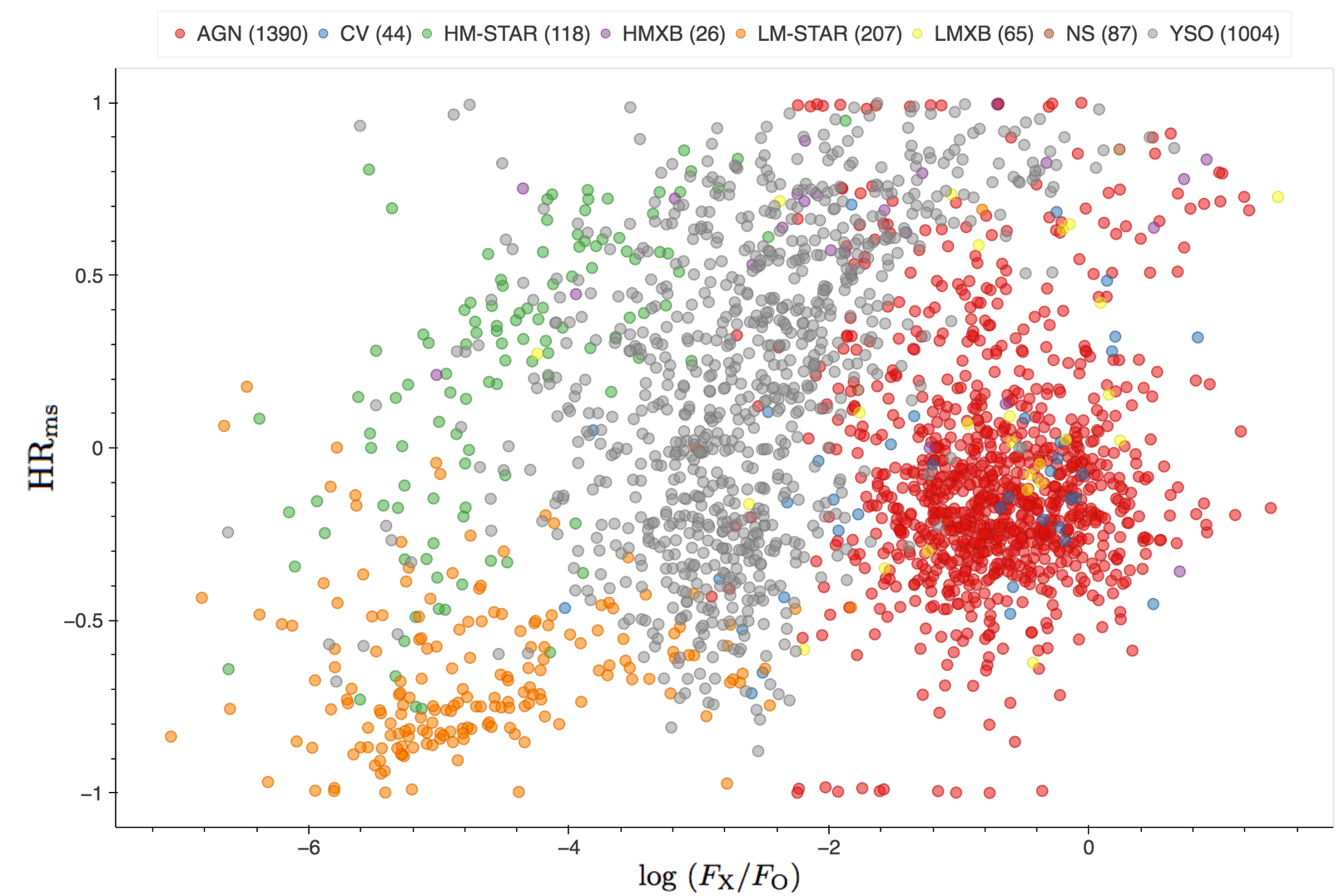}
\includegraphics[width=350pt,trim=0 0 0 0]{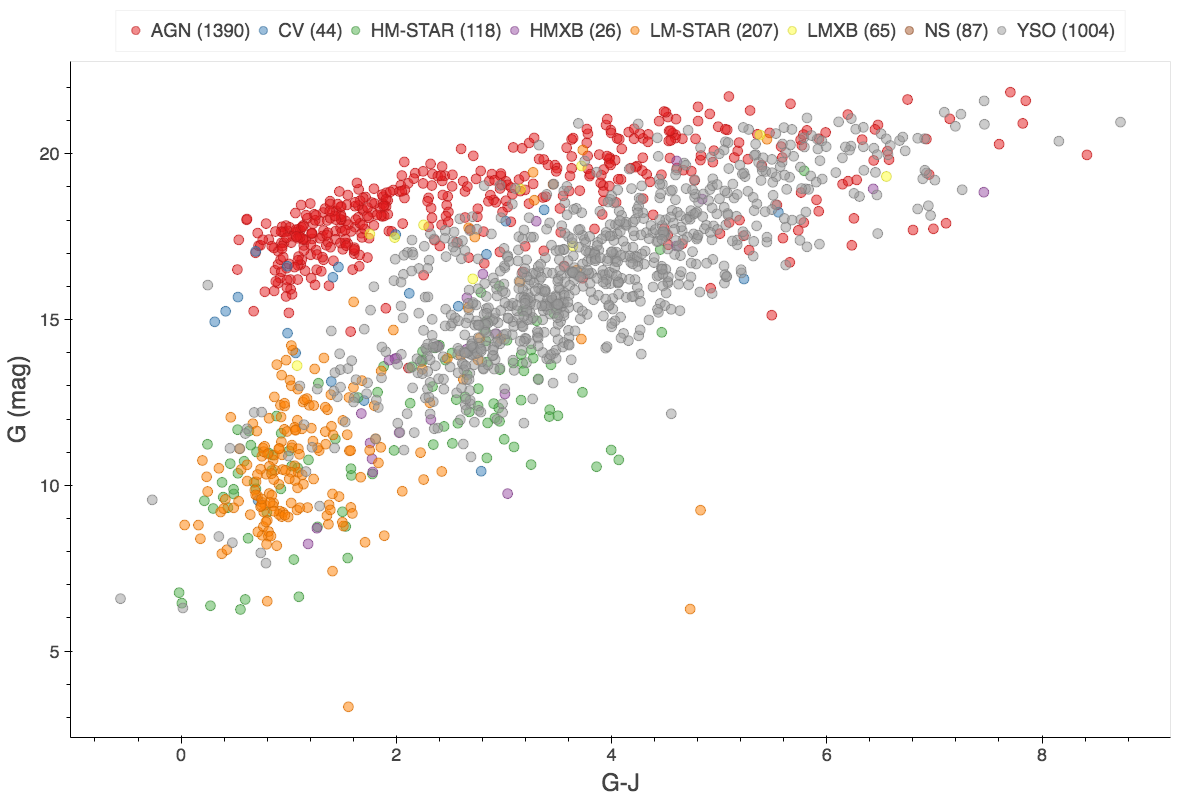}
\caption{
Examples of 2D slices of the multidimensional feature space  showing the TD content (see also \citealt{2021RNAAS...5..102Y}). The clustering of sources belonging to certain classes (see the legend on top of the plots) makes these  features a popular choice for non-ML classification. Note, however, that not all classes separate well with just these features, and there is a substantial overlap between the sources from different classes.}
\label{fig:2paramplot}
\end{center}
\end{figure*}

We have also developed an interactive web-based plotting tool to visualize the TD's content through various 2D slices of the multidimensional feature space \citep{2021RNAAS...5..102Y}\footnote{The interactive plots are available at \url{https://home.gwu.edu/~kargaltsev/XCLASS/}.}. 
Two examples of such plots, shown in Figure \ref{fig:2paramplot}, demonstrate a good degree of separation between sources of some classes for the choice of features shown in the plots. 
The top panel shows the plot of the X-ray HR (see its definition in Section \ref{sec:Xdata}) versus the X-ray to optical flux ratio, $\log (F_{\rm X}/F_{\rm O})$, which is often used in traditional classification methods. The X-ray to optical flux ratio is calculated by dividing the broadband X-ray flux $F_{\rm b}$ in the $0.5-7$\,keV energy range by the {\sl Gaia} $G$-band flux (see the conversion of the $G$-band magnitude to energy flux in Section \ref{sec:standard}). Note, however, that AGNs in the TD come from high-latitude surveys and, hence, are weakly extincted or absorbed by the intervening interstellar medium. If viewed through the Galactic plane, the AGNs would not show such a good degree of separation from other classes (e.g., YSOs).

\begin{deluxetable}{ll}
\tablecaption{List of MW Features Used for Classification\label{table:MWfeat}}
\tablewidth{0pt}
\tablehead{
\colhead{Feature} & \colhead{Description} 
}
\startdata
$F_{\rm s}$ & Flux in the 0.5$-$1.2 keV band\\
$F_{\rm m}$ & Flux in the 1.2$-$2 keV band\\
$F_{\rm h}$& Flux in the 2$-$7 keV band\\
$F_{\rm b}$ & Flux in the 0.5$-$7 keV band\\
HR$_{\rm ms}$ & Medium-soft hardness ratio $(F_{\rm m}-F_{\rm s})/(F_{\rm m}+F_{\rm s})$  \\
HR$_{\rm hm}$ & Hard-medium hardness ratio $(F_{\rm h}-F_{\rm m})/(F_{\rm h}+F_{\rm m})$ \\
HR$_{\rm h(ms)}$ & Combined hardness ratio $(F_{\rm h}-F_{\rm m}-F_{\rm s})/(F_{\rm h}+F_{\rm m}+F_{\rm s})$\\
$P_{\rm inter}$ & Inter-observation variability probability \\
$P_{\rm intra}$ & Intra-observation variability probability \\
\hline
G & Gaia EDR3 G-band magnitude \\
BP & Gaia EDR3 BP-band magnitude\\
RP & Gaia EDR3 RP-band magnitude\\
J & 2MASS J-band magnitude \\
H & 2MASS H-band magnitude \\
K & 2MASS K-band magnitude \\
W1 & WISE W1-band magnitude \\
W2 & WISE W2-band magnitude \\
\hline
G--BP & G--BP color\\
G--RP & G--RP color\\
G--J & G--J color\\
G--H & G--H color\\
G--K & G--K color\\
BP--RP & BP--RP color\\
J--H & J--H color\\
J--K & J--K color\\
H--K & H--K color\\
W1--W2 & W1--W2 color\\
W1--W3 & W1--W3 color\\
W2--W3 & W2--W3 color\\
\enddata
\tablecomments{The three sections of the table are the X-ray properties based on the CSCv2 followed by MW properties from {\sl Gaia} EDR3, 2MASS, and several WISE surveys (see text for details), and the important colors. W3-band magnitude is dropped after the feature selection (see Section \ref{sec:feature-selection} for details).}
\end{deluxetable}

\subsection{X-Ray Features}
\label{sec:Xdata}

After crossmatching the TD of literature-verified sources to the CSCv2 and applying a few cleaning steps (see Section \ref{sec:TDCats}), we extract the {\em per-observation} information from CSCv2 for all TD sources. This allows us to use sources with missing master-level\footnote{In the CSCv2, master-level products combine information from multiple {\sl Chandra} observations, if they are available.} fluxes and to  calculate an additional (to the CSCv2) inter-observation variability metric.

Detections with off-axis angles $>$10$\arcmin$ are dropped to reduce the degree of confusion during MW crossmatching (see Section \ref{sec:MWdata}), since these sources have larger PUs due to the larger and asymmetric off-axis point-spread function (PSF) of {\sl Chandra}. 
The source detections are filtered based on the per-observation source flags to avoid pileup\footnote{Detections with {\tt pileup\_warning} $>$ 0.3 counts\,frame$^{-1}$\,pixel$^{-1}$ are dropped.}, saturation, and readout streak contamination. 

For each detection, the CSCv2 provides the mode ($F_{\rm mode}$), as well as the lower and upper limits at 1$\sigma$ confidence ($F_{\rm lo}$ and $F_{\rm hi}$) of the X-ray flux distributions at the soft (0.5--1.2 \,keV), medium (1.2--2\,keV), and hard (2--7\,keV) bands. 
We assume the flux distribution to be the Fechner distribution, also known as the split normal distribution, consisting of two half-normal distributions with the same mode. 
We calculate the mean, and the variance of the Fechner distribution with the equations from \citet{2019Metro..56d5009P}: 

\begin{subequations}
\begin{align}
\mu = & F_{\rm mode} +  \sqrt{2/\pi} (F_{\rm hi} +  F_{\rm lo} - 2 F_{\rm mode}) \\
\sigma^2 =  & (1-2/\pi) (F_{\rm hi} +  F_{\rm lo} - 2 F_{\rm mode})^2 \nonumber\\ 
  & +  (F_{\rm hi} - F_{\rm mode})(F_{\rm mode} - F_{\rm lo})
\end{align}
\label{eq:CSC-2}
\end{subequations} 
for the soft, medium, and hard band,  respectively. 
Next we calculate the mean and variance of the broadband (0.5--7 keV) flux by combining the soft, medium, and hard bands: 

\begin{subequations}
\begin{align}
\mu_{\rm b} & = \mu_{\rm s} + \mu_{\rm m} + \mu_{\rm h} \\
\sigma_{\rm b}^2 & = \sigma_{\rm s}^2  + \sigma_{\rm m}^2 + \sigma_{\rm h}^2
\end{align}
\label{eq:CSC-3}
\end{subequations} 
where the subscripts $b$, $s$, $m$, and $h$ indicate the broad, soft, medium, and hard band respectively. 

The weighted average flux for multiple observations (whenever available) can be then  expressed as

\begin{subequations}
\begin{align}
\overline{\mu}  & =\frac{\sum_{i=1}^N (\mu_i / \sigma_i^2)}{\sum_{i=1}^N (1 / \sigma_i^2)} \\
\overline{\sigma}^2 & =   \frac{1}{\sum_{i=1}^N (1 / \sigma_i^2)} 
\end{align}
\label{eq:CSC-4}
\end{subequations} 
for the broad, soft, medium, and hard bands where $i$ indexes multiple observations of the same source, and $N$ is the number of observations available. 
If there is only one observation for a source, then the weighted average of the mean flux and the variance of the mean  will be equal to the mean and the variance of the flux distribution of that single observation. 
If the weighted average flux and its variance for a band are having null values, a mode of 0 and an upper limit of $10^{-17}$\,erg\,s$^{-1}$\,cm$^{-2}$ are used to replace them with a mean and a variance calculated from Equation \ref{eq:CSC-2}.
Sources with all band fluxes having null values are dropped. From the fluxes, we calculate three HRs, which are HR$_{\rm ms}$, HR$_{\rm hm}$, and HR$_{\rm h(ms)}$ respectively with their definitions in Table \ref{table:MWfeat}.

We also calculate the inter-observation variability probability parameter from the cumulative probability distribution of the chi-square statistic by fitting a constant model to broadband fluxes from multiple detections of the same source:

\begin{subequations}
\begin{align}
\chi^2 & =  \sum_{i=1}^N \frac{(\mu_{{\rm b,} i} - \overline{\mu_{{\rm b}}} )^2}{\sigma_{{\rm b,} i}^2} \\
P_{\rm inter} & = \int_0^{\chi^2} \frac{x^{\nu/2-1} \exp (-x/2)}{2^{\nu/2} \Gamma(\nu/2)}  d x.
\end{align}
\label{eq:CSC-6}
\end{subequations} 
Here $\nu=N-1$ is the number of degrees of freedom, $\mu_{{\rm b,} i}$ is the per-observation broadband mean flux, $\overline{\mu_{{\rm b}}}$ is the weighted average of the broadband mean flux, and $\sigma_{{\rm b,} i}^2$ is the variance of the broadband flux of each observation.

For the intra-observation variability parameter, we adopt the highest value of Kuiper's test probability of variability across all observations available in the CSCv2.

We also note that, with the help of an improved method for Markov Chain Monte Carlo (MCMC) sampling for CSCv2 aperture photometry (CSC team member Rafael Martinez-Galarza, private communication), a substantial fraction of fluxes missing (i.e., having ``Null'' values) in the current CSCv2 release (due to the lack of convergence in the MCMC calculation) have been recovered.

\subsection{MW Data}
\label{sec:MWdata}

We crossmatch the master-level CSCv2 source coordinates to the MW catalogs including {\sl Gaia} EDR3 \citep{2018A&A...616A...1G}, {\sl Gaia} EDR3 Distances \citep{2021AJ....161..147B},  Two Micron All-Sky Survey  (2MASS; \citealt{2006AJ....131.1163S}), AllWISE \citep{2014yCat.2328....0C}, CatWISE2020 \citep{2021ApJS..253....8M}, and unWISE \citep{2019ApJS..240...30S}. 
Initially, we use a large search radius of 10$\arcsec$ not to miss any possible counterparts (this also enables the estimation of local field source density, see Appendix \ref{sec:chance-match}). 
If a source has more than one MW match located within the $10''$ radius, we consider only the nearest match. 
Next, for each remaining match (potential counterpart), we co-add (in quadrature) the PUs from the CSCv2 source and the MW counterpart\footnote{The semi-major axis is taken as the uncertainty if the PU is given as an error ellipse. It is converted to a 2$\sigma$ level under the assumption of Gaussian distribution.}. 
Since the proper motion is not accounted for in the CSCv2, uncertainties of X-ray positions of fast moving sources with multiple observations in the CSCv2 may be underestimated. 
The combined 2$\sigma$ PU circle radius is used to filter out any MW matches that lie outside of it.

To search for the optical counterparts, we use the {\sl Gaia} EDR3 catalog. If the counterpart is found, we extract {\sl Gaia}'s  $G$-, $BP$-, and $RP$-band magnitudes and add them to the MW features to be used in the X-ray source classification. 
The {\sl Gaia} EDR3 catalog, complete down to $G=21$, provides an all-sky coverage together with an excellent positional accuracy of around 0.5 mas at $G=20$. 

In the near-infrared (NIR), the 2MASS's $J$, $H$, and $K$ magnitudes are used. The corresponding limiting magnitudes are $J=$15.8, $H=$15.1, and $K=$14.3 at 10$\sigma$ detection level, with slightly less sensitive limits (by $\sim$1 mag) in the Galactic plane due to confusion between sources \citep{2006AJ....131.1163S}.

In the IR, we use the W1, W2, and W3 bands from AllWISE, CatWISE2020, and unWISE catalogs. The 90\% completeness depth is achieved at W1=17.7 and W2=17.5 for the CatWISE2020 catalog, and AllWISE achieves  signal-to-noise ratio (S/N)=5 with flux at 54, 71, 730, and 5000 mJy (16.9, 16.0, 11.5 and 8.0 mag) in W1, W2, W3 and W4, respectively.
 
We do not use the W4 band due to the shallower depth of this band along with the larger PUs (in comparison with the W1 band\footnote{see Table 1 at \url{http://wise2.ipac.caltech.edu/docs/release/allwise/expsup/sec2_5.html}}), which could lead to the increased confusion between the IR and CSCv2 sources. 
For W1- and W2-band magnitudes, which are available from all three WISE catalogs, we only use the CatWISE2020 catalog when the magnitudes are missing from the AllWISE catalog and the unWISE when both AllWISE and CatWISE2020 are lacking the magnitude measurements. W3-band magnitudes are only available from the AllWISE catalog. 
We note that the W3-band magnitude is dropped after the feature selection while two colors involving the W3 band are selected as important features used for the classification pipeline (see Section \ref{sec:feature-selection} for details).

\section{Classification Methods and Procedures}
\label{sec:methods}

\subsection{MUWCLASS Pipeline Workflow}
\label{sec:workflowchart}

\begin{figure*}
\begin{center}
\hspace*{-0.9cm} 
\includegraphics[width=300pt,trim=0 0 0 0]{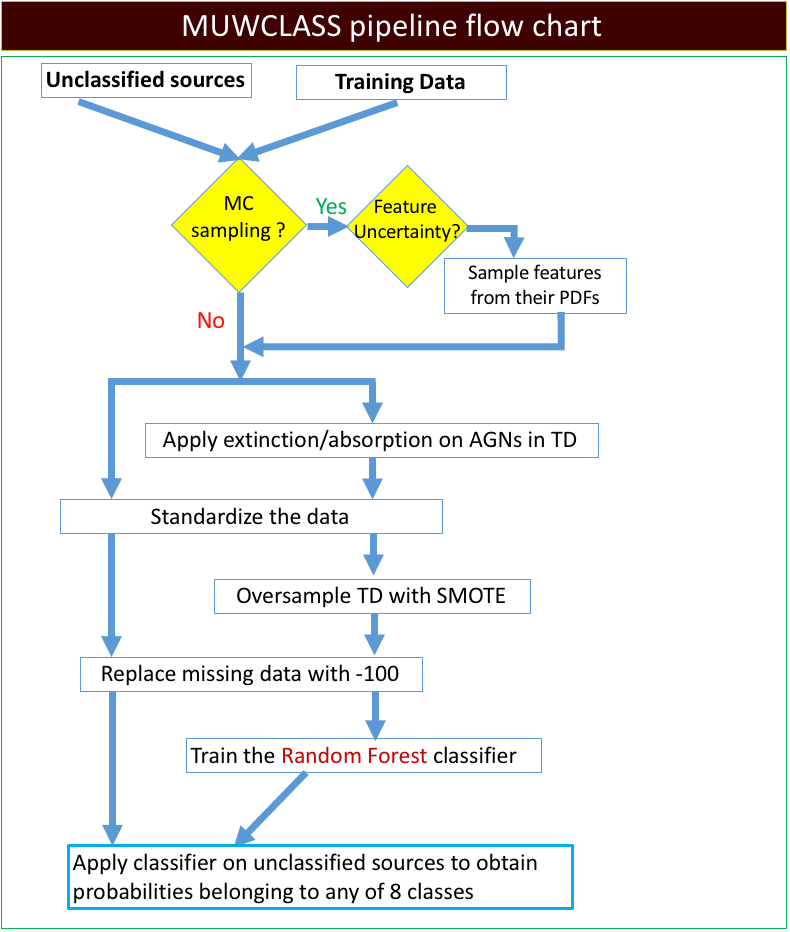}
\caption{Workflow chart of MUWCLASS pipeline.}
\label{fig:flowchart}
\end{center}
\end{figure*}

The MUWCLASS (see the workflow chart in Figure  \ref{fig:flowchart}) is designed to handle simultaneously the TD and the unclassified data. 
This allows us to account for the measurement uncertainties of various features (see Table \ref{table:MWfeat}) by randomly sampling the feature values (Monte Carlo (MC) sampling  box in Figure \ref{fig:flowchart}) from the assumed probability distribution function (PDF) of each feature (see Section \ref{sec:mc-sampling}). 
In addition, for each unclassified source, the direction-specific (to the source) reddening (absorption and extinction) is applied to all AGNs from the TD. This is done because an AGN outside of the Galactic plane will appear very different from the same AGN viewed through the Galactic plane (see Section \ref{sec:reddening}), and nearly all AGNs in the TD are located off the plane. 
MUWCLASS also calculates the derived features (including HRs and colors) after applying the reddening to AGNs, and then standardizes the data to provide a scaled distance metric across each feature (see Section \ref{sec:standard}). 
To mitigate the imbalance between different classes in the TD, an oversampling algorithm is applied (see Section \ref{sec:imbal}). 
The missing data are replaced with a large negative flag value (see Section \ref{sec:missingdata}) before passing them to the RF classifier (see Section \ref{sec:algorithm}).

\subsection{Monte Carlo (MC) Sampling}
\label{sec:mc-sampling}

The MC method is used to account for the uncertainties of the feature measurements by repeatedly randomly sampling the feature values from their PDFs. 
This method has also been introduced recently in \citet{2022AJ....164....6S}, and we incorporate the suggested factor of $\sqrt{2}$ to account for the underestimated measurement uncertainties. 
This is done for all features that have their uncertainties available and both for TD sources and  sources to be classified. 
With many samplings, we obtain a large number of classification results for each unclassified source, accounting for the uncertainties in both the TD and unclassified source's features.  
We then calculate the PDFs of the classification outcomes (i.e., vectors with probabilities for a source to belong to each of the predefined classes). Thus, the output of the classification is a PDF of probabilities of each class as opposed to a single probability obtained from traditional RF classification without accounting for uncertainties. 

For each unclassified source, we run 1000 times MC samplings such that a reasonable convergence of the classification can be achieved (see Appendix \ref{appendix:convergence-uncertainty}). It is extremely computational expensive, which we try to mitigate with the help of The George Washington University high-performance computing cluster \citep[PEGASUS;][]{GWUHPCCluster}.

\subsection{Absorption--Extinction on AGNs}
\label{sec:reddening}

Most ($> 99\%$) of the AGNs in our TD are located away from the Galactic plane ($|b|>10^\circ$), because surveys for AGNs are typically conducted outside of the plane, where the absorption and extinction are much lower than in the Galactic plane. This can potentially bias classification results because AGNs observed through the plane will look different from those in the TD, i.e., they will be dimmer, redder in optical and NIR, and have harder X-ray spectra.
In order to compensate for this bias, we redden all of the AGNs in our TD. The amount of reddening applied is determined by the location of the regions for which the classification is being performed. As far as we know, this step has not been performed in previous ML-based classification studies.

TD AGN X-ray fluxes are artificially absorbed using the X-ray photoelectric absorption cross sections from \cite{2000ApJ...542..914W} and the direction-specific hydrogen column density. 
The latter is calculated using the relation between the optical extinction and the hydrogen column density in the Galaxy $N_{\rm H}$ (cm$^{-2})= (2.21\pm 0.09) \times 10^{21} A_{\rm V}$ (mag) from \cite{2009MNRAS.400.2050G} in the direction toward the unclassified source, where $A_V$ is calculated from the  Schlegel, Finkbeiner, and Davis (SFD) extinction map \citep{1998ApJ...500..525S,2011ApJ...737..103S}, which has an angular resolution of 6.1$\arcmin$, using the standard $A_{\rm V}=3.1\times E(B-V)$ \citep{1999PASP..111...63F}.
The absorption correction, $C_{\rm abs}$, is calculated by integrating the absorbed energy flux density within the CSCv2 energy bands:

\begin{equation}
    C_{\rm abs} = \frac{\int f(E) \exp(- N_{\rm H}  \sigma(E)) d E }{\int f(E)  d E}
    \label{equa:F-absrobed}
\end{equation}
where the energy flux density $f(E)$ is assumed to be a power-law function $f(E) \propto E^{1-\Gamma}$ with the  photon index $\Gamma=2$, $N_{\rm H}$ is the estimated  hydrogen column density, and $\sigma(E)$ is the photoelectric absorption cross section. 
The X-ray fluxes of AGNs in the TD are multiplied by $C_{\rm abs}$  corresponding to the direction toward the source to be classified prior to training the RF model (see Section \ref{sec:algorithm}). We are using $N_{\rm H}$ derived from the extinction maps instead of the HI maps because the latter may underestimate the absorption by not taking into account  molecular hydrogen, and they also have coarser angular resolutions.

Similarly, the optical, NIR, and IR magnitudes are reddened using the same extinction map, and the amount of reddening is calculated at the effective wavelength of each band (see Table \ref{tab:limmag}). This is achieved using the {\sl extinction} Python package\footnote{\url{https://github.com/kbarbary/extinction}}. 
In some places of the Galactic plane, the absorption is very large, and if the AGN optical-NIR counterpart is faint, the reddening correction may push the AGN magnitudes beyond  the survey detection limit. Therefore, we remove any magnitudes that are larger than the limiting magnitude of each survey (see Table \ref{tab:limmag}).

To speed up the process when classifying many sources distributed across the sky,  we split the sources up into bins within which the $E(B-V)$ is assumed to be constant. The bin size is 25\,mmag, which is similar to the reddening uncertainty of the SFD extinction map \cite{2014ApJ...786...29S}. Since $E(B-V)$ ranges from 0 to $\sim$50, we have around 2000 bins. We calculate the mean value of $E(B-V)$ for all sources in each bin and apply it to redden the AGNs in the TD before classifying them. 

\begin{deluxetable}{l c c c c}
\tablecaption{Properties of Photometric Optical--NIR--IR Surveys Used for Classification.\label{tab:limmag}}
\tablewidth{0pt}
\tablehead{
\colhead{Band} & \colhead{${\rm Depth}^{\rm a}$} &  \colhead{$f_{\rm zp}^{\rm b}$} & \colhead{$\lambda_{\rm eff}^{\rm c}$} & \colhead{$W_{\rm eff}^{\rm d}$} \\
\colhead{} & \colhead{(mag)} &  \colhead{(erg\,s$^{-1}$\,cm$^{-2}$\,\AA$^{-1}$)} & \colhead{(\AA)} & \colhead{(\AA)} 
}
\startdata
G & 21.5 & 2.5$\times10^{-9}$ & 5822.39 & 4052.97\\
BP & 21.5 & 4.08$\times10^{-9}$ & 5035.75 & 2157.50\\
RP & 21.0 & 1.27$\times10^{-9}$ & 7619.96 & 2924.44\\
J & 18.5 & 3.13$\times10^{-10}$ & 12350 & 1624.32 \\
H & 18.0 & 1.13$\times10^{-10}$ & 16620 & 2509.40 \\
K & 17.0 & 4.28$\times10^{-11}$ & 21590 & 2618.87\\
W1 & 18.5 & 8.18$\times10^{-12}$ & 33526 & 6626.42 \\
W2 & 17.5 & 2.42$\times10^{-12}$ & 46028 & 10422.66 \\
W3 & 14.5 & 6.52$\times10^{-14}$ & 115608 & 55055.71 \\
\enddata
\tablecomments{$^{\rm a}$Corresponds to the faintest sources from the respective surveys found in our counterpart matches. 
$ ^{\rm b}$Zero-point spectral fluxes.  $ ^{\rm c}$The effective wavelength of the corresponding filter.  $ ^{\rm d}$The effective bandwidth of the corresponding filter. More details can be found at the VO Filter Profile Service website.}
\end{deluxetable}

\subsection{Preprocessing and Standardization of the Data}
\label{sec:standard}

After applying field-specific extinction and absorption to magnitudes and fluxes of AGNs from the TD, we calculate the colors and HRs for both TD and unclassified sources. If the necessary magnitude is missing, the color is also considered to be missing (all calculated colors can be found in Table \ref{table:MWfeat}). Additionally, optical--IR magnitudes are converted to energy fluxes (in erg\,s$^{-1}$\,cm$^{-2}$) using the conversion $F = f_{\rm zp} \exp(-{\rm mag}/2.5) W_{\rm eff}$ to enable calculation of flux ratios involving the division by the broadband X-ray flux. The corresponding zero points ($f_{\rm zp}$) and the effective bandwidths ($W_{\rm eff}$) are taken from the Virtual Observatory (VO) Filter Profile Service\footnote{\url{http://svo2.cab.inta-csic.es/theory/fps/}} \citep{2012ivoa.rept.1015R} and are given in Table \ref{tab:limmag}. We then divide the fluxes in all bands (except $F_{\rm b}$) by the 0.5--7\,keV X-ray flux to help mitigate the impact of varying distances to the sources since we currently do not use them in our classification. Finally, we take the base 10 logarithm of all flux quantities described in this paragraph, i.e., X-ray fluxes and optical--NIR--IR fluxes.

In order to allow our TD and unclassified data to be used with other ML algorithms (the provided Python--Jupyter notebooks  allow for this flexibility) and to address the imbalance problem in our TD (see Section \ref{sec:imbal}), we then standardize our data. The standardization is performed as follows:

\begin{equation}
X_i=\frac{x_i-\mu_i}{\sigma_i}
\label{equ:standard}
\end{equation}

where $x_i$ is the value of the $i$th feature for a particular source while, $\mu_i$ and $\sigma_i$ are the corresponding feature's mean and standard deviation across the entire TD. The same standardization is applied to both the TD and any sources that we classify. 
The standardization allows the use of a distance metric for algorithms rely on clustering as a means of classification (e.g., $K$-nearest neighbors, hereafter KNN). Note that the RF algorithm, primarily used in our study, does not rely on the distance metric.  

\begin{figure*}
\begin{center}
\includegraphics[width=350pt,trim=0 0 0 0]{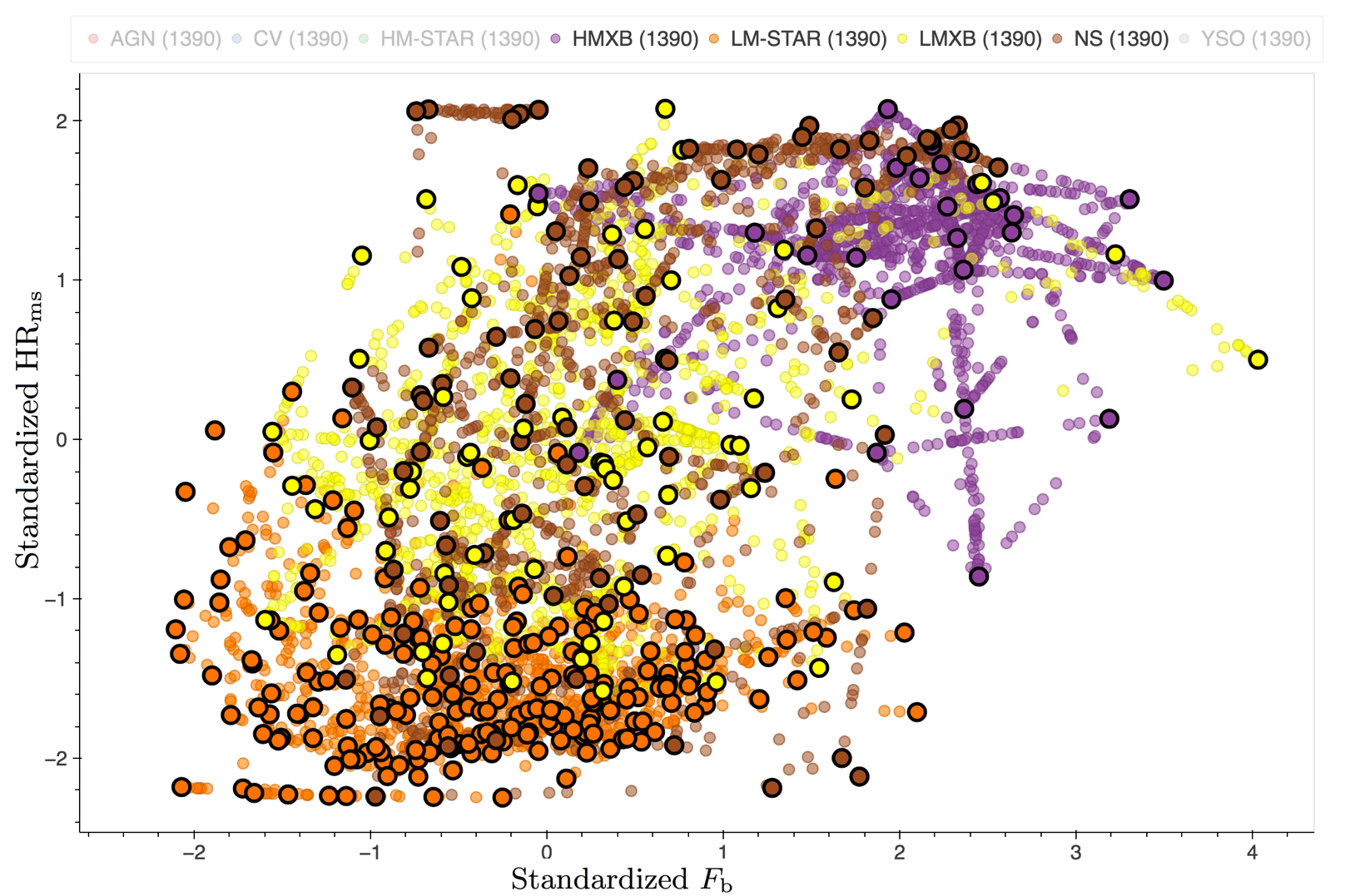}
\includegraphics[width=350pt,trim=0 0 0 0]{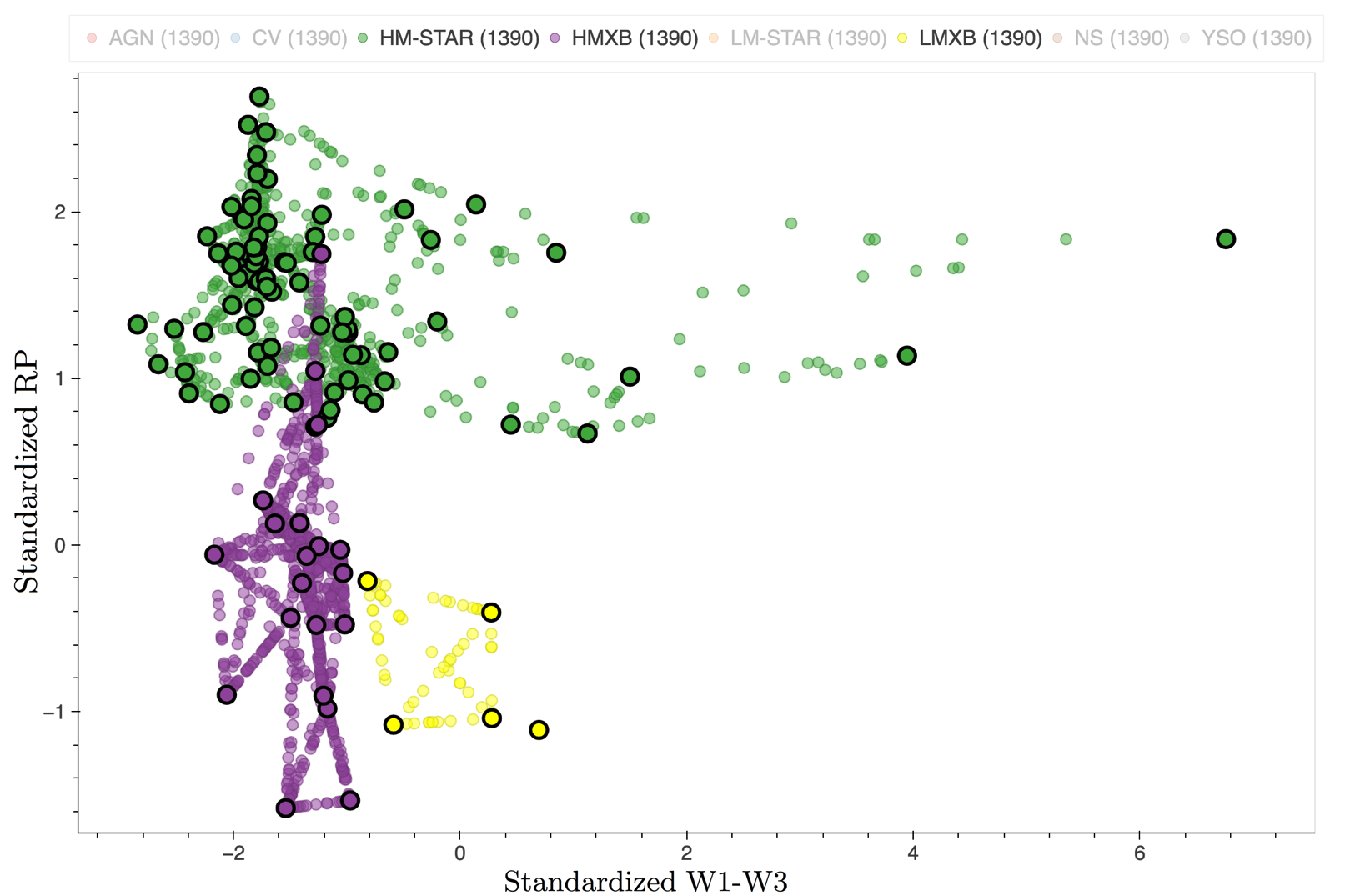}
\caption{Examples of 2D slices of the multidimensional feature space (see Table \ref{table:MWfeat}) showing the TD content after applying SMOTE to create synthetic sources and balance the numbers of sources across all classes. The original (real) TD sources have black borders while the synthetic ones do not have them. Note that the SMOTE procedure results in artificial linear structures when the original (real) sources are far apart.  The features have be preprocessed and standardized, as described in Section \ref{sec:standard}, prior to applying SMOTE.} 
\label{fig:SMOTE}
\end{center}
\end{figure*}

\subsection{Imbalanced Data}
\label{sec:imbal}

One major limitation of our TD is that it is heavily imbalanced. 
AGNs and YSOs substantially outnumber most other classes, which can skew the performance of the classifier in favor of choosing the majority class when classifying unidentified sources. 
There are several ways to partly remedy this problem. The simplest one is to weigh the source classes to punish the algorithm more heavily for misclassifying sources as the most populous classes. 
There are two predefined weights in the scikit-learn package that can be used, but customized weights can also be implemented\footnote{\url{http://scikit-learn.org/stable/modules/generated/sklearn.ensemble.RandomForestClassifier.html}}.

Another way to handle imbalanced data is to use oversampling. 
A popular implementation of this technique is the synthetic minority oversampling echnique (SMOTE; \citeauthor{2011arXiv1106.1813C} \citeyear{2011arXiv1106.1813C}) written in Python\footnote{ \url{https://imbalanced-learn.org/stable/references/generated/imblearn.over_sampling.SMOTE.html}}. 
This method creates a new synthetic source by choosing at random 1 out of 4 (a setting that can be changed) nearest neighbors from the same class for a given real source.
Next, a synthetic source is created at a randomly selected point between the two real sources in the high-dimensional feature space (see \citealt{2011arXiv1106.1813C} for details). 
Using this method, synthetic sources are added to all underpopulated classes  until each class has as many objects as the most populous class (AGNs in our case). Figure \ref{fig:SMOTE} shows examples of applying SMOTE to different classes of sources in our TD for two selected 2D slices of the multidimensional feature space. The features shown in Figure \ref{fig:SMOTE} have already been preprocessed and standardized as described in  Section \ref{sec:standard}. As one can see, the SMOTE procedure is not ideal as it leads to creation of artificial linear structures and also can have significant negative impact when one of the synthetic sources in the TD is an outlier with a wrong class assigned to it. Therefore, it is important to have a clean TD, for sources from classes with the smallest memberships.

The SMOTE procedure has also influenced our choice of flag parameter for missing data (see Section \ref{sec:missingdata}). We set the flag value to --100 for missing data so that sources with missing data will be offset far away from those with existing data. This ensures that for any source the algorithm will not choose a neighboring source with missing data.

\subsection{Missing Data}
\label{sec:missingdata}

Our TD is fairly complete for X-ray features (all sources have flux values in at least one band, 99\% in two bands, 94\% in three bands), but the MW features can be missing for a large fraction of sources (optical magnitudes are missing for 24\%, NIR magnitudes for 38\%, IR magnitudes for 23\%, and all of these magnitudes are missing for 8\% of the TD sources). 
The data can be missing for several reasons, such as the insufficient survey depth, confusing environment (e.g., gas clouds or other diffuse emission), or the actual lack of emission in a particular band. In the latter case, the lack of MW counterparts carries useful information (e.g., isolated NSs seen in X-rays are often identified by the lack of MW counterparts). There are multiple ways to deal with the missing data before classifying sources. One way, called imputation, replaces missing values with the mean values of each given feature from the TD. We disfavor this method as some sources, particularly solitary NSs, are extremely faint at optical--IR wavelengths \citep[see, e.g.,][]{2011AdSpR..47.1281M}, and their optical magnitudes are typically much fainter than the limiting magnitudes of the surveys we use \citep[see, e.g.,][]{1997ApJ...487L.181S}. 

An alternative method is to replace all missing data with a large negative flag value. This approach is frequently used 
(e.g., \citealt{2014ApJ...786...20L}, \citealt{2015ApJ...813...28F}), and it 
ensures that the sources with missing data will be offset far away from those with no missing data. 
Hence, we adopt this approach and set all missing data in our TD to $-100$. In the future, the use of more sensitive optical--IR surveys should  play the key role in the identification of X-ray sources from  optical--IR faint classes (e.g., isolated NSs) as long as accurate positions for X-ray sources can be obtained to combat the confusion, which is expected to increase with increasing survey depth.

\subsection{Random Forest Algorithm}
\label{sec:algorithm}

Our MUWCLASS pipeline uses a supervised ensemble decision-tree algorithm, RF \citep{BreimanML}, which is implemented via the scikit-learn Python package \citep{2012arXiv1201.0490P}. In short, this classifier constructs an ensemble of decision trees from bootstrapped samples of the TD. It constructs the trees by using a randomly selected subset of features at each node, finding the optimal feature and the optimal splitting associated with this feature for this subset, and then repeating the process until all sources in a node are of the same class (at which point the node becomes a leaf). Optimal features (and splittings) are determined by minimizing the Gini impurity criterion, which is defined as follows:
\begin{equation*}
I_{\rm Gini}=\sum_{\rm class}f_{\rm class}(1-f_{\rm class})
\end{equation*}
where $f_{\rm class}$ is the fraction of sources belonging to a specific class, which is separated by the selected feature and splitting. Unclassified sources are then fed through this ensemble of trees where each tree votes on the classification of these objects. RF also provides classification probabilities by counting the votes from each decision tree in the ensemble. 
While algorithms using a single decision tree \citep[e.g., C4.5;][]{1993cpml.book.....Q} can be prone to overfitting, an ensemble of decision trees is more resistant to overfitting \citep{BreimanML}.

The RF algorithm can be easily replaced in our pipeline with some other algorithms from the scikit-learn Python package. This allows one to explore the whole suite of ML algorithms provided by the scikit-learn algorithm library. 
Also, since our data are already standardized (see Section \ref{sec:standard}), algorithms that rely on distance metrics (e.g., clustering-based algorithms such as KNN) can be readily used.

\section{Optimization and Performance Evaluation}
\label{sec:algorithm_performance}

An important part of the ML approach to classification is the evaluation of how well the trained model (classifier) performs on data with known labels (classes) that were not used during the training. Below we describe several performance checks and explore the dependencies on various choices of parameters for the scikit-learn RF algorithm and MUWCLASS pipeline.

\subsection{Cross-Validation Method}
\label{sec:cross-validation}

For validation we take a subset of the original (i.e., prior to SMOTE) TD to train the classifier and predict the labels (classes) for the rest of the TD sources. 
Then the predicted labels and the true labels are compared.

To validate the performance of our pipeline, we use the leave-one-out cross-validation (LOOCV) procedure. At each iteration of the procedure, we remove one source, which is considered as the validation set, use the remaining ($N-1$) sources as the TD, and predict the classification of the left-out source. We iterate this procedure for all of the sources in the TD. This cross-validation procedure is an extreme case of $k$-fold cross-validation where $k=N$. 
It is computationally expensive to perform, but it provides a reliable and unbiased estimate of model performance. 
Another reason we use LOOCV is that the numbers of sources for some classes are quite small (e.g., 26 for HMXBs), and hence, setting aside a sizable fraction of these sources will substantially degrade the training of the classifier
and the performance of the pipeline while the cross-validation will no longer reflect the true performance of the TD (with all of the sources included).

\subsection{Hyperparameter Tuning}

During the cross-validation process, we tune the hyperparameters\footnote{See \url{https://scikit-learn.org/stable/modules/generated/sklearn.ensemble.RandomForestClassifier.html} for details.} of the RF algorithm, which could potentially improve the trained model performance. 
The RF algorithm has three main tunable hyperparameters, namely, the total number of trees (n\_estimators), the maximum number of levels in each decision tree (max\_depth), and the maximum number of features used at each node in the tree (max\_features). 
We evaluate the dependence of the pipeline performance on the values of these hyperparameters using LOOCV on the TD while varying one hyperparameter and fixing the others at their default settings. 
We find that at the default settings of the hyperparameters,  i.e., n\_estimators = 100, max\_depth=None\footnote{The nodes are expanded until all leaves are pure or until all leaves contain less than 2 samples.}, and max\_features = $\sqrt{\rm (n\_features)}$\footnote{max\_features = 6 for n\_features=29 in our case.}, the pipeline performs well achieving an overall accuracy of 88.6\%, a balanced accuracy of 70.2\%, a macro F-1 score of 68.1\%, and a Matthews correlation coefficient (MCC) of 82.9\% (see Appendix \ref{sec:performance-metrics} for definitions). The run time increases linearly with n\_estimators while the performance remains nearly unchanged (within $\sim$0.5\%) for n\_estimators$\ge$50. The run time and the performance are insensitive to the other hyperparameters around their default settings. Therefore, we choose to keep the hyperparameters at their default values.

\subsection{Feature Selection}
\label{sec:feature-selection}

Feature selection is the process of keeping the most important features while dropping the less important ones to reduce the number of total input variables (features) used to fit (train) the model (classifier). It is desirable to reduce the number of 
the features to both reduce the computational costs and to avoid redundant (correlated) information.

\begin{figure*}
\begin{center}
\hspace*{-0.9cm} 
\includegraphics[width=400pt,trim=0 0 0 0]{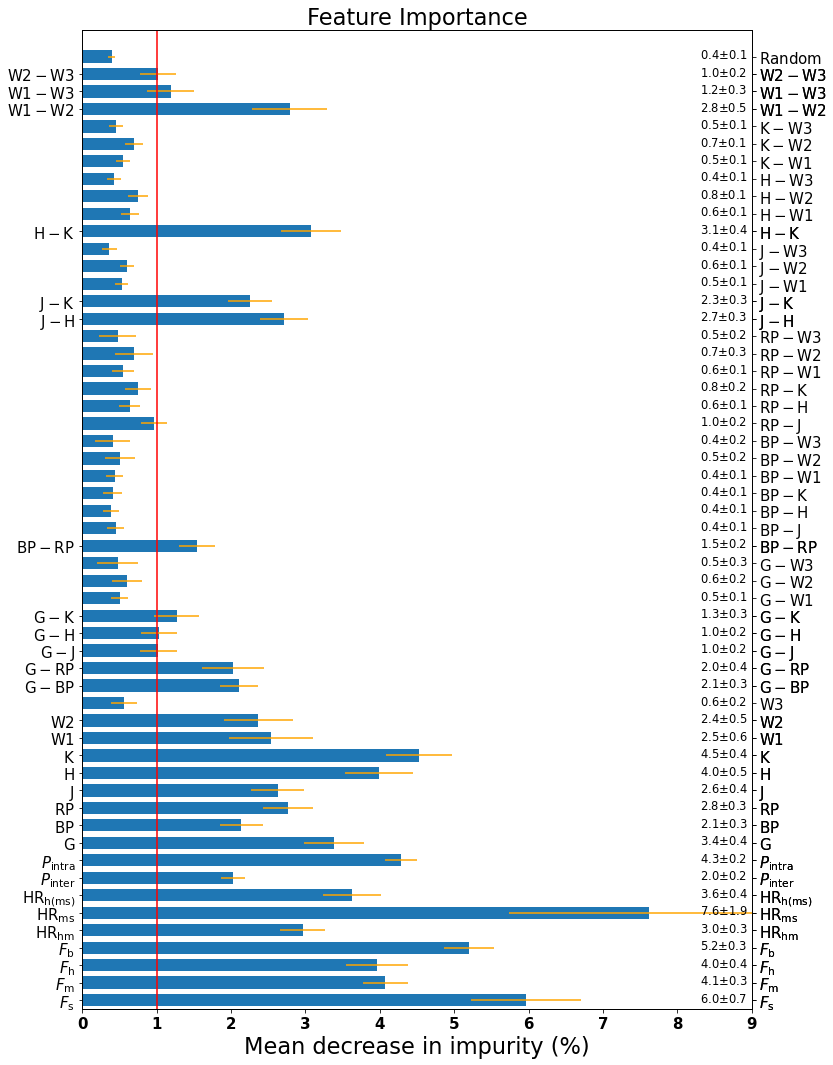}
\caption{
Feature importance (blue bars) and their 1$\sigma$ uncertainties (orange errorbars) for all considered features with their names specified next to the vertical axis on the right.  The values and their uncertainties are also shown next to the feature names. The subset of features  used in the ML classification (see Table \ref{table:MWfeat}) has the importance exceeding that shown by the red line (1\%) with their names shown next to the vertical axis on the left. We also added a dummy (random) feature and evaluated its importance (the top feature). 
}
\label{fig:feature-selection}
\end{center}
\end{figure*}

To select the most important features, we train the RF classifier with all 54 features including 4 X-ray fluxes, 3 X-ray HRs, 2 X-ray variability features, 3 optical-band magnitudes, 3 NIR-band magnitudes, 3 IR-band magnitudes and 36 colors\footnote{We note that the optical--NIR--IR magnitudes are converted to energy fluxes, and all fluxes (including three X-ray fluxes at soft, medium, and hard bands and optical--NIR--IR fluxes, except for the broadband X-ray fluxes) are then divided by the broadband fluxes, before taking the base 10 logarithm (see Section \ref{sec:standard}).}. 
We run LOOCV on the TD 1000 times with MC sampling of the feature uncertainties and calculate the feature importance by accumulating the impurity decrease within each tree for each classification. 
The importance values of all features are shown in Figure \ref{fig:feature-selection} where we calculate the mean and the standard deviation of each feature importance from the 1000 runs. 
We also add a random feature, sampled from a uniform distribution, for which we find the importance to be 0.40\%$\pm$0.05\%. Since any random feature carries no information, those real features that have similar (or smaller) importance values  are regarded as completely noninformative. Thus, we conservatively use a threshold of 1\% (2.5 times larger than the random feature), which leaves us with 29 features that are listed in Table \ref{table:MWfeat}. 
Unsurprisingly, all X-ray features are relatively important with importance $>$2\%. This is primarily due to the fact that, by construction, all TD sources have at least one measured X-ray flux value, while MW features are missing for some fraction of the TD sources.  The most important feature is HR$_{\rm ms}$ with $\sim$8\% importance.
There are other features that are also quite important, such as H and K magnitudes.  
Although the W3 magnitude by itself is relatively unimportant (dropped with  our importance cut), the  colors involving W3 mag ($W1-W3$ and $W2-W3$) turn out to be more informative, so we keep those two colors.

\subsection{Algorithms Comparison}
\label{sec:algorithm-comparison}

\begin{figure*}
\begin{center}
\hspace*{-0.9cm} 
\includegraphics[width=400pt,trim=0 0 0 0]{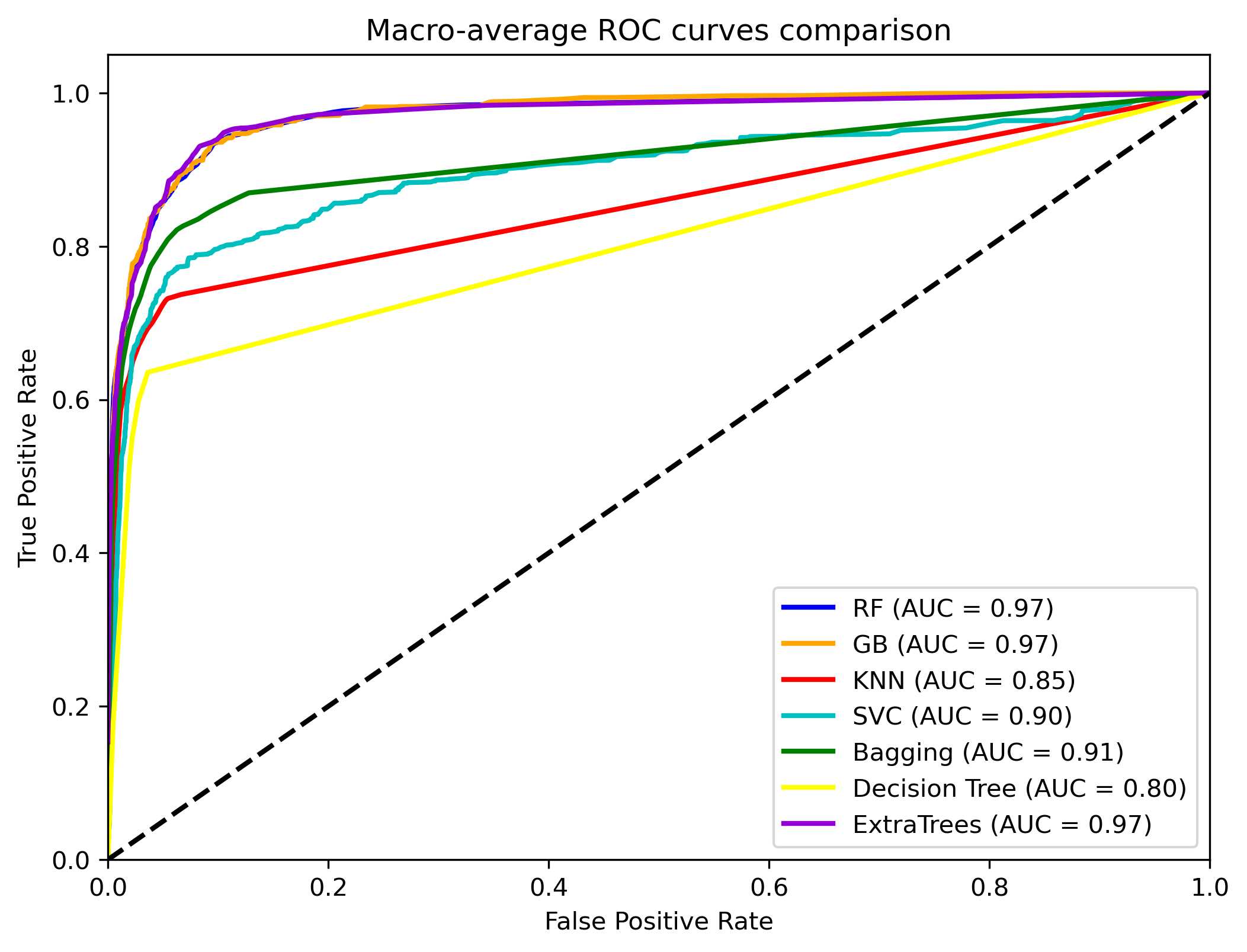}
\caption{Macro-average ROC curves comparing different supervised ML algorithms: the RF classifier (blue), the GB classifier (orange), the KNN classifier (red),  the SVC (cyan),  the bagging classifier (green), the decision-trees classifier (yellow), and the extra trees classifier (purple). The values listed in parentheses in the legend  correspond to the areas under the ROC curves. The larger the area, the better the performance of the algorithm.}
\label{fig:ROCcomp}
\end{center}
\end{figure*}

We compare a number of supervised ML algorithms using the receiver operating characteristic (ROC) curves (see Figure \ref{fig:ROCcomp}) and a number of other metrics (see Table \ref{tab:methods-comp} and Appendix \ref{sec:performance-metrics} for the definitions of the performance metrics). Specifically, we compare the RF classifier, the gradient boosting (GB) classifier\footnote{Gradient boosting is an ensemble algorithm that fits boosted decision trees by minimizing a differentiable loss function. See \url{https://scikit-learn.org/stable/modules/generated/sklearn.ensemble.GradientBoostingClassifier.html} for details.}, the KNN (with $k=5$) classifier, the support vector classifier (SVC; with a radial basis function kernel and a kernel coefficient $\gamma$= 1/n\_features), the bagging classifier, the decision-tree classifier, and the extra trees classifier.  
We choose to use a macro averaged ROC curve (see Appendix \ref{sec:performance-metrics}), which treats each class with the same weight, to measure the success of the classifier for underpopulated classes. The area under the curve (AUC) values (see Appendix \ref{sec:performance-metrics}) for each algorithm are listed in the legend of Figure \ref{fig:ROCcomp}. The larger is the AUC value, the better is the algorithm performance. Other metrics given in Table \ref{tab:methods-comp} include the accuracy, the balanced accuracy, the macro F1 score, the MCC, and the average run time per LOOCV. 
Across the tested algorithms, the performance differences between the three ensemble-based decision tree algorithms (i.e., RF, extra trees, and GB algorithms) are marginal except that the GB algorithm is computationally expensive. Other algorithms perform worse by a few percentage points for each metric with the SVC method being the most computationally expensive. 
Therefore, we decide to adopt the RF algorithm because it is widely used  (including the field of astronomy) and is also one of the best performing algorithms (see Figure \ref{fig:ROCcomp}). 
However, in our MUWCLASS pipeline, any user can easily replace the RF algorithm with some other algorithms available from the scikit-learn package (including those mentioned above).

\begin{deluxetable*}{lccccccc}
\tablecaption{ML Algorithm Comparison\label{tab:methods-comp}}
\tablewidth{0pt}
\tablehead{
\colhead{Algorithms} & \colhead{RF} &  \colhead{Extra Trees} & \colhead{GB} & \colhead{Bagging} & \colhead{SVC} & \colhead{KNN} & \colhead{Decision Tree}
}
\startdata
Accuracy       & 0.886 & 0.882   & 0.874    & 0.867    & 0.856 & 0.835 &  0.882  \\
Balanced Accuracy & 0.702 & 0.680   & 0.720    & 0.668    & 0.651 & 0.651 &  0.680 \\
Macro F1 Score & 0.681 &  0.664  & 0.665    & 0.628    &  0.619 & 0.593 & 0.592 \\
MCC & 0.829 & 0.822 & 0.815 & 0.803 & 0.787 & 0.760 & 0.766 \\ 
Run Time for LOOCV (s)  & 967 & 770 & 5080 & 841 &  6240 & 733 &  752    \\
\enddata
\end{deluxetable*}

\subsection{The Final Pipeline Performance Evaluation} 
\label{sec:performance}

We evaluate the pipeline performance by running LOOCV after performing the hyperparameter tuning and feature selection. 
We run 1000 MC samplings for each TD source to account for feature uncertainties, which equals to training the model 1000$\times N$  times, where $N$ is the number of sources in the TD. Besides the MC sampling of feature uncertainties, there are other processes involving randomness. One is the SMOTE algorithm, which creates synthetic sources, and the other is the RF algorithm itself, which randomly selects a subset of the TD for each tree and a subset of the features at each split. Each of these contributes to the variance in the classification (i.e., the classification uncertainty), but the main contribution comes from the feature measurement uncertainties (see Appendix \ref{appendix:convergence-uncertainty}).

From 1000 MC samplings, we calculate the mean probability ($P_{\rm class}$) for a source to belong to each of eight  X-ray source classes and the standard deviation ($\Delta P_{\rm class}$; hereafter the classification probability uncertainty), which characterizes the width of the $P_{\rm class}$ distribution. 
The predicted class of the source is the class with the largest $P_{\rm class}$.  
We define a classification confidence threshold (CT) as 

\begin{equation}
{\rm CT} = \min_{\rm class}(\frac{P_{\rm predicted\,class}-P_{\rm class}}{ \Delta P_{\rm predicted\,class}+ \Delta P_{\rm class}})
\end{equation}
where class index runs through all 7 classes that are different from the the predicted class. We replace the classification uncertainty with 10$^{-5}$ when the source is classified with zero uncertainty to avoid the zero division error.

To find the CT value which provides an optimal performance, we calculate the metrics including the accuracy, the balanced accuracy, the macro average F-1 score and the MCC as well as the completeness (defined as the fraction of sources that remain after the classification CT) and the balanced completeness (which is the average completeness per class) as a function of CT in Figure \ref{fig:Confidence-Sigma}. We find that metrics improve as we increase the CT while the completeness drops. We choose CT$=2$ since it achieves a good balance between accuracy and completeness, providing an accuracy = 97.0\%, balanced accuracy = 79.0\%, F-1 score = 79.5\%, MCC = 95.0\% completeness = 81.0\%, and balanced completeness = 56.9\%. 
We note that there are some fluctuations of the F-1 score and the balanced accuracy at higher CT values, which are caused by small number statistics in the minority classes. 
The users may choose a different confidence cut (e.g., choosing a different CT value or cutting $P_{\rm predicted\,class}$ at a specific value). In such case, the performance of MUWCLASS should be reevaluated using the TD provided.

For each classified source, we  obtain the distribution of the classification probability for each of the eight classes. Examples of the classification probability distributions for several randomly selected sources are shown in Figure \ref{fig:probability-histogram} for different levels of CT.

Figure \ref{fig:LOOCV} presents the normalized recall (upper row) and precision (lower row) confusion matrices  (CMs; see Appendix \ref{sec:performance-metrics} for definitions) for all sources in the TD (left panel) and confident classifications with ${\rm CT}\ge2$ (right panel) based on 1000 LOOCV runs. The source numbers (for each class) are shown below the class name on the vertical axis in the left panel, while the completeness of each class after applying the confidence cut ${\rm CT}\ge2$ is shown on the vertical axis in the right panel. 
The recall and precision CMs are obtained by normalizing a CM with raw counts of sources in each box by the total number of sources in a true class (recall CM), or by the total number of sources in a predicted class (precision CM). 
Below we summarize the performance metrics of MUWCLASS:

\begin{figure*}
\begin{center}
\hspace*{-0.9cm} 
\includegraphics[width=400pt,trim=0 0 0 0]{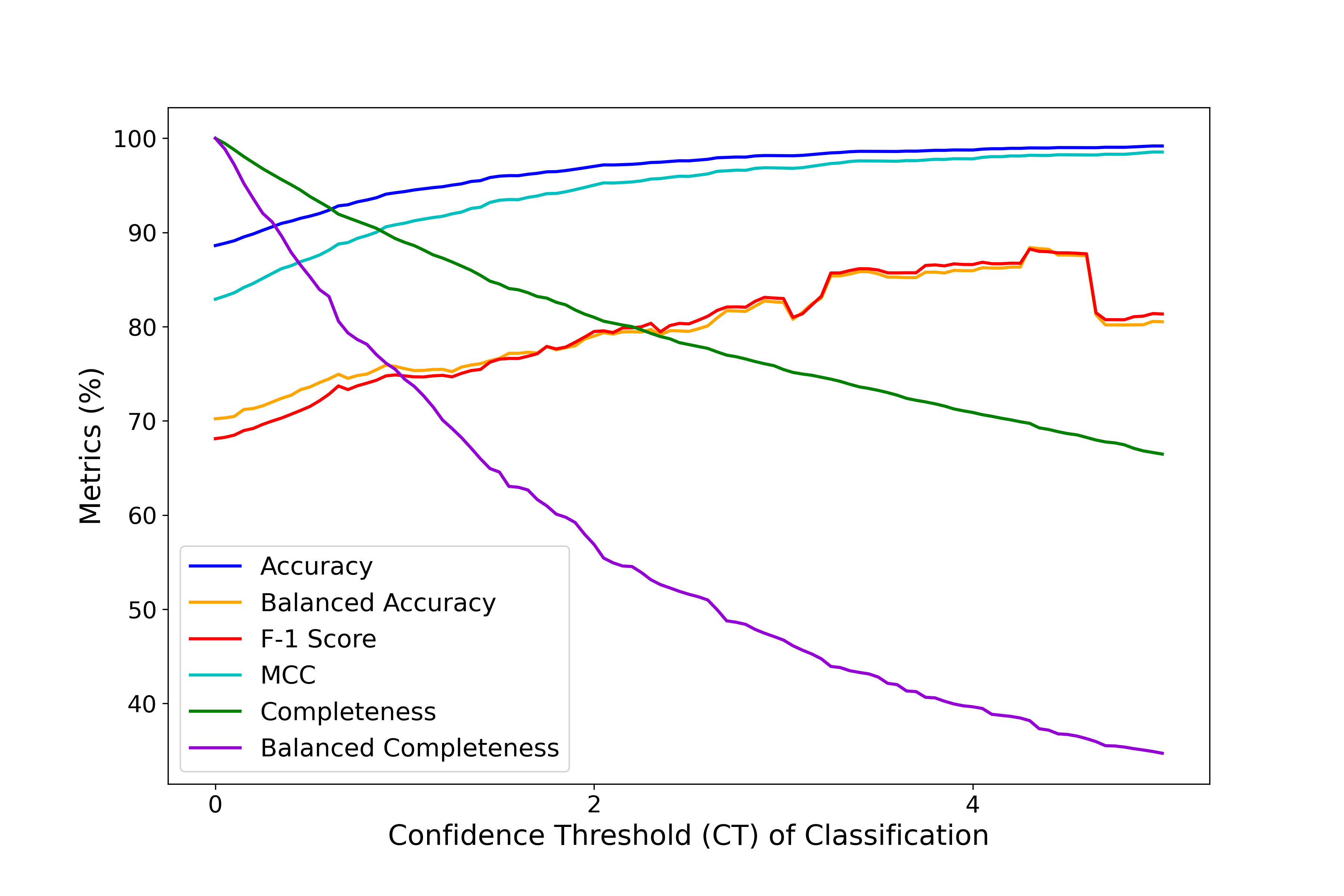}
\caption{The metrics (including the accuracy, the balanced accuracy, the F-1 score, and the MCC) as well as the completeness and the balanced completeness plotted as a function of the classification confidence threshold. The values are computed  with the TD using LOOCV method.}
\label{fig:Confidence-Sigma}
\end{center}
\end{figure*}

\begin{figure*}
\begin{center}
\includegraphics[width=500pt,trim=0 0 0 0]{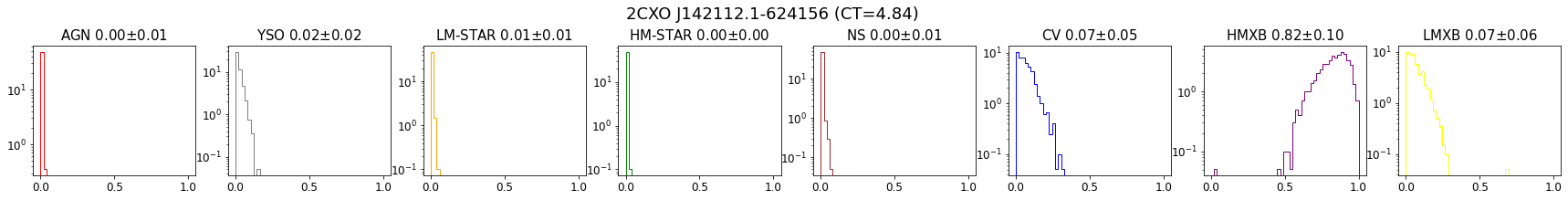}
\includegraphics[width=500pt,trim=0 0 0 0]{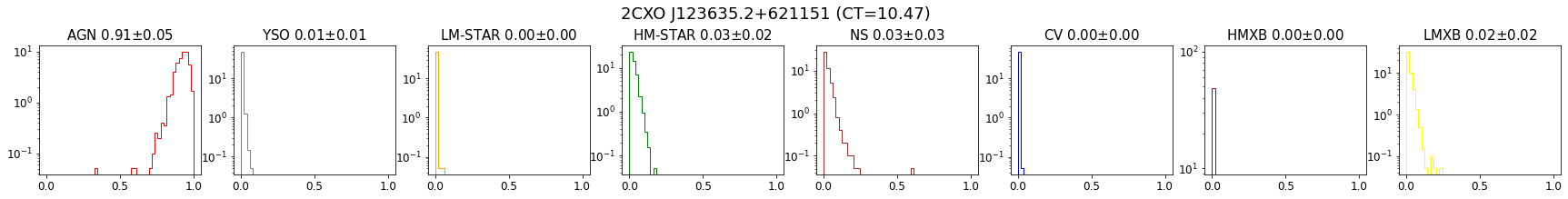}
\includegraphics[width=500pt,trim=0 0 0 0]{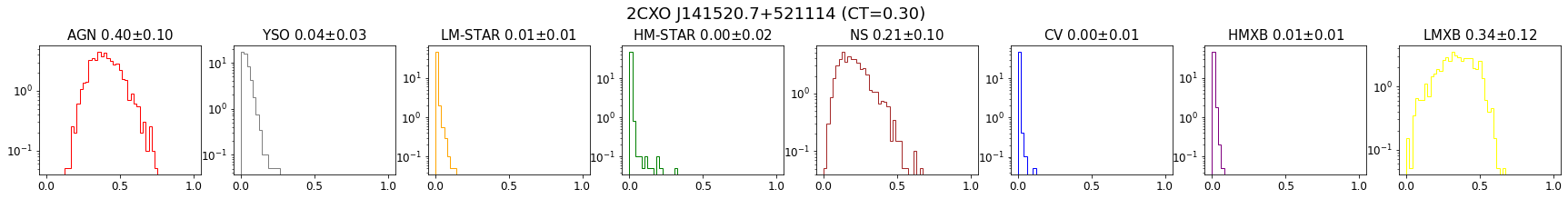}
\includegraphics[width=500pt,trim=0 0 0 0]{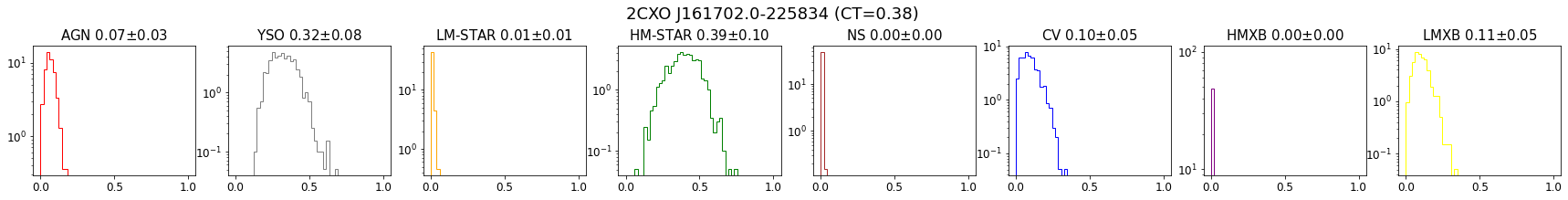}
\caption{Examples of  probability distributions per class (shown in the columns) for a few CSCv2 sources that we classify (one source  per row). The name of the source and the confidence threshold are shown at the top of each row, and the class and its probability with the $1\sigma$ uncertainties (see Sec.~\ref{sec:performance} for the definition) are shown at the top of each plot in the row.} 
\label{fig:probability-histogram}
\end{center}
\end{figure*}

\begin{enumerate}
    \item The accuracy for all classifications is 88.6\%, and it improves to 97.0\% for confident classifications. 
    \item The balanced accuracy for all classifications is 70.2\%, and it improves to 79.0\% for confident classifications. 
    \item All classes improve their performance after the confident classification filter except for the recall rate for LMXBs.
    \item The overall completeness (the fraction of classified sources after the confidence cut) drops from 100\% to 81.0\%, and the balanced completeness drops from 100\% to 56.9\% after the confidence cut.
    \item The completeness of confidently classified sources is high for populous classes like AGNs (87\% for true AGNs, 93\% for predicted AGNs) and YSOs (88\%) and relatively high for LM-STARs with 70\% while for other classes it is around or below 60\%. 
    \item The performance on AGNs and YSOs is extremely good with a recall rate and precision over 98\% for confident classifications. 
    \item The performance on LM-STARs is good with a recall rate of 96\% and precision of 92\% for confident classifications although they are sometimes confused with YSOs and HM-STARs.
    \item CVs, HM-STARs, HMXBs, and NSs perform reasonably well, with a recall rate and a precision of around 80\% for confident classifications. 
    \item LMXBs (including noninteracting binaries) perform the worst, and they are often confused with NSs, AGNs, and CVs.
\end{enumerate}

\begin{figure*}
\begin{center}
\includegraphics[scale=0.35,trim=0 0 0 0]{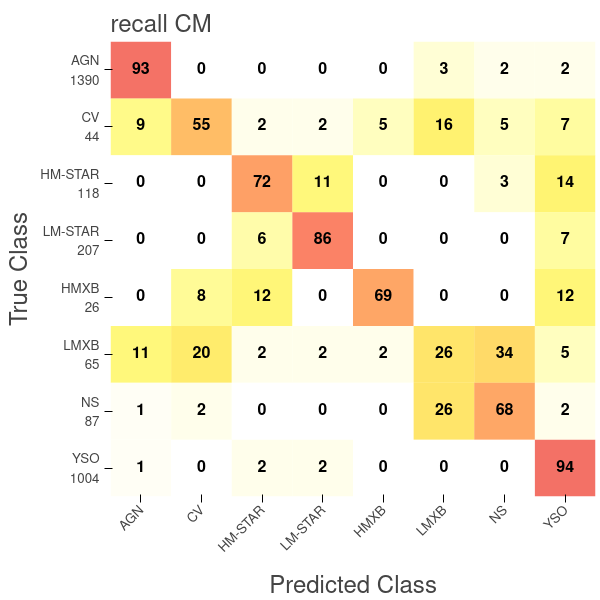}
\includegraphics[scale=0.35,trim=0 0 0 0]{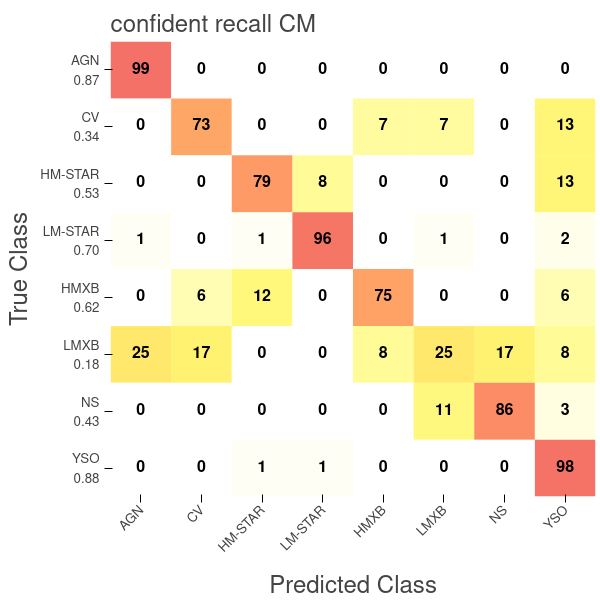}
\includegraphics[scale=0.35,trim=0 0 0 0]{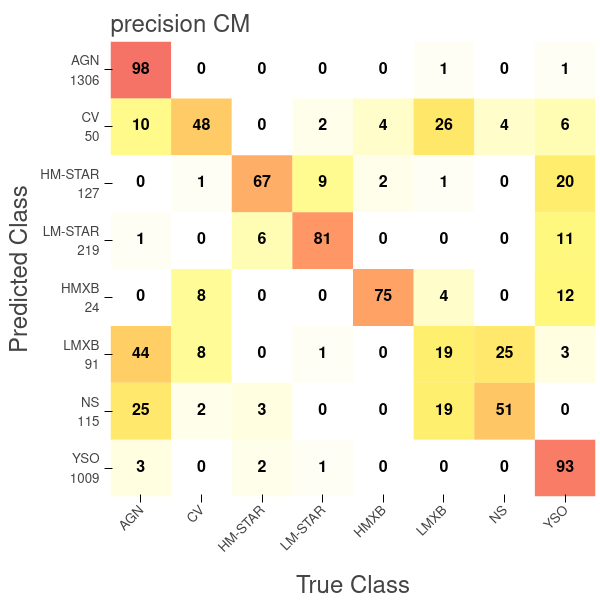}
\includegraphics[scale=0.35,trim=0 0 0 0]{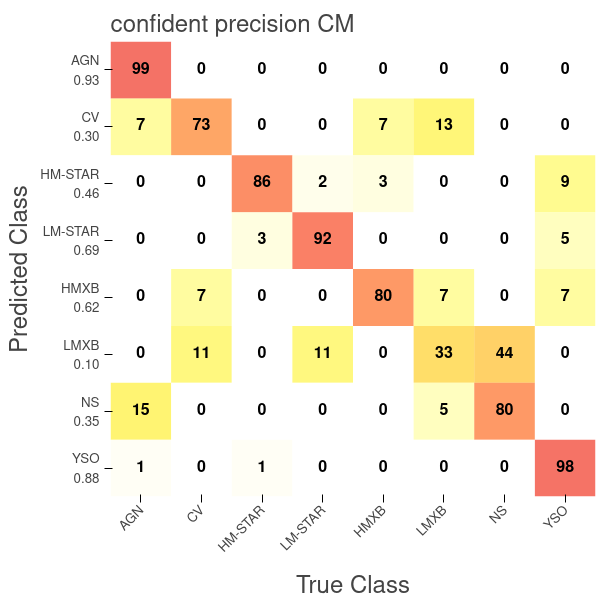}
\caption{The left panels shows the normalized confusion matrices (CMs) of all classifications while the right panels shows the normalized CMs for the confident classifications (${\rm CT}\ge2$). The upper panels show the normalized recall CMs, and the lower panels show the normalized precision CMs. 
The value within each element of the CM is the percentage of sources in a predicted (true) class, shown on the horizontal axis, that are from the true (predicted) class, shown on the vertical axis, for the normalized recall (precision) CMs.
The values under the class labels along the vertical axis in the left panels are the total numbers of the sources in the corresponding classes while in the right panels these values are the fractions of the sources surviving the confidence cut (CT$\ge2$) for each class. 
The darker the color, the higher the percentage is. } 
\label{fig:LOOCV}
\end{center}
\end{figure*}

We provide both recall and precision CMs for convenience because they can serve different purposes. For example, if one is interested in the fraction of a class that were retrieved (the ratio of the number of sources that are predicted as a particular class to the total number of sources of the same class) from the classification, it can be found on the diagonal of the recall CM (e.g., 86\% of true LM-STARs will be classified correctly by MUWCLASS). To estimate the fraction of accurately classified sources among all confident classifications, one can look at the confident precision CM (e.g., if there are 100 sources that are confidently classified as CVs, 
only 73 of them are expected to be true CVs). 
However, one has to keep in mind that the actual performance of MUWCLASS can be worse than indicated by these CMs if the TD does not represent well the sample of  sources to be classified  (i.e., there are selection biases). In reality, these biases are often present, and we discuss several of them in the paper. This difference between the expected and the actual performance is usually difficult to evaluate.

\section{Results}
\label{sec:results}

\begin{figure*}
\begin{center}
\includegraphics[width=450pt,trim=0 0 0 0]{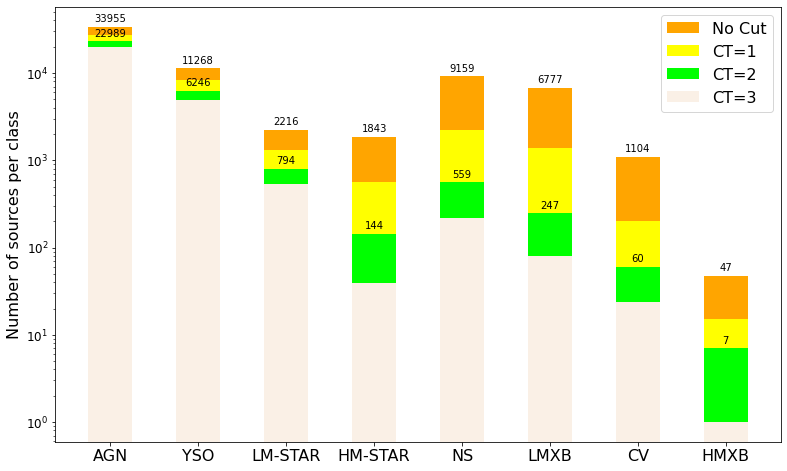}
\caption{
Summary of the classification outcomes for GCS with S/N$\ge$3 shown as a function of classification confidence for different values of CT. The values on top of the histogram bars show the numbers of sources for each class without (larger value) and with (smaller value) the CT=2 cut. Notice that for AGN, LM-STAR, and YSO classes (which are well represented in the TD) the relative change in the number of classified sources with the CT cut is relatively small compared to other source classes. 
} 
\label{fig:CSC-classification-breakdown}
\end{center}
\end{figure*}

In order to select well characterized point sources with reliably measured X-ray features, we define a {\sl good} CSCv2 sample (hereafter GCS) detected with significance S/N$\ge$3, PU$<$1$\arcsec$. 
Also we ensure that GCS sources do not have CSCv2 confused and extended source flags raised (i.e., {\sl conf\_flag} and {\sl extent\_flag} in CSCv2 source flags). 
Such filters leave 66,369 sources in GCS (constituting 21\% of all CSCv2 sources). 
The classifications of GCS (with a source detection threshold at S/N=3) are presented by histograms in Figure \ref{fig:CSC-classification-breakdown}, with different CT cuts. A summary table of the classification numbers for GCS sources with 2 different source detection thresholds (S/N=3 and S/N=6) per class is also shown in Table \ref{tab:CGS-classification-summary}, together with the fractions of classifications after the confidence cut (CT$\ge2$) and the fractions of GCS sources having at least one MW counterpart.  
The classifications of GCS are also available in the electronic (machine-readable) format with a few sources shown in Table \ref{tab:GCSMRT} as a subset of the entire table (see Appendix \ref{sec:MRTs}). 
   
In addition to classifying a large number of previously unclassified sources, we apply several statistical tests to the confidently classified GCS (hereafter CCGCS) of 31,046 sources with CT$\ge2$ to compare them with those from the TD to look for possible biases. An investigation on some individual interesting sources and fields will be discussed in Section \ref{sec:examples}.

As one can see, we find a large number of AGNs and YSOs. As the CT is increased, the number of classified sources in these classes does not drop as much (in percentage) as in other classes (e.g., NSs). Also, recalling that the performance evaluation based on the TD (see Figure \ref{fig:LOOCV}) shows that nearly all (99\%) of AGNs and YSOs are correctly classified, we expect the confident AGN and YSO classifications to be reliable, unless there are strong biases making AGNs and YSOs from the TD poorly represent those that are classified, or the classified sources are in the crowded environments (see Section \ref{sec:limitation}). The results for some other classes do not look nearly as good, e.g., for NSs and LMXBs only 80\% and 33\% of confident classifications are expected to be true. However, even after this correction, the number of identified NSs is surprisingly large, suggesting that an unaccounted bias may be affecting the classifications for this class (see Section \ref{sec:limitation}). 

From our experience with {\sl Chandra} data and CSCv2, most of the S/N$=3$ detections are very solid. However, it is still interesting to see what happens when the detection threshold is increased, i.e., how the numbers of classified sources per class  change for brighter sources. Comparison of the first row and the third row in Table \ref{tab:CGS-classification-summary} (S/N$\ge$3 versus\ S/N$\ge$6) shows that, although the numbers of sources in each class drop by a factor of 2--3, there is no particular class whose classification efficiency would be disproportionately affected by the increase in the detection significance. This suggests that the bias associated with the TD being on average brighter than the GCS sources (see Section \ref{sec:X-ray-flux}) may not be having a large impact on the classifications. Also, sources with higher detection significance are expected to have more accurate measurement for their X-ray features, which should be translating to narrower distributions for their classification probabilities. Therefore, a larger fraction of classified sources would generally be expected to pass a particular CT threshold compared to that for the sources with lower detection significance. Indeed, as one can see from the second row and the fourth row in Table \ref{tab:CGS-classification-summary}, this fraction increases for all sources but HM-STAR,
with only marginal increases for LM-STAR and YSO and particularly large increases for NS, LMXB, and CV classes. This is understandable because most of sources classified as stars (including HM-STARs, LM-STARs, and YSOs) have optical counterparts while for NS, LMXB, and CV classes the fraction of sources with MW counterparts is much lower (see the fifth row in Table \ref{tab:CGS-classification-summary} and Section \ref{sec:X-ray-flux}), and hence, the accuracy of measurements of their X-ray features matters more.

\begin{deluxetable*}{lcccccccc}
\tablecaption{Numbers and Fractions of Classifications Per Class\label{tab:CGS-classification-summary}}
\tablewidth{0pt}
\tablehead{
\colhead{} & \colhead{AGN} &  \colhead{YSO} & \colhead{LM-STAR} & \colhead{HM-STAR} & \colhead{NS} & \colhead{LMXB} & \colhead{CV} & \colhead{HMXB}
}
\startdata
 GCS$^{\rm a}$  & 33,955 & 11,268 & 2216 & 1843 & 9159 & 6777 &  1104 & 47 \\
Confident fraction$^{\rm b}$ & 0.677 & 0.554 & 0.358 & 0.078 & 0.061 & 0.036 & 0.054 & 0.149 \\
GCS (with S/N$\ge6$)$^{\rm c}$  & 14,529 & 4988 & 1137 & 690 & 3484 & 2556 &  487 & 24 \\
Confident fraction (for S/N$\ge6$)$^{\rm d}$ & 0.775 & 0.585 & 0.383 & 0.075 & 0.122 & 0.088 & 0.119 & 0.250 \\
MW counterpart fraction$^{\rm e}$ & 0.663 & 0.999 & 0.997 & 1.000 & 0.062 & 0.293 & 0.895 & 1.000 \\
\enddata
\tablecomments{$^{\rm a}$Number of GCS sources classified as the corresponding class. $^{\rm b}$Fractions of confident classifications (CT$\ge2$) for GCS sources per class. $^{\rm c}$Number of  GCS sources with S/N$\ge6$ classified as the corresponding class. $^{\rm d}$Fractions of confident classifications (CT$\ge2$) of  GCS sources with S/N$\ge6$ per class. $^{\rm e}$Fractions of classified GCS sources having at least one MW counterpart per class.}
\end{deluxetable*}

\subsection{Comparison to Catalogs with Known Source Classes}

We compare the classification results of CCGCS to several publicly available catalogs of classified sources (which we did not use in our TD for various reasons). 
To aid the comparison, we merge our LM-STAR and HM-STAR classes into a STAR class and LMXB and HMXB classes into an XRB class. The comparison of the classifications is summarized in Table \ref{tab:classification-matching} and briefly discussed below.

Crossmatching (by coordinates) our CCGCS with the FIRST-NVSS-SDSS AGN catalog \citep{2009ApJ...699L..43S} gives 17 matches, all of which are classified by us as AGNs. A comparison to the combined WISE and SDSS spectroscopic data catalog  \citep{2014ApJ...788...45T} results in 146 AGN matches and 7 YSO matches in our classification of CCGCS. The crossmatching to the two catalogs corresponds to the 
recall rates of 100\% and 95.4\%, which are consistent with that estimated from the CM (99\%) for confidently classified AGNs.

We also crossmatch CCGCS to the TD of a recent large-scale ML-based classification study of 4XMM-DR10  \citep{2022AA...657A.138T}. 
The recall rates calculated from the crossmatching are 93.6\%, 92.0\%, 36.3\% for AGNs, STARs, XRBs, which are comparable to those estimated from the confident recall CM in Figure \ref{fig:LOOCV} while the recall rate for CVs is 25\% with only 8 sources crossmatched. 
We check those sources  that have discrepant classifications from our results and \cite{2022AA...657A.138T}, and we find that most of them are from nearby, resolved galaxies or globular clusters, which are complicated and crowded environments where our MUWCLASS pipeline is not expected to work primarily due to the limitations of the MW surveys we are currently using (i.e., primarily confusion when crossmatching sources). There may also be some misclassificaitons in these complex environments in the TD compiled by \cite{2022AA...657A.138T}. For instance, two sources (2CXO J031818.8--663230 and 2CXO J013647.4+154744) claimed to be XRBs by \cite{2022AA...657A.138T} are classified by MUWCLASS as LM-STARs and actually do appear to be foreground stars (based on their parallaxes and/or proper motions\footnote{Note, we do not currently use this information for ML classification with MUWCLASS, and we checked it manually.} in {\sl Gaia} EDR3), which are coincident by chance with two resolved, nearby galaxies. We intentionally do not add the sources from nearby galaxies (or globular clusters) to our TD since crossmatching to MW counterparts in such crowded environments can easily result in false matches.

The SIMBAD catalog is overall accurate and has been used carefully to verify source classifications previously (e.g., \citealt{2022MNRAS.512.3858L,2022arXiv220400346D}). 
While constructing our TD, we find that some sources have wrong SIMBAD classifications. For example, 2CXO J203213.1+412724 is a $\gamma$-ray binary with a known young pulsar, and belongs to our HMXB class \citep{2015MNRAS.451..581L}, but is labeled as a Be star in SIMBAD; 2CXO J112401.1--365319 is a black widow system, and belongs to our LMXB class \citep{2014ApJ...783...69G}, but is labeled as an NS in SIMBAD. Although we cannot be sure that all SIMBAD classifications are accurate, we expect that most of them still are and proceed with the comparison under this assumption.
By crossmatching CCGCS to the SIMBAD catalog,  we find an estimated recall rate of 98.1\% for AGNs, 97.3\% for YSOs, 75\% for NSs, which are again comparable to those indicated by the confident recall CM in Figure \ref{fig:LOOCV}. The recall rate for STARs is 29.7\%, where the most confusion is from the YSOs (with 62.9\% of STARS classified as YSOs). This is not very alarming since, when YSOs are sufficiently evolved, they are not too different from STARs, and the SIMBAD's definition of YSO class may differ from ours. The agreement for CVs and XRBs is rather poor. We check the CVs and XRBs that are classified differently (from SIMBAD) by MUWCLASS and find that most of them are located in globular clusters or resolved galaxies, where our classification results are unreliable due to the inability to identify the true MW counterpart in such dense environments with  the MW surveys we currently use.

\begin{deluxetable}{lcc}
\tablecaption{Comparison of Classifications to Other Catalogs\label{tab:classification-matching}}
\tablewidth{0pt}
\tablehead{
\colhead{Other Catalogs} & \colhead{MUWCLASS} &  \colhead{Matches}  
}
\startdata
\hline 
FIRST-NVSS-SDSS AGN  & AGN & 17 (100\%)\\
\hline 
WISE-SDSS Galaxy & AGN & 146 (95.4\%)\\
                      &  YSO & 7 (4.6\%) \\
\hline 
4XMM-DR10 TD & & \\
AGN & AGN & 626 (93.6\%) \\
    & non-AGN & 43 (6.4\%)\\ 
STAR & STAR & 172 (92.0\%\\
    & non-STAR & 15 (8.0\%)\\
XRB & XRB & 101 (36.3\%) \\
    & NS & 56 (20.1\%) \\
    & others & 121 (43.5\%)\\
CV  & CV & 2 (25\%) \\
    & non-CV & 6 (75\%) \\
\hline 
SIMBAD & & \\
 Galaxy & AGN & 5089 (98.1\%) \\
       & non-AGN & 101 (1.9\%) \\
YSO & YSO & 2423 (97.3\%) \\
    & non-YSO & 68 (2.7\%) \\
STAR & STAR & 479 29.7\% \\
     & YSO & 1014 (62.9\%) \\
    & others & 120 (7.4\%) \\
XRB & XRB & 30 (11.7\%) \\
 & non-XRB & 226 (88.3\%) \\
NS & NS & 3 (75\%) \\
    & non-NS &  1 (25\%) \\
CV & CV & 1 (5.3\%) \\
 & non-CV & 18 (94.7\%) 
\enddata
\tablecomments{STAR class consists of LM-STAR and HM-STAR classes while XRB class  consists of LMXB and HMXB classes. We had to introduce merged classes in this table because  of the limitations (or differences) of the catalogs used for comparison. The percentages in brackets in matches column correspond to the fractions of sources classified as listed in the second (MUWCLASS) column among the sources belonging to the class listed in the first (other catalogs) column.}
\end{deluxetable}

\subsection{X-Ray Flux Distribution and MW Counterpart Fraction}
\label{sec:X-ray-flux}

\begin{figure*}
\begin{center}
\includegraphics[width=500pt,trim=0 0 0 0]{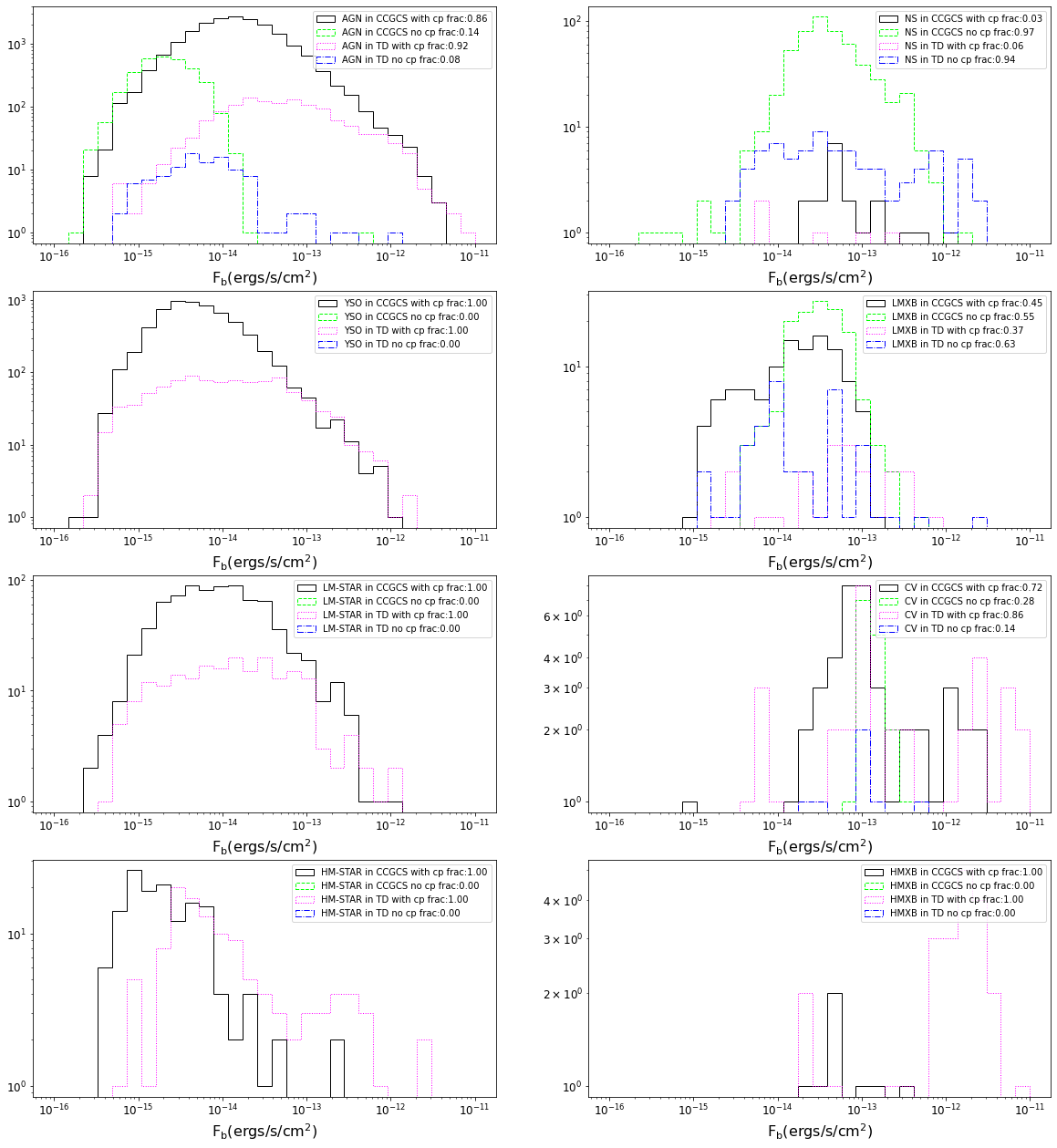}
\caption{The distributions of the number of sources vs.  broadband X-ray fluxes of CCGCS sources and sources from the TD. The solid black and dashed green histograms show the distributions of CCGCS with and without MW counterparts. The dotted magenta and dashed-dotted blue histograms show the distributions of sources from the TD with and without MW counterparts. The fractions of sources with and without MW counterparts are also listed in the legends.} 
\label{fig:Fb_distribution}
\end{center}
\end{figure*} 

We compare the broadband X-ray flux distribution of CCGCS and the TD in Figure  \ref{fig:Fb_distribution}. Ideally, the distributions should closely resemble each other while the substantial difference would indicate a possible bias.  We also compare the fraction of sources that have MW counterparts for each of the eight classes in both CCGCS and the TD. For an X-ray source that has at least one MW counterpart matched, the source will be added to the group of {\sl with cp}. Sources with no MW counterpart matched will be placed in the {\sl no cp} category. 

For populous classes, where a substantial fraction of sources is confidently classified (i.e., AGNs, YSOs and LM-STARs), the TD sources have systematically higher X-ray fluxes than those of CCGCS. The X-ray flux distribution of the TD sources tends to flatten below 10$^{-13}$\,erg\,s$^{-1}$\,cm$^{-2}$ until the detection limit, while that of CCGCS continuously grows below 10$^{-13}$\,erg\,s$^{-1}$\,cm$^{-2}$. 
This probably reflects the fact that brighter X-ray sources are easier to study and classify using traditional approaches (e.g., spectroscopy, periodicity detection, characteristic variability patterns, etc.).

The discrepancy of the X-ray flux distributions represents a selection bias between CCGCS and the TD because the broadband flux is one of the features participating in our ML classification. 
However, all other fluxes are divided by the broadband flux before they are used as features, which should help to mitigate the bias.

From Figure \ref{fig:Fb_distribution}, one can also see that brighter sources have a higher chance of crossmatching to MW counterparts, 
resulting in a higher fraction of sources with MW matches for the TD. 
We also calculate the fractions of sources crossmatched to each MW survey at  optical, NIR, or IR band for CCGCS, which are 34\%, 24\% and 47\%. This is also lower than those calculated for the TD, which are 76\%, 62\% and 77\%. 
This makes sense as brighter sources are often more nearby, thus brighter at all wavelengths. 
This is another selection bias that may potentially skew the classification results (see discussion in Section \ref{sec:limitation}). Confidently classified YSOs, STARs, and HMXBs from CCGCS all have counterparts, which is not surprising
since the TD sources of these classes are required to be bright enough in the optical--IR bands so that the stellar spectral classification can be applied. 
On the other hand, since CCGCS sources are systematically fainter, we would expect that a fraction of them should be missing MW counterparts entirely as they are too faint to be detected by the surveys we are currently  using. This aspect is currently not learned by our MUWCLASS pipeline, thus leading to 
 a potential classification bias.

\subsection{Galactic Latitude Distribution}
\label{sec:latitude}

\begin{figure*}
\begin{center}
\includegraphics[width=500pt,trim=0 0 0 0]{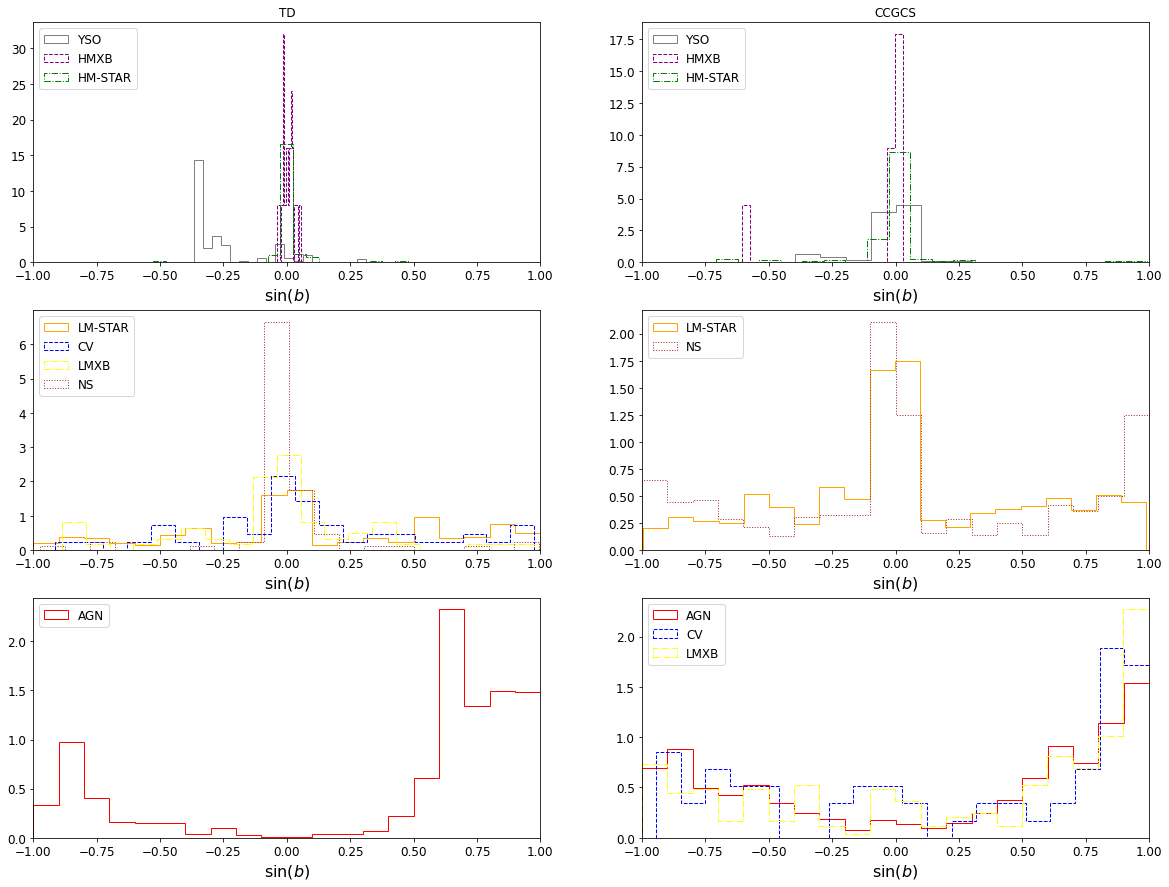}
\caption{The distributions of the normalized number of sources vs. $\sin (b)$, where $b$ is the Galactic latitude, for the TD (left panel) and CCGCS (right panel). Top panels show sources from YSO, HMXB, and HM-STAR classes whose distributions exhibit  
strong concentration toward the Galactic plane. Middle panels show sources from those classes whose distributions exhibit intermediate concentration toward the Galactic plane.  Bottom panels show  sources from those classes whose distributions exhibit weak or no concentration toward the Galactic plane.}
\label{fig:b_hist}
\end{center}
\end{figure*} 

The Galactic latitude parameter $b$ has been used for classifications in several previous ML classification studies 
\citep[e.g., ][]{2004ApJ...616.1284M,2022AA...657A.138T}. 
We do not use $b$ as a feature in our work, 
since the distribution of $b$ in the TD is dependent on the design of the surveys for some classes (e.g., AGNs); thus using $b$ as a feature would cause an additional  bias. 
Still it is useful to look at the distribution of $b$ in the classified sources, as it provides an independent test of our classifications. 
Figure \ref{fig:b_hist} shows the normalized histograms of $\sin (b)$ 
(the $\sin$ function transforms $b$ into a uniform space of solid angles) 
of the TD (on the left panel) and CCGCS (on the right panel). We find that HMXBs, YSOs, and HM-STARs are highly concentrated on the Galactic plane (in both the TD and CCGCS). 
This is in agreement with the fact that most of these types of sources are located in the Galactic plane. 
We notice that there is another concentration around  $b=-20^\circ$ or $\sin (b)=-0.34$ for YSOs in the TD, which presents a selection bias associated with the fact that many TD YSOs come from of the Orion SFR \citep{2012AJ....144..192M}. 
LM-STARs exhibit weaker concentrations toward the Galactic plane for both the TD and CCGCS, which is consistent with the fact that LM-STARs are Galactic sources with soft X-ray spectra and have low X-ray luminosities, so they can only be detected in X-rays when they are relatively nearby (unless there is a large coronal flare).  
For nearby LM-STARs whose distances are comparable to the local width of the Galactic disk, the distribution should not be strongly concentrated toward the Galactic plane. 
As expected, AGNs lack any concentration toward the Galactic plane for both the TD and CCGCS. 
In fact, the high reddening in the Galactic plane makes many of the extragalactic sources too dim to be detected causing the apparent deficit of AGNs in the plane. 
For CVs and LMXBs, we find that they also show reduced concentrations toward the Galactic plane for CCGCS in comparison with the TD. 
We also note that the spatial distribution of both TD and CCGSC sources are affected by the nonuniformity of the CSCv2's sky coverage and varying depth of {\sl Chandra} observations.  

\begin{figure*}
\begin{center}
\includegraphics[width=500pt,trim=0 0 0 0]{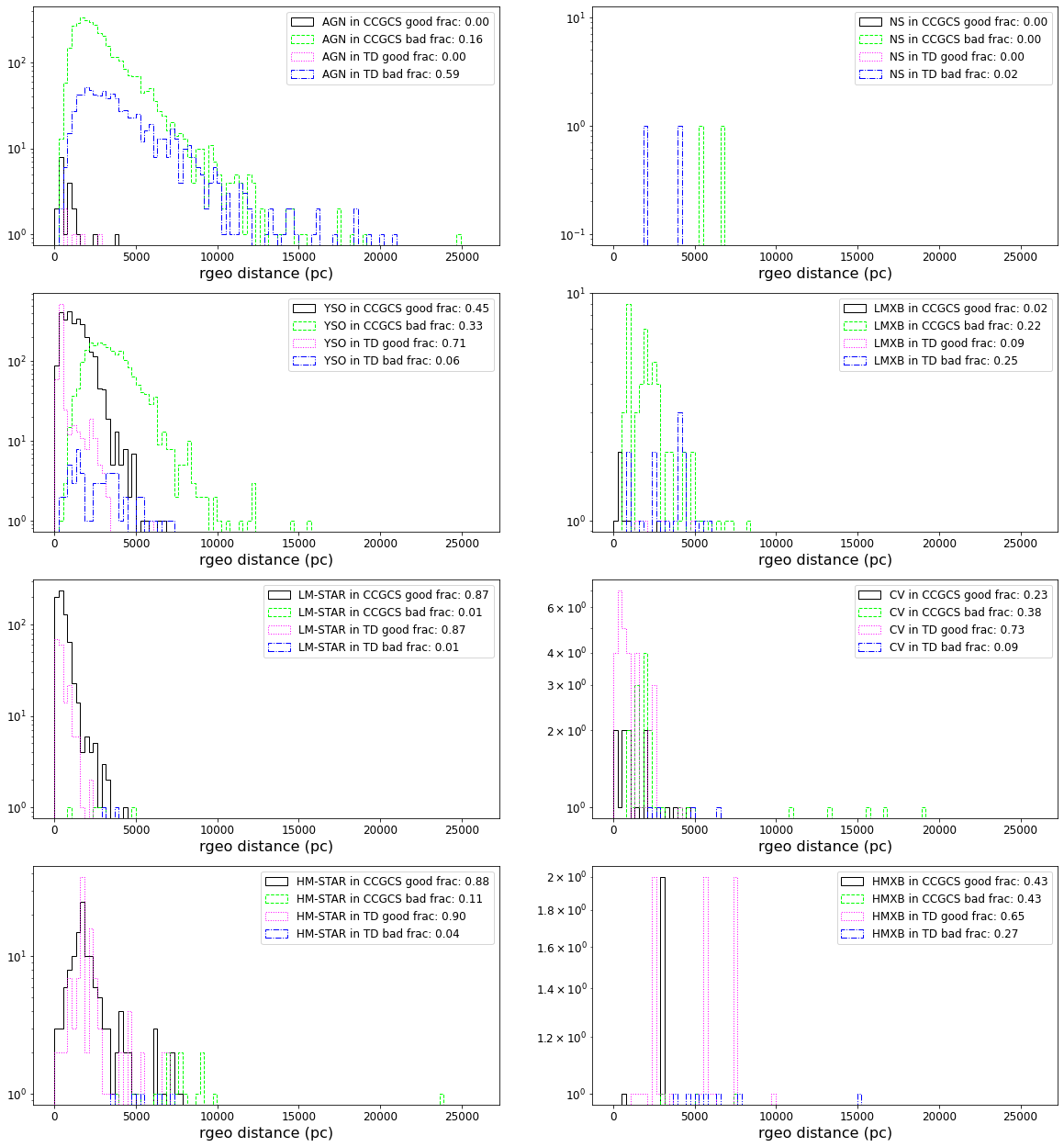}
\caption{The distributions of the number of sources vs. {\sl rgeo} distance for CCGCS and the TD. The black solid histogram shows the distribution for distance measurements of CCGCS with good quality parallax measurements while the dotted magenta histogram shows that for the TD. The histograms with bad quality parallax measurements are shown as green dashed and blue dotted-dashed lines for CCGCS and TD, respectively. The fractions of sources with good or bad distance measurements with respect to all sources of the same class in CCGCS or TD are listed in the legend.}
\label{fig:distance_dsitribution}
\end{center}
\end{figure*} 

\subsection{Distance Distribution} 
\label{sec:distance}

We also look at the distribution of the {\sl rgeo} distances (from {\sl Gaia} EDR3 distance catalog; \citealt{2021AJ....161..147B}) for both CCGCS and TD. 
The parallax to parallax error ratio  $\overline{\omega}/\sigma_{\overline{\omega}}$ is used to characterize the robustness of distance measurements. We define sources with    $\overline{\omega}/\sigma_{\overline{\omega}}>=3$ as those with good quality distance measurements. 
In Figure \ref{fig:distance_dsitribution}, the distributions of the distances for sources with good quality ($\overline{\omega}/\sigma_{\overline{\omega}}>=3$) and bad quality ($\overline{\omega}/\sigma_{\overline{\omega}}<3$) parallax measurements are plotted for both CCGCS and TD.

We see that very few AGNs have good quality distance measurements. Since {\sl Gaia} can only reliably measure distances up to a few kiloparsecs,  all extragalactic sources are expected to have unreliable distance measurements.
HM-STARs and YSOs appear to have systematically higher values of {\sl rgeo} compared to LM-STARs, which is consistent with the fact that LM-STARs in the TD and CCGCS have to be relatively nearby to be detected in X-rays and/or to be reliably classified via their optical spectra (for the inclusion into the TD). 

\begin{figure*}
\begin{center}
\includegraphics[scale=0.5,trim=0 0 0 0]{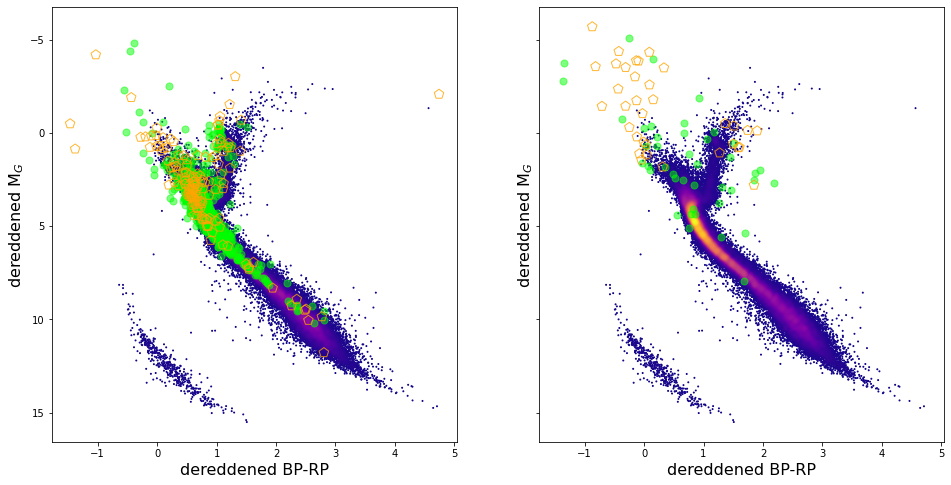}
\caption{The left panel shows the extinction-corrected color-magnitude diagram (CMD) for CCGCS of LM-STAR class as green circles and LM-STARs from the TD as open orange pentagons. 
The right panel shows a similar plot for the HM-STARs. Both plots are overlaid on the CMD of the Gaia DR2 low-extinction stars from Figure 5 in \cite{2018A&A...616A..10G}. Color of the points indicates the local density.} 
\label{fig:H-R_stars}
\end{center}
\end{figure*}

\subsection{The Color-Magnitude Diagrams of Stars}

After applying the same data filtering as described in Section 2.1 of \cite{2018A&A...616A..10G} to ensure reliable astrometric solutions as well as accurate photometric measurements, 595 out of 794 LM-STARs and 66 out of 144 HM-STARs from CCGCS remained. We also apply the same data filtering for the TD leaving 157 out of 207 LM-STARs and 81 out of 118 HM-STARs. The distances are based on {\sl rgeo} \citep{2021AJ....161..147B}, and the extinction correction (dereddening) is applied using $E(B-V)$ inferred from  the {\sl combined19} model from the MWdust 3D extinction map \citep{2016ApJ...818..130B} using the extinction Python package\footnote{\url{https://github.com/kbarbary/extinction}} at the effective wavelength of each band.

The dereddened CMDs for both LM-STARs and HM-STARs from CCGCS and the TD are shown in Figure \ref{fig:H-R_stars} on top of the density plot of the CMD of a large number of low-extinction stars ($E(B-V)<0.015$ mag) from Figure 5 in \cite{2018A&A...616A..10G}. In these plots, we  exclude the sources from the Galactic center (within 1 deg$^2$) and a few known Galactic SFRs \citep{2002ARep...46..193A} with poorly known (and/or strongly variable) extinction (matched within 1$\arcmin$).

Most X-ray sources classified as LM-STARs align well with the main sequence while the HM-STARs exhibit a larger scatter in the upper left part of the CMD with higher luminosities and bluer colors. 
One interesting source is 2CXO J084621.1+013755, which is the notable outlier to the far right in the CMD. It is a Mira Cet type variable star in SIMBAD, which is a pulsating star characterized by a very red color.

\section{Examples of Classification Applications}
\label{sec:examples}

The rapid classification of a large number of X-ray sources enables many different applications. For instance, these classifications could be used to systematically search for X-ray-emitting compact objects in SNRs (see, e.g., \citealt{2017ApJ...839...59P}). They can also be used to search for rarely found source types (e.g., quiescent HMXBs, LMXBs, NSs) or to search for potential X-ray counterparts to the numerous unidentified $\gamma$-ray (TeV and/or GeV) sources (see e.g., \citealt{2016ApJ...816...52H,2017ApJ...841...81H}). Below we use our classification results of CSCv2 sources to perform a more detailed exploration of 5 sources classified as HMXBs with high confidence (see Section \ref{sec:HMXBs}). 
We also explore a sample of unidentified H.E.S.S. sources (see Table \ref{tab:HESSGPS}) with potential X-ray counterparts having interesting  classifications (particularly, NS class; see Section \ref{sec:HESS}). 

\subsection{HMXBs} \label{sec:HMXBs}

There are two known HMXBs that were not included in our TD and hence were classified by the pipeline.  
2CXO J142112.1--624156 has the highest probability (among all classified sources) of being an HMXB ($P_{\rm HMXB}=$82\%$\pm$10\%). This source, also known as 2S 1417--642, is, in fact, an HMXB comprised of an accreting pulsar with a 17\,s spin period orbiting a Be star with 42 days orbital period. It was observed by {\sl Chandra} in quiescence \citep{2017MNRAS.470..126T} and by Neutron Star Interior Composition Explorer during an outburst \citep{2021arXiv210313444M}. There is another known HMXB (2CXO J171556.4--385154) correctly classified as an HMXB with $P_{\rm HMXB}=$57\%$\pm$16\% and somewhat lower CT=1.6 than  the threshold CT = 2 adopted above. 
This source is also known as XTE J1716--389 and is an obscured HMXB (see \citealt{2010MNRAS.408.1866R}). Both sources were not included in our TD because their coordinates in the \cite{2006A&A...455.1165L} HMXB catalog were inaccurate and offset from their {\sl Chandra} positions by $>5\arcsec$. 
These two sources will be added to the TD in future MUWCLASS pipeline releases.

2CXO J193309.5+185902 is classified as an HMXB with a probability of $P_{\rm HMXB}=$69\%$\pm$14\%. 
Interestingly, the source lies within the radio SNR G54.4--0.3, which is estimated to be at a distance of $d=3-7$\,kpc \citep{1992A&AS...96....1J,1998ApJ...504..761C} although more recent estimates favor $d=5-7$\,kpc \citep{2017ApJ...843..119R,2020AJ....160..263L}. The radio-quiet PSR J1932+1916  is a middle-aged (spin-down age 35 kyr) GeV  pulsar, which  lies just outside of the northern rim of the SNR. It is still unclear whether or not the pulsar is related to the SNR \citep{2013ApJ...779L..11P,2017MNRAS.466.1757K,2019JPhCS1400b2018M}. 
The {\sl Gaia} EDR3 counterpart of 2CXO J193309.5+185902 has Gmag=12.6 and a parallax distance of $d=3.14^{+0.15}_{-0.12}$\,kpc, with significant astrometric noise ($>6$) and a renormalized unit weight error of 1.12, which suggests that the optical source is in a binary. Note that the distance estimate may be inaccurate given the large astrometric noise. Additionally, the optical counterpart is likely a known H$\alpha$ emission line star, but no spectral type for the star has been published\footnote{The coordinates of the H$\alpha$ emitting star published in  \cite{1999A&AS..134..255K} appear to be $\approx8\arcsec$ offset from those of the {\sl Gaia}'s optical counterpart to the X-ray source. However, because there are no other relatively bright stars within 10$\arcsec$ of 2CXO J193309.5+185902,  the position listed in \cite{1999A&AS..134..255K} is likely inaccurate.} \citep{1999A&AS..134..255K}. The 3D extinction maps of \cite{2016ApJ...818..130B} give $E(B-V)=1.26$ at the source's distance. Assuming the distance is correct, the dereddened {\sl Gaia} absolute magnitude, $M_{\rm G}=-3.1$ and color $BP-RP=-0.1$, place this source in the O--B star region in the {\sl Gaia} CMD, supporting the HMXB classification. At the parallax distance of 3.1\,kpc, the source would have an observed X-ray luminosity of $L_{\rm X}=2\times10^{32}$\,erg\,s$^{-1}$. 
An absorbed power-law model fit\footnote{All spectra fit in this section were taken from the CSCv2 data products \citep{2020AAS...23515405E}.} to the source's spectrum gives $N_{\rm H}=(1.0\pm0.5)\times10^{22}$\,cm$^{-2}$, and a photon index $\Gamma=1.7\pm0.4$. 
The spectral slope is compatible with  those of a quiescent Be XRB, high-mass $\gamma$-ray binary (HMGB; \citealt{2013A&ARv..21...64D}), or perhaps a $\gamma$ Cas type analog, which have comparable luminosities (see, e.g., \citealt{2016AdSpR..58..782S}). Future spectroscopic follow-up of this source can help to further elucidate its nature.

2CXO J085910.9--434343 has an HMXB classification probability $P_{\rm HMXB}$ = 64\%$\pm$9\%, with its second-highest classification probability being a YSO ($P_{\rm YSO}=$30\%$\pm$8\%). The source position on the sky overlaps with the young open cluster RCW 36, which has an age of $\sim$1 Myr and is located at a distance of 700\,pc (see,  e.g., \citealt{2013A&A...558A.102E}). The source, which is highly variable (i.e., $P_{\rm intra}\approx1.0$), has a {\sl Gaia} EDR3 parallax distance of $d=755^{+244}_{-139}$\,pc, placing it in the RCW 36 cluster. Recently, \cite{2021ApJ...916...32G} studied this source and found that it is a flaring pre-main-sequence star, which showed a ``superflare'' reaching a peak X-ray luminosity of $L_{\rm X}\approx10^{32}$\,erg\,s$^{-1}$. The source is also strongly absorbed, with $A_V=9.1$ \citep{2021ApJ...916...32G}. It is likely that the large absorbing column density and strong variability led to the misclassification as an HMXB. According to the confident precision CM (Figure \ref{fig:LOOCV}, right bottom panel), a small fraction of sources confidently classified as HMXBs is expected to be actually YSOs.

2CXO J201641.4+370925 is classified as an HMXB with $P_{\rm HMXB}=$61\%$\pm$11\% 
and its second-highest classification probability being a YSO with $P_{\rm YSO}=$17\%$\pm$6\%. The source has a counterpart in {\sl Gaia} EDR3 with Gmag=18.4. 
The parallax measurement places it at a distance of about 3\,kpc \citep{2018A&A...616A...1G,2021AJ....161..147B}. At this distance, the source has an observed luminosity of $L_{\rm X}=4\times10^{31}$\,erg\,s$^{-1}$. The 3D extinction maps of \cite{2016ApJ...818..130B} give an absorption of $E(B-V)=0.8$ at this distance. After correcting for absorption, the optical counterpart's absolute {\sl Gaia} G-band magnitude is $M_{\rm G}=$3.76, and the color is $BP-RP=0.75$. This places the source on the blue side of the main sequence in the {\sl Gaia} CMD. We note that the parallax distance for this source has rather large uncertainties, spanning distances of $d\approx2-5$\,kpc. At the largest distance of 5\,kpc, the absorption is larger ($E(B-V)=1.8$), and the source is consistent with an HM-STAR. Additionally, the X-ray luminosity becomes $L_{\rm X}\approx1\times10^{32}$\,erg\,s$^{-1}$, which is consistent with those of  HMXBs in quiescence, HMGBs, or $\gamma$ Cas type Be binaries. However, the source is faint with very few counts having energies below $1.8$\,keV. An absorbed power-law fit to the spectrum gives a photon index $\Gamma=0.4\pm0.8$, suggesting a hard spectrum albeit with very large uncertainty. The source is relatively bright at NIR wavelengths having a 2MASS $K$-band mag of 14.3 \citep{2006AJ....131.1163S}, thus spectroscopic observations of the source could help to better understand its nature. 

\begin{deluxetable*}{lll}
\tablecaption{Summary of Confidently Classified Significant (S/N$>$6) X-Ray Sources within the Fields of 20 Unidentified H.E.S.S. Sources \label{tab:HESSGPS}}
\tablewidth{0pt}
\tablehead{
\colhead{HESS Source Name} &  \colhead{Summary of Confident Classifications} &  \colhead{Possible Association} 
}
\startdata
 HESS J1912+101 & 2LM-STAR, 1NS & 2CXO J191237.9+101044 (NS)  \\
 HESS J1843--033 & 1NS, 8YSO & 2CXO J184335.8--034653 (NS) \\
 HESS J1813--126 & 1AGN, 1YSO &  PSR J1813--1246 / 2CXO J181303.0--124907 (AGN) \\
 HESS J1018--589B & 1AGN, 1HMXB, 1LM-STAR & PSR J1016--5857 / 2CXO J101812.9--585930 (AGN)  \\
 HESS J1614--518 & 2NS & unknown SNR? / 2CXO J161610.8--515545 (NS)? \\
 HESS J1841--055 & 1NS, 2YSO & 2CXO J184201.9--052823 (NS) \\
 HESS J1800--240 & 1NS, 5YSO & 2CXO J180010.0--240129 (NS) \\
 HESS J1857+026 & 1LM-STAR, 1NS, 1YSO & PSR J1856+0245 / 2CXO J185643.6+021921 (NS) \\
 HESS J1848--018 & 1NS, 1YSO & \nodata \\
 HESS J1023--575 & 5HM-STAR, 1LM-STAR, 29YSO & \nodata \\
 HESS J1458--608 & 1YSO & \nodata  \\
 HESS J1616--508 & 1LM-STAR, 4YSO & \nodata  \\
 HESS J1632--478 & 2YSO & \nodata  \\
 HESS J1634--472 & 3YSO & \nodata  \\
 HESS J1713--381 & 1YSO  & \nodata  \\
 HESS J1718--385 & 4YSO & \nodata  \\
 HESS J1745--303 & 1LM-STAR & \nodata  \\
 HESS J1826--130 & 4YSO & \nodata \\
 HESS J1844--030 & 1YSO & \nodata \\
 HESS J1923+141 & 6YSO & \nodata \\
\enddata
\tablecomments{In the possible association column, we list a possible association based on our classifications and (in some cases) previously proposed association. Note that for extended or multicomponent TeV sources more than one association is possible. In this table, we restrict ourselves to sources that pass the detection  S/N$\ge6$ and CT$\ge2$ cuts. The supplementary products from the GitHub repository contain all  classifications for S/N$\ge6$ as well as for S/N$\ge3$.   Note that the fields of many TeV sources are only partly covered (see Figure~\ref{fig:HESSGPS}). Also, the analysis is restricted to data existing in CSCv2  (which means we did not use any Chandra observations after 2014). For 2CXO J161610.8--515545 (NS), the question mark is added because the X-ray source is very far from the TeV source center.}
\end{deluxetable*}

\subsection{Unidentified TeV Sources from the H.E.S.S. Galactic Plane Survey} \label{sec:HESS} 

\begin{figure*}
\begin{center}
\includegraphics[width=500pt]{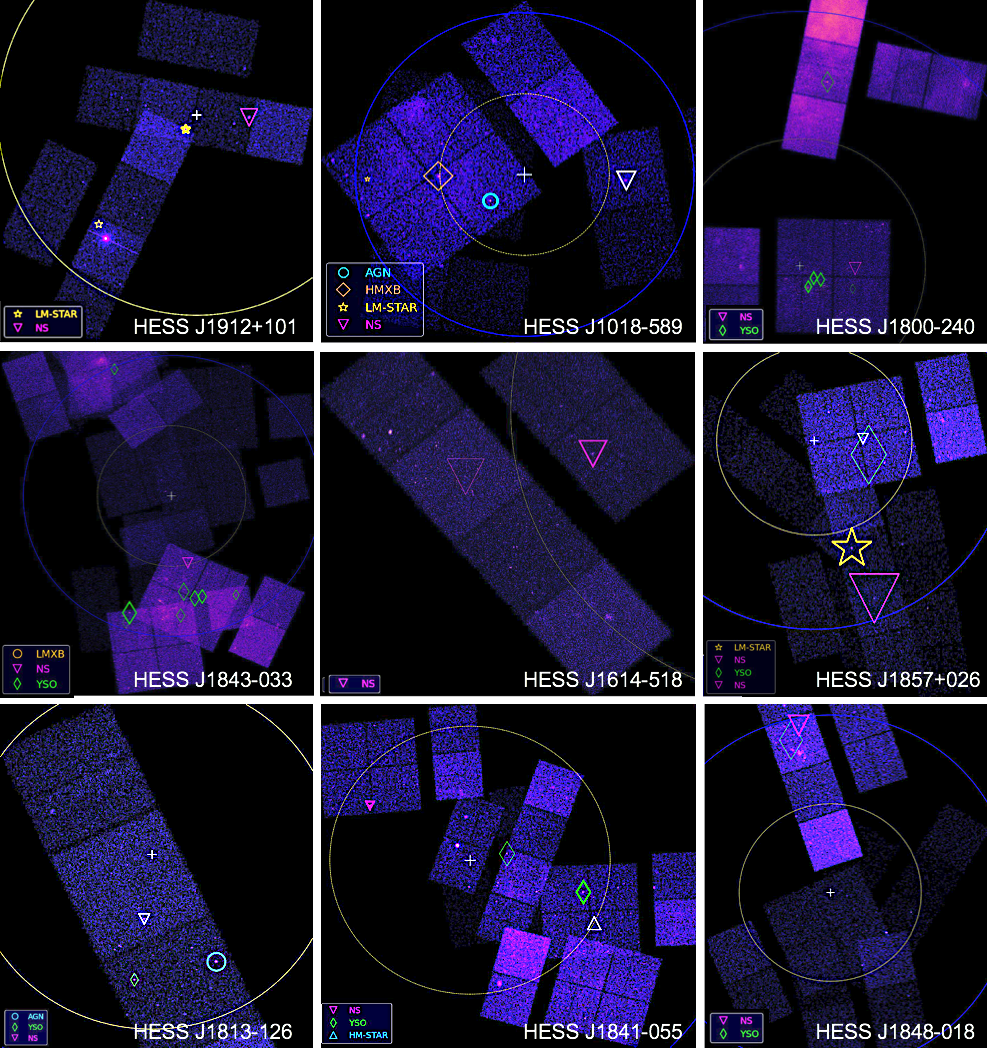}
\caption{The Chandra images of nine H.E.S.S. fields of those we investigated in Section \ref{sec:HESS}. The names of the unidentified H.E.S.S. sources are shown in the lower right corner of each image. Each Chandra image consists of ObsIDs made before the end of 2014, which were used for CSCv2. 
White crosses mark the center of the H.E.S.S. sources, and yellow dashed circles indicate the size for resolved H.E.S.S. sources or the 68\% PUs for point H.E.S.S. sources, and blue solid circles are twice the size of the yellow dashed circles, which we used to search for potential X-ray counterparts of the TeV emission. Symbols show significant (S/N$\ge$6) X-ray sources with confident classifications (CT$>2$): gold stars$-$LM-STARs, magenta inverted triangles$-$NSs, cyan circles$-$AGNs, orange diamonds$-$HMXBs, green thin diamonds$-$YSOs, orange circles$-$LMXBs, and cyan triangles$-$HM-STARs. White inverted triangles mark NSs, and white triangles mark HM-STARs from the TD. The widths of the symbol lines increase with the classification probability while the symbol sizes increase with the X-ray detection significance. Similar pictures are available at the GitHub repository (see Section \ref{sec:HESS}) for all unidentified TeV sources listed in Table \ref{tab:HESSGPS}, including less confident classifications.
} 
\label{fig:HESSGPS}
\end{center}
\end{figure*}

Here we explore some interesting X-ray sources (detected with  S/N$\ge6$) within the extent of unidentified TeV sources from the HGPS (\citealt{2018A&A...612A...1H}) that were confidently classified by MUWCLASS pipeline. 
Since many pulsars move fast, we classify all X-ray sources within 2 times the TeV source  ``size" value, corresponding to the actual source size for resolved sources, or 2 times the 68\% positional uncertainty values for unresolved sources in HGPS \citep{2018A&A...612A...1H}. Unlike GCS, we do not perform filtering on the X-ray PUs and CSCv2 source flags to allow for extended sources to be also classified because pulsars surrounded by compact (and relatively faint) PWNe are often expected as counterparts to Galactic TeV sources (see, e.g., \citealt{2013uean.book..359K}). Individual X-ray source classifications (with all confidences and for all source detected with S/N$\ge3$) are provided in the electronic (machine-readable) format (a few {\sl Chandra} sources discussed below are shown in Table \ref{tab:HESSMRT}, which represents a part of that larger table; see Appendix \ref{sec:MRTs}). 
Additional data products (e.g., source class visualizations on top of X-ray images; see Figure \ref{fig:HESSGPS} for examples) are available at the GitHub repository. 
The summary table of all {\em confident} (i.e., CT$\ge2$) classifications for  X-ray sources detected with S/N$\ge6$ is given in Table \ref{tab:HESSGPS}. 
Below we  discuss some interesting cases in more detail.

The field of the unidentified source HESS J1912+101 contains a relatively bright  ($F_{\rm b}\approx3\times10^{-13}$\,erg\,s$^{-1}$\,cm$^{-2}$) X-ray source 2CXO J191237.9+101044. 
It is confidently classified as an NS ($P_{\rm NS}=$78\%$\pm$15\%). Two potential scenarios have been suggested for HESS J1912+1901. 
In the first scenario, the TeV emission is produced by a PWN from the energetic pulsar PSR J1913+1011; however, no X-ray emission was detected from the pulsar or its putative PWN \citep{2008ApJ...682.1177C}. The second scenario is that there is a low surface brightness SNR producing the TeV emission (see, e.g., \citealt{2015ICRC...34..886P}), but no clear SNR has yet been detected (see, e.g., \citealt{2017ApJ...845...48S,2019RAA....19...45R}). 2CXO J191237.9+101044 is labeled as an extended source in the CSCv2, but its 10$'$ off-axis location makes it difficult to differentiate between extended emission or multiple point sources, as {\sl Chandra}'s PSF degrades with increasing off-axis angle. The X-ray spectrum of the source is  well fit by an absorbed power-law model with $N_{\rm H}=(0.7\pm0.4)\times10^{22}$\,cm$^{-2}$, and $\Gamma=1.3\pm0.4$.  Additionally, there are no NIR sources coincident with  the {\sl Chandra} source in the deeper UKIDDS Galactic plane survey data \citep{2008MNRAS.391..136L}.
If the source is truly extended, it may be a pulsar surrounded by a PWN, in which case it could be responsible for the TeV emission from HESS J1912+1901. A deeper on-axis {\sl Chandra} observation of this source could help to resolve the extended emission and better constrain the X-ray spectrum.

Another unidentified TeV source, HESS J1843--033, is resolved into three peaks in most recent HGPS images. The most separated peak was labeled as a new TeV source (HESS J1844--030), while HESS J1843--033 consists of two ``hotspots'', which we refer to as HGPSC 083 and HGPSC 084 (see Figure 28 in \citealt{2018A&A...612A...1H}). Given the presence of  hotspots and  the diffuse structure of HESS J1843--033, it is likely that the TeV emission comes from multiple sources, including faint emission from SNR G28.6--0.1 \citep{1989ApJ...341..151H,2003ApJ...588..338U,2018A&A...612A...1H}. Interestingly, the HGPSC 084 region contains a source 2CXO J184335.8--034653, which is confidently classified as an NS, with $P_{\rm NS}=$87\%$\pm$16\%. 
The source is also flagged by the CSCv2 as being extended, which may suggest that it is a pulsar surrounded by a PWN. 
However, the source is rather faint ($F_{\rm b}=2.7\times 10^{-14}$\,erg\,cm$^{-2}$ s$^{-1}$) and is imaged $\sim 7.6'$ off-axis. 
The source's spectrum is not well constrained, but appears to be fairly hard, having an absorbed power-law photon index $\Gamma=1.1\pm0.9$, which is consistent with a PSR-PWN scenario. There are no MW counterparts to the source, even in deeper surveys that we checked.
A deeper on-axis observation is needed to confirm the source classification and probe the extended nature of the source.

HESS J1813--126 is an unidentified TeV source spatially coincident with middle-aged GeV pulsar PSR J1813--1246 (included in our TD), which has been proposed as the X-ray counterpart to HESS J1813--126 \citep{2009Sci...325..840A}. However, \cite{2014ApJ...795..168M} found that this fairly energetic pulsar does not have an X-ray PWN, which casts some doubt on a PWN scenario. 
2CXO J181303.0--124907 lies within the extent of HESS J1813--126 and is offset from PSR J1813-1246 by $\sim6'$. This source is confidently classified by our pipeline as an AGN with $P_{\rm AGN}=$95\%$\pm$4\%. Based  on the HR of the source, \cite{2014ApJ...795..168M} also concluded that this source was likely an AGN. The source's X-ray spectrum is well fit by an absorbed power-law model with $N_{\rm H}=(1.2\pm0.7)\times10^{22}$\,cm$^{-2}$, and $\Gamma=1.0\pm0.4$, supporting the AGN classification. Additionally, the X-ray source position is coincident, within 2$\sigma$, with a bright (181$\pm6$\,mJy at 1.4 GHz) radio point source detected in the NVSS survey \citep{1998AJ....115.1693C}, further supporting its AGN nature. Future NIR--IR spectroscopic observations can help to constrain the AGNs redshift and determine whether or not it is plausible that the AGN contributes significantly to the TeV emission. 

HESS J1018--589 is composed of two components, A and B, with A having a confirmed X-ray counterpart and B having a plausible X-ray counterpart. HESS J1018--589A has been shown to have TeV emission produced by the HMGB 1FGL J1018.6--5856 (\citealt{2015A&A...577A.131H}; 2CXO J101855.6--585645 in CSCv2). This source is not in our TD as it was not yet discovered when the HMXB catalog we use was produced.
Therefore, it serves as a further test of the performance of our pipeline, which correctly and confidently classified 1FGL J1018.6--5856  as an HMXB with a probability $P_{\rm HMXB}=$74\%$\pm$15\%. The plausible counterpart to HESS J1018--589B is  PSR J1016--5857, which has an X-ray PWN  and is possibly associated with SNR G284.3--1.8 \citep{2012A&A...541A...5H,2022arXiv220413167K}. However, we also confidently classify an AGN, 2CXO J101812.9--585930, which lies in between 1FGL J1018.6--5856 and PSR J1016--5857. Interestingly, this source is also reported in the 40 months NuSTAR serendipitous source catalog \citep{2017ApJ...836...99L}, as it has a hard spectrum with a photon index $\Gamma=1.3\pm0.2$. Unfortunately,  the source is highly absorbed; thus no redshift was measured, so the distance to the AGN is unconstrained. An NIR--IR spectrum should be taken to constrain the AGN's redshift to determine if it can contribute to the TeV emission in this complex region.

HESS J1614--518 is a very extended ($\approx0.8^{\circ}$ in diameter) TeV source with shell-like (SNR?) morphology,  which is mostly located outside of the field of view of  the {\sl Chandra} ACIS observation. Out of two NS candidates identified in the {\sl Chandra} image, only one (2CXO J161610.8--515545) overlaps with the eastern outskirts of the TeV source. The X-ray source lacks any optical or NIR counterpart and is confidently classified as an NS   with $P_{\rm NS}=$85\%$\pm$16\%. However, due to its off-center location and a very marginal coverage of HESS J1614--518 by {\sl Chandra}, the case for the actual association with HESS J1614--518 is rather weak. A deeper on-axis {\sl Chandra}  observation of 2CXO J161610.8--515545 is needed to establish its nature reliably  as the source is rather faint ($F_{\rm b}\approx4.7\times 10^{-14}$\,erg\,s$^{-1}$\,cm$^{-2}$) and is currently imaged $\approx 11'$ off-axis (which excludes the source from GCS). The other two NS candidates identified in the same {\sl Chandra} observation are located  significantly farther from HESS J1614--518 and hence are even less likely to be related to it. 

HESS J1841--055 is a similarly large source, which is somewhat better covered by {\sl Chandra} thanks to the multiple ACIS observations, one of which (observation ID, hereafter ObsID, 7552) is both fairly close to the TeV source center and relatively deep (18.9 ks). In this observation, we classified 2CXO J184201.9--052823 as a confident NS candidate with $P_{\rm NS}=$84\%$\pm$17\%. However, the source, which lacks any MW counterparts, is rather faint  ($F_{\rm b}\approx5.1\times 10^{-14}$\,erg\,s$^{-1}$\,cm$^{-2}$) and is located about $7'$ off-axis, which limits any further analysis with the existing {\sl Chandra} data. 

HESS J1800--240 consists of hotspots A and B. The latter hotspot coincides with the massive SFR G5.89--0.39 (analyzed by \citealt{2016JHEAp..11....1H}). It was covered by {\sl Chandra} ACIS observation where an X-ray source  2CXO J180010.0--240129 is classified as an NS with $P_{\rm NS}=$73\%$\pm$11\%. This X-ray source is offset by about $5\arcmin$ from HESS J1800--240B position, which would not be unusual for an NS born in the SFR, because NSs can get a substantial kick velocity at birth, and a relic PWN, responsible for the TeV emission, may be offset from the current pulsar position. The classified source is fairly faint in X-rays ($F_{\rm b}\approx4\times 10^{-14}$\,erg\,s$^{-1}$\,cm$^{-2}$) and also has a faint ($J$=15) NIR (2MASS) source  located within the X-ray source’s PU.  However, given the large stellar density in this region, a chance coincidence probability is not negligible. We note that in our TD we have several NSs with accidental optical--NIR matches, which we intentionally did not remove to allow sources with accidental matches still to be classified as NSs rather than another class. Of course, NIR emission from a real isolated NS is not expected to be detectable with the surveys we use. Deeper optical observations and spectroscopy can help to determine if  2CXO J180010.0--240129  really has a counterpart and verify its  classification.  

HESS J1857+026 has a fairly complex TeV morphology, which has been modeled with two Gaussians in HGPS.  The brightest part of  the TeV source is likely associated with the known (included in our TD) PSR J1856+0245, which is likely responsible for the TeV emission. However, some contribution to the overall TeV emission of HESS J1857+026 can come from a different source. In the {\sl Chandra} ACIS ObsID 10513, which imaged the southwestern outskirts of the TeV source, we find 2CXO J185643.6+021921 confidently classified as an NS with $P_{\rm NS}=$83\%$\pm$17\%. The source has no MW counterparts and is relatively bright ($F_{\rm b}\approx3\times 10^{-13}$\,erg\,s$^{-1}$\,cm$^{-2}$). However, the short  5\,ks  {\sl Chandra} exposure precludes any detailed analysis, and a deeper observation may be warranted. 

Finally, we would like to mention HESS 1848--018, which is located in the direction of the massive SFR W43. We did not find any NS candidates within the central part of this extended TeV  source in {\sl Chandra} data represented in CSCv2. The only NS candidate  (2CXO J184850.3--012509; classified with $P_{\rm NS}=$83\%$\pm$15\%) is located $\approx 32'$ north of the TeV source center in the vicinity of a massive old (likely globular) cluster GLIMPSE C01. Therefore, it is not likely to be related to HESS 1848--018. The NS candidate source is relatively faint ($F_{\rm b}\approx4.7\times 10^{-14}$\,erg\,s$^{-1}$\,cm$^{-2}$) and has no MW counterparts. More recent {\sl Chandra} data (not in CSCv2) that exist for this field must be analyzed to further explore the nature of this source.

In many of the H.E.S.S. fields we have also confidently classified a number of YSOs, as well as LM-STARs and HM-STARs, which rules out a large number of X-ray sources as potential TeV counterparts, as these types of sources are not expected to produce TeV emission. 
We caution that currently our pipeline is not capable of identifying some classes of sources (such as NSs and HMXBs, which are typically associated with Galactic TeV sources) with high certainty (see the validation results in  Figure \ref{fig:LOOCV}). For example, even after the confidence cut (CT$\ge 2$; bottom right panel in Figure~\ref{fig:LOOCV}), 15\% of sources identified as NSs are actually AGNs (some of AGNs may also be TeV sources).  However, the pipeline classifications can be used to substantially reduce the number of sources that can be potential TeV emitters and to identify  promising targets for future in-depth investigations.     
We also note that some  H.E.S.S. fields  overlap with particularly crowded environments, such as HESS J1023--575, which is coincident with Westerlund 2. 
The classification results  for these fields are less reliable because of the potential confusion in optical--NIR.

\section{Limitations and Future Developments}
\label{sec:discuss}

MUWCLASS pipeline and the CSCv2 classification results presented here have a number of limitations, which users must be aware of. Below we will discuss these limitations and outline possible solutions, which we may implement in the future pipeline releases. We also encourage the users to clone the GitHub repository, independently run the pipeline, and try to improve it on their own.

\subsection{Limitations and Caveats}
\label{sec:limitation}

One of the biggest limitations is the uneven performance of MUWCLASS for different source classes that is, to a large degree, caused by the unbalanced TD although it can also be due to the heterogeneous astrophysical nature of sources from certain classes as well as due to the strong interstellar absorption in many regions of the Galactic plane. 
As one can see from the CM plots in Figure \ref{fig:LOOCV}, in most cases, we reliably classify AGNs, LM-STARs, and YSOs, but the classification does not work so well for other classes with a particularly bad performance for the LMXB class, which is strongly mixed with NSs. Aside from having few NS and LMXB sources in the TD, about 63\% of LMXBs does not have any MW counterparts (likely due to large distances and reddening) making them appear similar to NSs. On the other hand, 5 NSs in the TD have spurious optical counterparts, which we did not remove since some spurious counterparts are expected in the dense Galactic fields.  With these limitations in mind, one can use the pipeline to weed out AGNs, LM-STARs, and YSOs if the goal is to look for other kinds of sources, or to create a combined class (e.g., NS+LMXB) to improve the performance if finding either of these is of interest. In the end, a manual investigation of a subset of interesting sources identified with the pipeline will likely still be needed (see Section \ref{sec:examples}).

Sources from the TD are systematically brighter than those from CCGCS (see Section \ref{sec:X-ray-flux}) as a result of the selection effect when building the TD. The literature-verified sources from the TD are classified using reliable traditional classification methods, which require the sources to be bright enough, e.g., NSs discovered by searching for the presence of a surface (pulsations, cyclotron resonance scattering feature, thermonuclear bursts, etc.), stars and AGNs classified based on the optical--IR--X-ray spectroscopic data.
This bias can lead to overestimation of the performance metrics that are evaluated using the TD. 
These selection effects can be mitigated by building a more complete TD by, e.g., creating synthetic fainter sources from the real ones by placing the latter  at larger distances with appropriate reddening applied to them.

The eight classes of X-ray sources used in this paper (see Section \ref{sec:TDCats}) represent a compromise between the desire to have a separate class for astrophysically different sources and the need to maintain statistically meaningful number of sources for each class. As a result, some of the classes are astrophysically very heterogeneous. For example, we group magnetars together with rotation-powered pulsars in our NS class, and we group red-back and black widow type systems with accreting LMXBs in our LMXB class. The HMXB class includes both accreting X-ray pulsars and gamma-ray binaries whose emission is likely powered by colliding winds. Moreover, some classes of sources are simply missing from our current TD. For example, we do not include GRBs, ULXs, tidal disruption events, or normal galaxies (not AGNs). 
Another way to deal with those rare type of classes that are not included in the current TD is to introduce an outlier detection algorithm to search for sources that have distinctly different properties and/or do not belong to any of the classes in the TD, which has been done in some other works \citep{2022AA...657A.138T}.

Although {\sl Chandra} source localizations are excellent compared to other X-ray observatories, the PUs of some sources can still be as large as a few arcsec (e.g., large off-axis angles, faint sources). This can lead to a large degree of confusion when looking for MW counterparts in the crowded regions of the Galactic plane. For this reason, we restrict PU$<1\arcsec$ for both GCS and the populous classes in the TD. However, even $1\arcsec$ uncertainty may be too large in some dense Galactic environments. We remove sources from the crowded environments (e.g., Galactic center, LMC/SMC, and globular clusters; see Section \ref{sec:TDCats}) in the TD, 
but we do not apply this filter apply to GCS. 
Therefore, one has to be cautious about any classification results for those dense fields. 
In general, PUs also include a systematic error that is associated with the accuracy to which the telescope pointing is known. In CSCv2 this systematic error is included by adding 0.71$\arcsec$ (95\% confidence, derived by comparing CSCv2 source positions to the SDSS data) in quadrature to the statistically averaged PUs\footnote{\url{https://cxc.cfa.harvard.edu/csc/columns/positions.html\#ra_dec}}. Currently, there is no attempt in CSCv2 
to correct for the imperfect astrometry by cross-correlating X-ray source positions with their {\sl Gaia} EDR3 counterparts' positions. This can be done in the future, but one has to be wary of parallaxes and proper motions, which may need to be taken into account when such astrometric corrections are performed (it is easy to do on a per-observation level, but then the X-ray position accuracy could be worse than that at the master level in CSCv2). Wherever the degree of confusion still remains large, a probabilistic crossmatching algorithm (e.g., Nway; \citealt{2018MNRAS.473.4937S}) can be used to take into account all possible counterparts within the positional error circle of an X-ray source.

The SFD extinction maps are less accurate at low Galactic latitudes (i.e., $|b|<5^{\circ}$; \citealt{1998ApJ...500..525S}), where many of GCS sources are located. The extinction maps from \cite{2014MNRAS.443.2907S} agree with the SFD maps within a factor of $\sim$2 (with most differences being within 20\%, see Figure 11 from \citealt{2014MNRAS.443.2907S}) in most places of the Galactic plane ($|b|<2^{\circ}$).
Due to the full sky coverage of the SFD extinction maps, we use them to select the $E(B-V)$ in a given direction to adjust the fluxes of AGNs from the TD to the region where the {\sl Chandra} sources are being classified. 
However, the extinction--absorption correction can be inaccurate if the dust column in a given direction is extremely large and/or there is a significant gradient  in the reddening on small angular scales.

\subsection{Future Developments}

This paper describes the first public release of MUWCLASS pipeline and provides some examples of its possible applications. 
We plan future releases of MUWCLASS pipeline as well as more detailed studies carried out with the help of the pipeline. By making the pipeline and the TD public and easily accessible, we hope to stimulate future progress in this direction, and to allow the broader community to contribute to the pipeline development. We discuss some of the possible future improvements below.    

To have a better balanced TD that is more adopted to the Galactic environment, it is important to increase the numbers of sources in currently underpopulated classes. We identified some resources that we plan to use in our future work, but we also hope that the community can contribute sources to an open database of classified X-ray sources, which will have the TD presented here as the starting point. The work on this database is ongoing, and a graphical user interface that visualizes the TD content is already available online\footnote{Our online visualization tool of the TD: \href{https://home.gwu.edu/~kargaltsev/XCLASS/}{https://home.gwu.edu/~kargaltsev/XCLASS/}} \citep{2021RNAAS...5..102Y}. Future additions can include (but are not limited to) various types of X-ray sources from the INTEGRAL reference source catalog \footnote{\url{https://www.isdc.unige.ch/integral/science/catalogue}} \citep{2003A&A...411L..59E}, LMXBs from The Swift Bulge Survey, CVs from \cite{2022MNRAS.511.4937S,2011AJ....142..181S,2009ApJ...696..870D}, and some other online catalogs and projects \citep{2005ApJS..160..319G,2007prpl.conf..313F,2009yCat.1280....0K}. 

In addition to increasing and balancing the TD, it is important to use deeper MW surveys and extend the MW coverage to the radio. The currently used 2MASS, WISE-based, and {\sl Gaia} surveys are all-sky allowing for a maximum coverage. It may be beneficial to use deeper new surveys in the Galactic plane, but the varying depth of the surveys needs to be taken into account to avoid potential biases. In particular, replacing 2MASS with the high-resolution sensitive NIR survey should be the top priority. This will greatly improve MUWCLASS's ability to differentiate between CVs, LMXBs, HMXBs, and NSs. The MW survey data that could be added in the future include the Dark Energy Survey \citep{2018ApJS..239...18A}, the Pan-STARRS Survey \citep{2016arXiv161205560C}, the VISTA surveys\footnote{\url{https://www.eso.org/public/teles-instr/paranal-observatory/surveytelescopes/vista/surveys/\#:~:text=Six\%20large\%20public\%20surveys\%20conducted,wide\%20range\%20of\%20scientific\%20questions.}}, the UKIRT Infrared Deep Sky Survey \citep{2007MNRAS.379.1599L}, and the VLT Survey Telescope Photometric H$\alpha$ Survey of the Southern Galactic Plane and Bulge \citep{2014MNRAS.440.2036D}.  
In addition, the latest generation of radio surveys is able to reach a depth and resolution which warrants including radio domain information among the MW features. As soon as ASKAP \citep{2021PASA...38....9H}, MeerKAT \citep{2016mks..confE...1J}, and VLASS surveys \citep{2020PASP..132c5001L} release their source catalogs, we plan to add radio fluxes to our TD. 

The time domain information in our current TD is limited to X-rays. The characterization of the variability time scale is also extremely coarse (i.e., intra and inter observation variability). In addition to determining how variable the source is, it would be beneficial to better to characterize the variability timescale within each observation using methods developed for the optical variability characterization, or to use per-observation data with extensive X-ray light curves. We also plan to add optical variability among the features in the future releases using such surveys as Transiting Exoplanet Survey Satellite \citep{2014SPIE.9143E..20R}, The Zwicky Transient Facility \citep{2019PASP..131a8002B}, All-Sky Automated Survey for Supernovae \citep{2017PASP..129j4502K}, which provide multiple observations and light curves, and {\sl Gaia} DR3 with extensive variability analysis \citep{2022arXiv220606416E}.

Thanks to a number of wide-area sensitive surveys, improved 3D extinction maps are becoming available (e.g., the 3D extinction maps of the northern Galactic plane based on IPHAS photometry; \citealt{2014MNRAS.443.2907S}). Besides the  reddening affecting AGNs when viewed through the Galactic plane, other sources inhabiting the Galactic plane also exhibit different reddening depending on their positions and distances. A better way to deal with the reddening bias between TD sources and unclassified sources in the Galactic plane is to use 3D extinction--absorption maps and distance information for the sources so that extinction--absorption corrections can be applied to all sources (not just AGNs) from the TD. A potentially better approach would be to deredden (deextinct) all sources in the TD and in the field to be classified before the classification. The reliability of this approach requires further investigation and testing because it may require filtering on the {\sl Gaia} EDR3 distance uncertainties causing different kinds of biases.

Finally, MUWCLASS is designed in such a way that it can be easily adapted to incorporate X-ray source catalogs from other existing X-ray observatories such as {\sl XMM-Newton} (in preparation), Swift-XRT, and eROSITA, as well as from future observatories like AXIS and Athena.

\section{Summary} 
\label{sec:summary}

We developed an automated MW classification pipeline (MUWCLASS) of X-ray sources, which relies on the supervised ML approach with the RF algorithm. The pipeline is released together with a large TD of literature-verified sources that we have constructed over a period of a few years. It is the first supervised ML classifier to be applied to a significant fraction of sources from the CSCv2, which takes into account the measurement uncertainties and augments the X-ray properties with rich MW properties from the {\sl Gaia} EDR3, 2MASS, and WISE catalogs. The main scientific products and the results are summarized below:

\begin{enumerate}
    \item A TD of 2941 literature-verified sources, consisting of 8 X-ray source classes, including AGNs, CVs, HM-STARs, HMXBs, LM-STARs, LMXBs, NSs and YSOs, has been compiled and made publicly available. It was supplemented by a convenient web-based viewing tool, which was released together with the TD\footnote{\url{https://home.gwu.edu/~kargaltsev/XCLASS/}}. 
    \item The pipeline combined several novel approaches that were not used previously at all or in combination. In particular, we included the MC method to account for the measurement uncertainties of the features and applied a field-specific reddening on the AGNs from the TD to account for the additional large reddening for AGNs viewed through the Galactic plane. 
    We also introduced a novel confidence classification definition (the CT parameter), which relies on our ability to calculate distributions for the classification probabilities given the MC approach. 
    \item We performed careful evaluation of the performance of MUWCLASS using LOOCV approach for the  TD and discussed various possible biases, limitations, and improvements. We found that the accuracy and balanced accuracy scores increase from 88.6\% and 70.2\%  to 97.0\% and 79.0\%  after the confidence cut ${\rm CT}=2$ while the overall completeness and balanced completeness  drop from 100\% to 81.0\% and 56.9\%. 
    \item We classified 66,369 GCS sources ($\sim$21\% of all CSCv2 sources) using MUWCLASS pipeline. We found $\sim$31,000 sources with reliable classification where most are classified as AGNs (74\%) and YSOs (20\%). 
    However, the classification results for other classes are less reliable and need further verification with manual investigations and follow-up studies. Although a large number of NSs have been identified, this may be the result of a selection bias related to the fact that the classified sources are generally fainter that the TD sources, thus are also more likely to be missing MW counterparts, similar to NSs.
    \item As an example of the scientific pipeline application, we studied a number of individual interesting sources, including several sources classified as HMXBs and X-ray sources within the extent of 
    20 unidentified HGPS sources. Several plausible NS candidates have been identified with some of them being possible particle accelerators associated with the TeV emission. 
    \item MUWCLASS pipeline and the main results of this work are made publicly available at the GitHub repository\footnote{\url{https://github.com/huiyang-astro/MUWCLASS_CSCv2}}. The machine-readable tables (MRTs) are also available for the properties and the classification results of GCS, TD and H.E.S.S. field sources (see Appendix \ref{sec:MRTs}). 
\end{enumerate}

Although the performance of the current MUWCLASS in many situations may be limited, it already can save substantial efforts by confidently identifying sources from some commonly occurring classes. Thus it can be used to efficiently shrink the sample of more exotic sources, which need follow-up investigations or in-depth analysis.  
We hope that, in the future, 
MUWCLASS will become a powerful exploration tool capable of making accurate predictions of the astrophysical source nature rapidly for a large number of sources. 
In addition to the pipeline development, we strongly believe that the expansion of the TD must be a community-driven effort with only limited mediation from the maintaining site.  We expect that future development of the pipeline will make it useful for classification of eROSITA and XMM-Newton catalogs. We also envision that the tool could be useful for rapid classification of archival sources in large fields associated with gravitational wave sources, neutrino sources, and ultra-high-energy cosmic ray sources. Although written with X-ray sources in mind, the pipeline can be easily modified to classify any other sources (optical, NIR, IR, radio, etc.)

\section{Acknowledgement}

We are grateful to Martin Durant, Rafael Martínez-Galarza, Bettina Posselt, and George Pavlov for  fruitful discussions that helped to advance this work. O.K. is also thankful to Derek Brehm who chose this topic for the BS degree thesis that served as a first step leading to this paper.

This work was supported by NASA Astrophysics Data Analysis Program (ADAP) award 80NSSC19K0576.
Partial support for this work was also provided by the National Aeronautics and Space Administration through Chandra Awards AR3-14017X, AR9-20005, and AR8-19008X   issued by the Chandra X-ray Observatory Center, operated by the Smithsonian Astrophysical Observatory for the National Aeronautics Space
Administration under contract NAS803060. 
J.H. acknowledges support from an appointment to the NASA Postdoctoral Program at the Goddard Space Flight Center, administered by the Universities Space Research Association through a contract with NASA.

Database: this work has made use of the Chandra Source Catalog, provided by the Chandra X-ray Center as part of the Chandra Data Archive \citep{2020AAS...23515405E}; the SIMBAD database, operated at Centre de Données astronomiques de Strasbourg (CDS), Strasbourg, France \citep{2000A&AS..143....9W}; the VizieR catalog access tool, CDS, Strasbourg, France \citep{2000A&AS..143...23O}; INTEGRAL General Reference Catalog; the BeSS database, operated at LESIA, Observatoire de Meudon, France\footnote{http://basebe.obspm.fr}; the TeVCat online source catalog\footnote{http://tevcat.uchicago.edu}.

Hardware: this work was completed in part with resources provided by the High Performance Computing Cluster at The George Washington University, Research Technology Services. 

Software: 
Astropy \citep{2013A&A...558A..33A}, Astroquery \citep{2019AJ....157...98G}, scikit-learn \citep{scikit-learn}, imbalanced-learn \citep{JMLR:v18:16-365}, hvplot\footnote{\url{https://hvplot.holoviz.org/}}.

\appendix

\section{Classification Performance Metrics}
\label{sec:performance-metrics}

For binary classifications, a comparison of the predicted label and true label can lead to four outcomes: true positive (TP; a test result that correctly indicates the presence of a condition or attribute), true negative (TN; a test result that correctly indicates the absence of a condition or attribute), false positive (FP; a test result that wrongly indicates that a particular condition or attribute is present), and false negative (FN; a test result that wrongly indicates that a particular condition or attribute is absent). The metrics commonly used to evaluate the performance of a classifier include accuracy, precision, recall, the F-1 score, the MCC, the ROC curve, and the CM, which are all built on the statistics of the four outcomes mentioned above.

Precision (also named as positive predictive value or efficiency) is the fraction of TP predictions among all predicted positive outcomes; 
recall (also known as TP rate or sensitivity) is the fraction of TP predictions among all true outcomes; specificity (also named as TN rate) is the fraction of actual negative outcomes that have been predicted as negative; F-1 score is the harmonic average of precision and recall:

\begin{eqnarray}
{\rm precision}  &= \frac{{\rm TP}}{{\rm TP} +{\rm FP}}, \\
{\rm recall}  &= \frac{{\rm TP}}{{\rm TP}+{\rm FN}}, \\
{\rm specificity} &= \frac{{\rm TN}}{{\rm TN}+{\rm FP}}, \\
\text{F-1}~{\rm Score}  &= \frac{2 \times {\rm precision} \times {\rm recall}}{{\rm precision} + {\rm recall}}.
\end{eqnarray}

In a multiclass classification problem, the metrics defined above for the binary classification can be utilized by treating one class as the positive class and all other classes as the negative class and repeating the calculations for each class. 
The so-called macro average is used to calculate the arithmetic average of the metrics for each class. For the imbalanced TD that we have, the macro average places the same weight on each class so that the metric will not be dominated by the contribution from the populous classes. 

For a multiclass case, accuracy is defined as the percentage  of outcomes that are correctly predicted as their true classes with respect to the  total number of outcomes. We also calculate the balanced accuracy which is the arithmetic average of the accuracy per class. 
The balanced accuracy is essentially a macro average of recalls. 

The MCC takes into account TP, FP, TN, and FN and is designed to be a classification metric resistant to class imbalance \citep{2020MCC}. The multiclass MCC is defined as follows\footnote{\url{https://scikit-learn.org/stable/modules/model_evaluation.html\#matthews-corrcoef}}:

\begin{equation}
    {\rm MCC} = \frac{c \times s - \sum_k^K p_k \times t_k}{\sqrt{(s^2-\sum_k^K p_k^2)(s^2-\sum_k^K t_k^2)}}
\end{equation}
where $t_k$ is the number of times class $k$ occurred, $p_k$ is the number of times class $k$ predicted, $c$ is the total number of samples correctly predicted and $s$ is the total number of samples.

The performance of the pipeline for each source class can also be visualized using a ROC curve. A ROC curve shows the performance (recall vs. $1-$ specificity) of a binary classifier as a function of classification confidence. A macro-averaged ROC curve can be calculated by averaging the ROC curves of each class, which can be treated as the performance on the whole TD taking into account the underpopulated classes. For each ROC curve, the AUC score represents an overall (averaged) performance of the classification pipeline. 
A perfect classifier has an AUC = 1 whereas a completely random classifier would have an AUC = 0.5.

The CM, also known as the error matrix, is another way to visualize the performance of an algorithm using a matrix layout where each row represents the number of sources in an actual (or predicted) class while each column provides the number in a predicted (or actual) class. Since our TD is highly imbalanced, we care more about the fraction of sources that are correctly predicted to each class rather than the absolute number. Therefore, we normalize the CM by the total number of sources in each row, such that each element represents the fraction of all true (predicted) sources of the row class, that are predicted (true) sources of the column class, to obtain the normalized recall (precision) CM. A more diagonal matrix represents a better performance.

\section{Convergence and Uncertainty Contributions in Classification}
\label{appendix:convergence-uncertainty}

While the MC method is conceptually straightforward and very flexible, it is recognized as lacking well-established convergence criteria \citep{2004WRR....40.4603B}. 
We implement a convergence analysis of the MC method that we use to account for the measurement uncertainties that are used to construct the classification probability distributions. 
Assuming the sample mean (arithmetic average) of the classification probability, $\overline{P_N} = \frac{1}{N}\sum_{i=1}^N P_i$, where $N$ is the number of samples used to calculate the mean,  will eventually converge to the true mean probability, we use a sample mean for a larger number ($N=3000$) of samples, 
$\overline{P_{3000}}$, as a pseudo true mean, for the convergence analysis. The sample variance of the classification probability with $N=3000$, ${\rm Var}[P_{3000}]$ is also calculated and 
used as a substitute for the true variance.

\begin{figure*}
\begin{center}
\includegraphics[scale=0.5,trim=0 0 0 0]{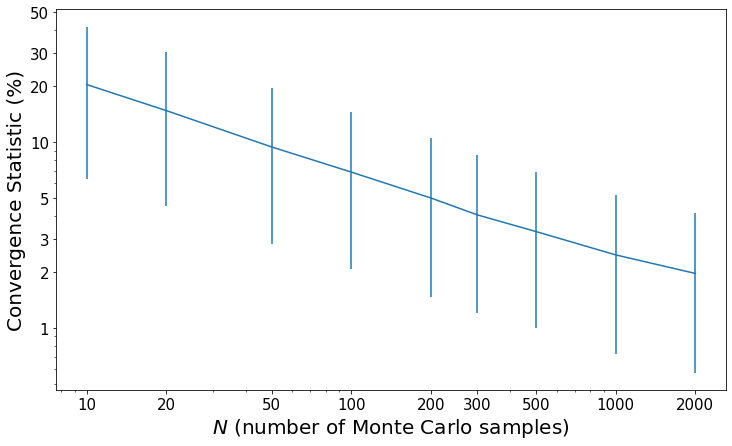}
\caption{The median of the convergence statistic $\frac{|\overline{P_N}-\overline{P_{3000}}|}{\sqrt{{\rm Var}[P_{3000}]}}$ calculated for GCS, with the errorbars indicating the 16th and 84th percentiles (corresponding to 1$\sigma$ uncertainty), as a function of the number of MC samples $N$. 
}
\label{fig:convergence}
\end{center}
\end{figure*}

We define the convergence statistic as the absolute deviation of the sample mean of the classification probability from the pseudo true mean $\overline{P_{3000}}$, divided by the square root of the pseudo true variance, $\sqrt{{\rm Var}[P_{3000}]}$, calculated as a function of the number of samples $N$, i.e., $\frac{|\overline{P_N}-\overline{P_{3000}}|}{\sqrt{{\rm Var}[P_{3000}]}}$. 
Since we are most interested in the reliability of the predicted class, only its classification probability (the highest classification probability among 8 classes) is used in the calculation. Figure \ref{fig:convergence} shows the median of the convergence statistics calculated for GCS, with errorbars indicating the 16th and 84th percentiles, as a function of the number of MC samples $N$. 
Given our computational resource limitations, we set the number of MC samples $N=1000$ such that 84\% of the convergence statistics values (for GCS) are less than 5\%. 
We note that, even if $N$ is reduced by a factor of 10 ($N=100$), 84\% of the convergence statistics values are still less than 14\%.

\begin{figure*}
\begin{center}
\includegraphics[width=350pt,trim=0 0 0 0]{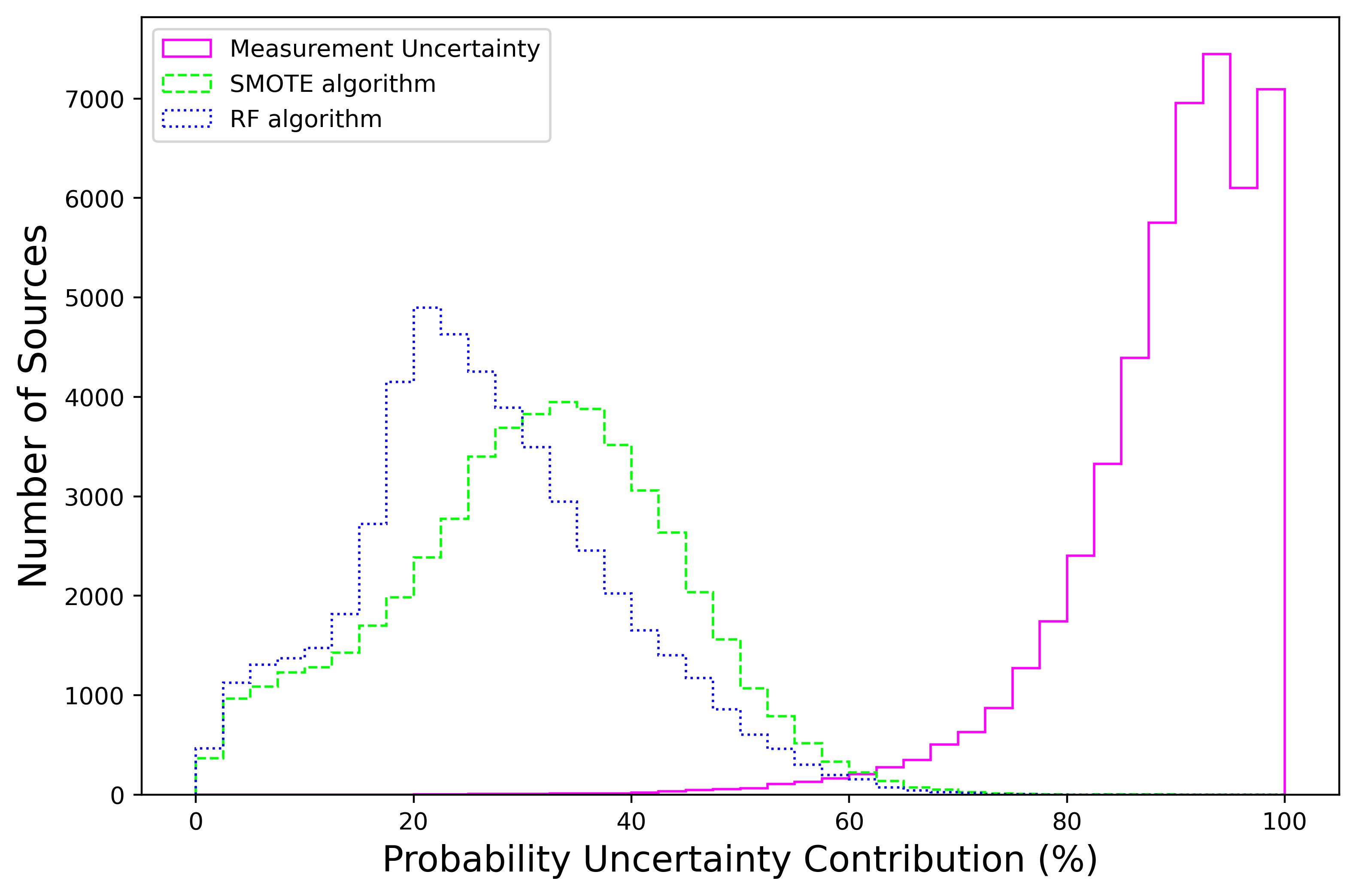}
\caption{The distributions of the probability uncertainty contributions for GCS demonstrating the relative importance of various uncertainty factors.} 
\label{fig:uncertainty}
\end{center}
\end{figure*}

We evaluate the contributions to the uncertainties of the classification probabilities from three processes that involve random sampling: (1) MC sampling to account for the feature measurement uncertainties; (2) oversampling with the SMOTE algorithm; and (3) the RF algorithm. To evaluate the relative contributions to the classification uncertainties from each process, we enable randomness in one process but not in the others and repeatedly run the classification pipeline 1000 times for each source in GCS. 
We then also run the classification pipeline with the randomness enabled for all three processes to be used as the benchmark.  
For each of these three sampling processes, we calculate the sample mean and the sample error (the square root of the sample variance) of the classification probability for the predicted class.
For each source, we evaluate three classification probability uncertainties (defined as the standard deviation of the classification probabilities characterizing the width of the distributions; see Figure \ref{fig:probability-histogram}) for the predicted class, which represent contributions to the total probability uncertainty from the measurement uncertainties, SMOTE algorithm, and RF algorithm, respectively. 
The individual contribution is calculated by dividing the probability uncertainty from one process by the combined probability uncertainty from three processes, which is calculated by adding the three probability uncertainties in quadrature.  
The distributions of the probability uncertainty contributions are shown in Figure \ref{fig:uncertainty}. 
The mean and the standard deviation of the distribution for the uncertainty contributions are 89\%$\pm$9\%,  31\%$\pm$13\%, and 27\%$\pm$12\% for the measurement uncertainties, SMOTE algorithm, and the RF algorithm respectively. 
The main uncertainty contribution is the measurement uncertainty. The SMOTE method and RF algorithm contribute comparable uncertainties, which are less than the contribution from the measurement uncertainty. This underscores the importance of including the measurement uncertainty in the classification procedure.

It is likely that increasing n\_estimators can decrease the uncertainty contribution from the RF algorithm. Thus, we recommend using larger n\_estimators (e.g., 1000) when a smaller number of MC samplings ($\le10$) is chosen, e.g., to reduce the computing time.

\section{Chance Coincidence Estimation for Positional crossmatching}
\label{sec:chance-match}

\begin{figure*}
\begin{center}
\includegraphics[width=350pt,trim=0 0 0 0]{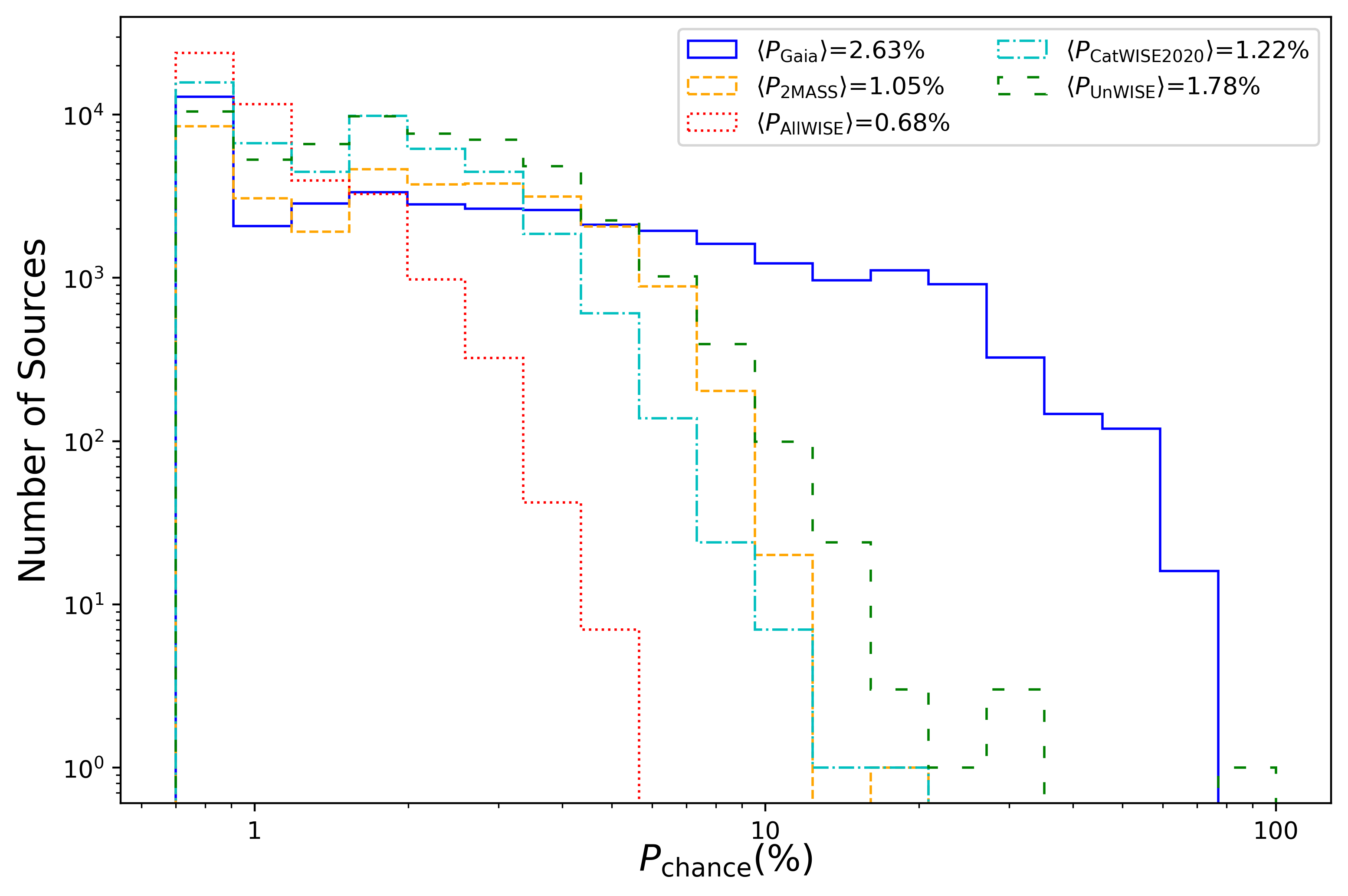}
\caption{The distribution of the chance coincidence probability for GCS. The average chance coincidence probability  calculated for each MW catalog is listed in the legend.} 
\label{fig:confusion-probability}
\end{center}
\end{figure*}

In this section, we estimate the potential rates of accidental matches of the MW counterparts to GCS. The X-ray sources are matched to MW catalogs (including {\sl Gaia} EDR3, 2MASS, AllWISE, CatWISE2020, unWISE) using the combined X-ray and MW PUs. 
The chance coincidence probability for a field with an average MW catalog source density, $\rho$ ($\rm arcsec^{-2}$), and 2$\sigma$ combined PU uncertainty radius, r ($\arcsec$), is given by $P_{\rm chance} = 1 - \exp(- \rho \pi r^2)$, which is simply the probability of having one or more MW catalog sources within a randomly placed circle of the chosen radius. 
We count all MW catalog sources within a larger  ($R=10\arcsec$) radius to calculate the field source density, $\rho = \frac{N}{A}$ where $N$ is the number of MW catalog sources within the $R=10\arcsec$ circle, and  $A=\pi R^2$ is the area.
We then calculate the chance coincidence probabilities for each source in GCS and plot their distributions in Figure \ref{fig:confusion-probability} with the average chance coincidence probability (per MW catalog) shown in the legend. 
The calculation presented here only offers a rough estimation and a variable radius (or other methods), for  measuring  the local source density, may be needed for more accurate estimates.

\section{Computational Efficiency}
\label{sec:computation}

A single classification within one extinction bin, requires $\sim$30 s with one core on The George Washington University Pegasus cluster. 
Classification of multiple sources sharing the same extinction bin takes almost the same time as for a single source. To classify all CSCv2 sources with $\sim$2000 extinction--absorption bins 1000 times using 20 nano nodes (each node with a Dual 20-Core 3.70\,GHz Intel Xeon Gold 6148 processor) on the Pegasus cluster running in parallel takes about 1 day. 
For comparison, a typical laptop (e.g., Acer Swift 3 with AMD Ryzen 7 4700U Processor, and an 8GB LPDDR4 memory) takes about 13 days to produce the same classification results for $\sim$66,000 GCS sources with 1000 MC samplings. Therefore, although it is very much possible to run the classification pipeline on the laptop, we recommend to reduce the number of MC samplings by a factor of 10  if the number of source to be classified is large.

\section{Examples of classified sources with their properties and column descriptions for the machine-readable table}
\label{sec:MRTs}

The properties and the classification results of GCS, TD, and of the X-ray sources within the unidentified H.E.S.S. sources are available in the form of three MRTs. A subset of each MRT is shown Tables \ref{tab:GCSMRT}--\ref{tab:HESSMRT}. 
We also provide the list and description of columns used in the three MRTs.
\\
{\it CSCv2}. This is CSCv2 source name in the form ``2CXO Jhhmmss.s\{+$|$--\}ddmmss". Column (1) is the shortened version of CSCv2 in Tables \ref{tab:GCSMRT}--\ref{tab:HESSMRT}. \\
{\it RAdeg, DEdeg}. These are the CSCv2 master source J2000 right ascension and declination, in degrees.\\
{\it PU}. This is the CSCv2 master source err\_ellipse\_r0, the major radius of the 95\% confidence level position error ellipse, in arcsecs, see column (2) in Tables \ref{tab:GCSMRT}--\ref{tab:HESSMRT}.\\
{\it S/N}. This is the CSCv2 master source X-ray significance; see column (3) in Tables \ref{tab:GCSMRT}--\ref{tab:HESSMRT}. \\
{\it Fb, Fs, Fm, Fh}. These are the average broad-, soft-, medium-, and hard-band fluxes (see their calculations in Section \ref{sec:Xdata}), in units of erg\,s$^{-1}$\,cm$^{-2}$. Column (4) is Fb in units of $10^{-13}$\,erg\,s$^{-1}$\,cm$^{-2}$ in Tables \ref{tab:GCSMRT}--\ref{tab:HESSMRT}. \\
{\it PIntra}.  This is the intra-observation variability probability, which is calculated from the highest value of Kuiper’s test variability probability across all observations available in CSCv2; see column (5) in Tables \ref{tab:GCSMRT}--\ref{tab:HESSMRT}. \\
{\it PInter}. This is the inter-observation variability probability (see its calculation in Section \ref{sec:Xdata}); see column (6) in Tables \ref{tab:GCSMRT}--\ref{tab:HESSMRT}. \\
{\it Gmag, BPmag, RPmag}. These are {\sl Gaia} EDR3 G-, BP-, and RP-band magnitudes. Column (7) is Gmag in Tables \ref{tab:GCSMRT}--\ref{tab:HESSMRT}. \\ 
{\it Jmag, Hmag, Kmag}. These are 2MASS J-, H-, and K-band magnitudes. Column (8) is Jmag in Tables \ref{tab:GCSMRT}--\ref{tab:HESSMRT}. \\ 
{\it W1mag, W2mag, W3mag}. These are W1-, W2-, and W3-band magnitudes from AllWISE, CatWISE2020, and unWISE catalogs (see their definitions in Section \ref{sec:MWdata}). Column (9) is W1mag in Tables \ref{tab:GCSMRT}--\ref{tab:HESSMRT}. \\ 
{\it plx}. This is {\sl Gaia} EDR3 absolute stellar parallax, in milliarcsecond. \\
{\it PM}. This is {\sl Gaia} EDR3 total proper motion, in milliarcsecond per year. \\
{\it rgeo, b\_rgeo, B\_rgeo}. These are median, 16th percentile, and 84th percentile of the geometric distance from {\sl Gaia} EDR3 Distances catalog, in parsecs. \\
{\it PAGN, PCV, PHM*, PHMXB, PLM*, PLMXB, PNS, PYSO}. These are the classification probabilities of the X-ray source to be classified as each of the eight classes in our TD. \\
{\it Class}. This is the predicted class of the source with the highest classification probability among eight classes; see column (10) in Tables \ref{tab:GCSMRT}--\ref{tab:HESSMRT}. \\
{\it ClassP}. This is the classification probability of the predicted class calculated from MUWCLASS; see column (11) in Tables \ref{tab:GCSMRT}--\ref{tab:HESSMRT}. \\
{\it CT}. This is the classification CT (see its definition in Section \ref{sec:performance}); see column (12) in Tables \ref{tab:GCSMRT}--\ref{tab:HESSMRT}. \\
{\it Flags}. This is the compilation of CSCv2 master source flags including conf\_flag (conf), extent\_flag (extent), and pileup\_flag (pileup), jointed by a ``$|$"; see column (14) in Table \ref{tab:TDMRT}. \\
{\it Catalog}. This is the name of the source from the literature-verified catalogs for the classification of TD sources. This column is only available for the TD MRT and the MRT of the H.E.S.S. field sources. \\
{\it TClass}. This is the true class of the source from the TD; see column (13) in Table \ref{tab:TDMRT}. This column is only available for the TD MRT and the MRT of the H.E.S.S. field sources.  \\ 
{\it ClassR}. This is the reference of the classifications of the TD sources. This column is only available for the TD MRT and the MRT of the H.E.S.S. field sources. \\
{\it HESS}. This is the H.E.S.S. source name that the CSCv2 source resides in. Column (13) is the shortened version of HESS\_name in Table \ref{tab:HESSMRT}. This column is only available for the MRT of the H.E.S.S. field sources.\\
Columns of 1$\sigma$ uncertainty are are available for Fb, Fs, Fm, Fh, Gmag, BPmag, RPmag, Jmag, Hmag, Kmag, W1mag, W2mag, W3mag, plx, PAGN, PCV, PHM*, PHMXB, PLM*, PLMXB, PNS, PYSO, ClassP columns which are e\_Fb, e\_Fs, e\_Fm, e\_Fh, e\_Gmag, e\_BPmag, e\_RPmag, e\_Jmag, e\_Hmag, e\_Kmag, e\_W1mag, e\_W2mag, e\_W3mag, e\_plx, e\_PAGN, e\_PCV, e\_PHM*, e\_PHMXB, e\_PLM*, e\_PLMXB, e\_PNS, e\_PYSO, e\_ClassP columns, respectively.

\begin{longrotatetable}
\begin{deluxetable*}{lccccccccccc}
\tablecaption{A Subset of GCS MRT\label{tab:GCSMRT}}
\tablewidth{700pt}
\tabletypesize{\scriptsize}
\tablehead{
\colhead{2CXO Name$^{\rm a}$} & \colhead{PU$^{\rm b}$} & 
\colhead{S/N$^{\rm c}$} & \colhead{$F_{\rm b}$} & 
\colhead{$P_{\rm intra}$} & \colhead{$P_{\rm inter}$} & 
\colhead{G} & \colhead{J} & 
\colhead{W1} & \colhead{Class} & \colhead{Class Prob} & \colhead{CT}  \\ 
\colhead{} & \colhead{(arcsec)} & \colhead{} &  \colhead{($10^{-13}$\,cgs)$^{\rm d}$} & \colhead{} & \colhead{} & \colhead{(mag)} & \colhead{(mag)} & \colhead{(mag)} & \colhead{} & \colhead{(\%)} & \colhead{}
} 
\startdata
J123635.2+621151 & 0.79 & 4.53 & $0.0038\pm0.0006$ & \nodata & 0.224 & \nodata & \nodata & \nodata & AGN & $91\pm5$ & 10.475\\
J141520.7+521114 & 0.86 & 3.33 & $0.02\pm0.005$ & \nodata & \nodata & \nodata & \nodata & $17.19\pm0.09$ & AGN & $40\pm10$ & 0.296\\
J031818.8--663230 & 0.72 & 7.87 & $0.12\pm0.01$ & 0.98 & 0.498 & $13.862\pm0.007$ & $12.22\pm0.02$ & $11.62\pm0.02$ & LM-STAR & $81\pm15$ & 2.543\\
J013647.4+154744 & 0.71 & 13.19 & $0.107\pm0.006$ & 1 & 0.996 & \nodata & $10.93\pm0.02$ & $9.95\pm0.02$ & LM-STAR & $91\pm6$ & 9.562\\
J161702.0--225834 & 0.72 & 11.03 & $0.25\pm0.03$ & 0.924 & \nodata & $14.796\pm0.008$ & \nodata & \nodata & HM-STAR & $39\pm10$ & 0.381\\
J201641.4+370925 & 0.93 & 9.16 & $0.33\pm0.04$ & \nodata & \nodata & $18.45\pm0.003$ & $15.59\pm0.07$ & $13.39\pm0.02$ & HMXB & $61\pm11$ & 2.587\\
J142112.1--624156 & 0.72 & 7.4 & $3.1\pm0.5$ & 0.979 & \nodata & $16.192\pm0.003$ & $13.3\pm0.02$ & $11.75\pm0.01$ & HMXB & $82\pm10$ & 4.836\\
J085910.9--434343 & 0.73 & 16.12 & $1.2\pm0.08$ & 1 & \nodata & $19.38\pm0.03$ & $14.48\pm0.03$ & $11.09\pm0.03$ & HMXB & $64\pm9$ & 2.106\\
J193309.5+185902 & 0.72 & 14.94 & $1.8\pm0.1$ & 0.906 & \nodata & $12.616\pm0.003$ & $10.32\pm0.02$ & $9.11\pm0.02$ & HMXB & $69\pm14$ & 2.785\\
J171556.4--385154 & 0.93 & 3.06 & $3\pm1$ & \nodata & \nodata & \nodata & 15.06 & \nodata & HMXB & $57\pm16$ & 1.557\\
\enddata
\tablecomments{The 1$\sigma$ uncertainties are shown whenever available. Only a portion of this table is shown here to demonstrate its form and content. A machine-readable version of the full table is available. $^{\rm a}$The shortened 2CXO name from the CSCv2 by removing the leading 2CXO string. $ ^{\rm b}$The 2$\sigma$ X-ray positional uncertainty. $ ^{\rm c}$The CSCv2 master source X-ray significance. $ ^{\rm d}$The energy flux in cgs units of erg\,s$^{-1}$\,cm$^{-2}$.}
\end{deluxetable*}
\end{longrotatetable}

\begin{longrotatetable}
\begin{deluxetable*}{lccccccccccccc}
\tablecaption{A Subset of TD MRT\label{tab:TDMRT}}
\tablewidth{700pt}
\tabletypesize{\scriptsize}
\tablehead{
\colhead{2CXO Name} & \colhead{PU} & 
\colhead{S/N} & \colhead{$F_{\rm b}$} & 
\colhead{$P_{\rm intra}$} & \colhead{$P_{\rm inter}$} & 
\colhead{G} & \colhead{J} & 
\colhead{W1} & \colhead{Class} & \colhead{Class Prob} & \colhead{CT} & \colhead{True Class} & \colhead{CSC flags} \\ 
\colhead{} & \colhead{(arcsec)} & \colhead{} &  \colhead{($10^{-13}$\,cgs)} & \colhead{} & \colhead{} & \colhead{(mag)} & \colhead{(mag)} & \colhead{(mag)} & \colhead{} & \colhead{(\%)} & \colhead{} & \colhead{} & \colhead{}
} 
\startdata
J084621.1+013755 & 0.71 & 21.31 & $5.3\pm0.3$ & 0.9 & \nodata & $6.27\pm0.007$ & $1.5\pm0.3$ & --1.8 & YSO & $36\pm10$ & 0.233 & LM-STAR & \nodata\\
J112401.1--365319 & 0.73 & 10.27 & $0.58\pm0.07$ & 0.973 & \nodata & \nodata & \nodata & \nodata & AGN & $57\pm8$ & 2.867 & LMXB & \nodata\\
J203213.1+412724 & 0.72 & 8.11 & $0.23\pm0.02$ & 0.943 & 1 & $11.28\pm0.01$ & $9.52\pm0.02$ & $8.48\pm0.02$ & HM-STAR & $57\pm8$ & 2.636 & HMXB & extent$|$conf\\
\enddata
\tablecomments{The 1$\sigma$ uncertainties are shown whenever available. The first 12 columns are the same as in Table \ref{tab:GCSMRT}. Only a portion of this table is shown here to demonstrate its form and content. A machine-readable
version of the full table is available.}
\end{deluxetable*}
\end{longrotatetable}

\begin{longrotatetable}
\begin{deluxetable*}{lcccccccccccc}
\tablecaption{A Subset of H.E.S.S. Field Source MRT\label{tab:HESSMRT}}
\tablewidth{700pt}
\tabletypesize{\scriptsize}
\tablehead{
\colhead{2CXO Name} & \colhead{PU} & \colhead{S/N} & \colhead{$F_{\rm b}$} & \colhead{$P_{\rm intra}$} & \colhead{$P_{\rm inter}$} & \colhead{G} & \colhead{J} & \colhead{W1} & \colhead{Class} & \colhead{Class Prob} & \colhead{CT}  & \colhead{HESS Name$^{\rm a}$} \\ 
\colhead{} & \colhead{(arcsec)} & \colhead{} &  \colhead{($10^{-13}$\,cgs)} & \colhead{} & \colhead{} & \colhead{(mag)} & \colhead{(mag)} & \colhead{(mag)} & \colhead{} & \colhead{(\%)} & \colhead{} & \colhead{} 
} 
\startdata
J191237.9+101044$^{\rm b}$ & 1.58 & 14.66 & $2.7\pm0.2$ & 0.762 & \nodata & \nodata & \nodata & \nodata & NS & $78\pm15$ & 2.585 & J1912+101\\
J161610.8--515545 & 2.02 & 7.15 & $0.47\pm0.07$ & 0.136 & \nodata & \nodata & \nodata & \nodata & NS & $85\pm16$ & 2.656 & J1614--518\\
J184201.9--052823 & 1.36 & 6.22 & $0.51\pm0.09$ & 0.629 & \nodata & \nodata & \nodata & \nodata & NS & $84\pm17$ & 2.275 & J1841--055\\
J180010.0--240129 & 0.78 & 11.39 & $0.4\pm0.04$ & 0.515 & \nodata & \nodata & $15.01\pm0.09$ & \nodata & NS & $73\pm11$ & 3.254 & J1800--240\\
J185643.6+021921 & 0.72 & 7.21 & $3.0\pm0.4$ & 0.026 & \nodata & \nodata & \nodata & \nodata & NS & $83\pm17$ & 2.185 & J1857+026\\
J184850.3--012509 & 0.92 & 11.51 & $0.47\pm0.05$ & 0.052 & \nodata & \nodata & \nodata & \nodata & NS & $83\pm15$ & 2.523 & J1848--018\\
J184335.8--034653$^{\rm b}$ & 1.23 & 8.44 & $0.27\pm0.03$ & 0.62 & \nodata & \nodata & \nodata & \nodata & NS & $87\pm16$ & 2.636 & J1843--033\\
J181303.0--124907$^{\rm b}$ & 0.77 & 16.68 & $1.39\pm0.09$ & 0.913 & \nodata & \nodata & $16.12\pm0.09$ & $13.68\pm0.04$ & AGN & $95\pm4$ & 12.802 & J1813--126\\
J101812.9--585930 & 0.72 & 35.62 & $3.8\pm0.1$ & 0.98 & 1 & \nodata & 15.96 & $13.191\pm0.009$ & AGN & $90\pm4$ & 13.824 & J1018--589\\
J101855.6--585645$^{\rm b}$ & 0.71 & 58.64 & $9.6\pm0.2$ & 0.872 & 1 & $12.269\pm0.003$ & $10.44\pm0.02$ & $9.9\pm0.02$ & HMXB & $74\pm15$ & 2.116 & J1018--589\\
\enddata
\tablecomments{The 1$\sigma$ uncertainties are shown whenever available. The first 12 columns are the same as in Table \ref{tab:GCSMRT}. Only a portion of this table is shown here to demonstrate its form and content. A machine-readable version of the full table is available. $^{\rm a}$The shortened HESS source from the HGPS by removing the leading HESS string.  $^{\rm b}$Those sources are extended sources according to CSC flags.}
\end{deluxetable*}
\end{longrotatetable}

\bibliography{MUWCLASS.bib}{}
\bibliographystyle{aasjournal}

\end{document}